\documentclass[lettersize,journal]{IEEEtran}
\usepackage{amsmath,amsfonts}
\usepackage{algorithmic}
\usepackage{algorithm}
\usepackage{array}
\usepackage{textcomp}
\usepackage{stfloats}
\usepackage{url}
\usepackage{verbatim}
\usepackage{relsize}
\usepackage{graphicx}
\usepackage{hhline}
\usepackage{xcolor}
\usepackage{cite}
\usepackage{bm}
\usepackage{dsfont}

\DeclareMathOperator{\arctantwo}{arctan2}
\usepackage{caption}
\usepackage{subcaption}
\usepackage{fancyhdr}
\begin{document}

\title{Viewport Prediction, Bitrate Selection, and Beamforming Design for THz-Enabled 360$^{\circ}$ Video Streaming}

\author{Mehdi~Setayesh,~\IEEEmembership{Graduate Student Member\textup{,}~IEEE},~and~Vincent~W.S.~Wong,~\IEEEmembership{Fellow\textup{,}~IEEE}

\thanks{This paper has been accepted for publication in part in the \textit{Proceedings of IEEE International Conference on Multimedia and Expo~(ICME)}, Brisbane, Australia, July 2023 \cite{setayesh2023PredFramework} and the \textit{Proceedings of the IEEE International Conference on Communications~(ICC)}, Denver, CO, June 2024 \cite{mehdiICC}.  \newline
The authors are with the Department of Electrical and Computer Engineering, The University of British Columbia, Vancouver, BC, V6T 1Z4, Canada (e-mail: setayeshm@ece.ubc.ca, vincentw@ece.ubc.ca).}
}

\maketitle

\begin{abstract}
360$^{\circ}$ videos require significant bandwidth to provide an immersive viewing experience. Wireless systems using terahertz (THz) frequency band can meet this high data rate demand. However, self-blockage is a challenge in such systems. To ensure reliable transmission, this paper explores THz-enabled 360$^{\circ}$ video streaming through multiple multi-antenna access points~(APs). Guaranteeing users' quality of experience (QoE) requires accurate viewport prediction to determine which video tiles to send, followed by asynchronous bitrate selection for those tiles and beamforming design at the APs. To address users' privacy and data heterogeneity, we propose a content-based viewport prediction framework, wherein users' head movement prediction models are trained using a personalized federated learning (PFL) algorithm. To address asynchronous decision-making for tile bitrates and dynamic THz link connections, we formulate the optimization of bitrate selection and beamforming as a macro-action decentralized partially observable Markov decision process~(MacDec-POMDP) problem. To efficiently tackle this problem for multiple users, we develop two deep reinforcement learning (DRL) algorithms based on multi-agent actor-critic methods and propose a hierarchical learning framework to train the actor and critic networks. Experimental results show that our proposed approach provides a higher QoE when compared with three benchmark algorithms.
\end{abstract}


\section{Introduction} 
In the realm of 360$^{\circ}$ video streaming, users delve into an immersive visual experience using a head-mounted display~(HMD) for video playback. Compared with conventional video streaming, 360$^{\circ}$ video offers high-resolution 360$^{\circ}$ visual field across three degrees of freedom. The broader and higher-definition visual field entails that a larger number of pixels must be transmitted, thereby necessitating a significantly higher data rate~\cite{zhang2021buffer}. The abundant bandwidth available in the terahertz~(THz) frequency band can potentially overcome this challenge, especially for transmitting 360$^{\circ}$ video streams that require a high data rate in the range of gigabits per second~(Gbps) and can provide a truly immersive user experience~\cite{chaccour2022can}.\par

Wireless systems operating in THz frequency band encounter a number of challenges, including a limited communication range, channel impairment due to molecular absorption, and susceptibility to blockage by obstacles \cite{shafie2021coverage}. Moreover, when a user turns around to view another part of a 360$^{\circ}$ video with its HMD, the THz link may be blocked by the user's own body, which is known as \emph{self-blockage} \cite{chaccour2022can}. The availability of a line-of-sight (LoS) link is crucial for reliable THz communication. To improve the reliability in a THz-enabled 360$^{\circ}$ video streaming system, multiple access~points~(APs) can jointly transmit 360$^{\circ}$ videos to the users~\cite{shafie2021spectrum}.\par

In a multi-user 360$^{\circ}$ video streaming system, delivering the entire 360$^{\circ}$ video with the highest quality to all users may exceed the available bandwidth. However, at any given time, a user is watching a 360$^{\circ}$ video only from one direction. The region of the video that a user is currently watching is called a \emph{viewport}~\cite{kan2021rapt360}. To efficiently utilize the network bandwidth, it is desirable that each user receives its viewport with the maximum possible quality, rather than the entire video~frame~\cite{yaqoob2021combined}. 
Viewport prediction is a key enabler for streaming 360$^{\circ}$ videos over wireless systems. \par

Viewport prediction is categorized into content-independent and content-based approaches. The content-independent approach relies on users’ historical head movements, while the content-based approach utilizes both the video content and users' historical head movements to predict future viewports. Thus, the content-based approach can achieve higher prediction accuracy~\cite{li2022spherical}. In multi-user 360$^{\circ}$ video streaming systems, users' different viewing patterns lead to data heterogeneity. Additionally, users may be reluctant to share their historical data due to privacy concerns. To address these challenges, a personalized federated learning~(PFL) algorithm can be employed to train the viewport prediction models~\cite{zhang2021buffer}. \par

For streaming 360$^{\circ}$ videos, each video is divided into chunks in the temporal domain, with each chunk containing a few seconds of video frames. In the spatial domain, each 360$^{\circ}$ video frame is divided into tiles~\cite{kan2021rapt360}. Prefetching is used to prevent video stalling during playback. Specifically, a prefetching scheme is employed to decide when and how the tiles for the subsequent video chunks should be sent to each user~\cite{qian2018flare}. Each user asynchronously requests a new video chunk based on its buffer status. Transmitting a set of tiles for each 360$^{\circ}$ video chunk based on the predicted viewport of a user can reduce bandwidth consumption and enable a more flexible transmission mechanism through bitrate selection for the tiles~\cite{maniotis2019tile}. Note that higher viewport prediction accuracy leads to better bitrate selection for tiles and an improved prefetching scheme, thereby resulting in higher bandwidth efficiency and better quality of experience~(QoE) for the users. Additionally, since users’ future head movements can be captured as an integral component of viewport prediction, proactive detection of self-blockage occurrences becomes possible in THz-enabled 360$^{\circ}$ video streaming systems.\par

In this paper, we propose a content-based viewport prediction framework that utilizes a PFL algorithm to train the head movement prediction model. This framework is an extension of the content-based approach we proposed in~\cite{setayesh2023PredFramework}. Our proposed framework incorporates a more practical saliency detection model that can be trained without requiring the saliency map of the video frames to be part of the training dataset. Furthermore, we describe how additional tiles that cover a marginal region of a viewport can be selected and sent to the users to account for prediction errors.\par
We study 360$^\circ$ video streaming in a multi-user THz wireless system with multiple multi-antenna APs. Users' requests for video chunks give rise to an optimization problem encompassing bitrate selection for the video tiles and beamforming design at the APs. Due to the asynchronous decision-making and hierarchical structure of this problem, we formulate it as a macro-action decentralized partially observable Markov decision process (MacDec-POMDP)~\cite{xiao2022asynchronous,lyu2022multi}. To solve this problem, we propose a hierarchical deep reinforcement learning~(DRL) framework comprising two multi-agent deep deterministic policy gradient~(DDPG) algorithms. The proposed DRL framework is an extension of the approach we proposed in~\cite{mehdiICC}. In particular, we combine the viewport prediction framework with the bitrate selection and beamforming design algorithms in a 360$^\circ$ video streaming system, which is not trivial. Furthermore, we replace the weighted minimum mean square error~(WMMSE) beamforming algorithm proposed in~\cite{mehdiICC} with a multi-agent DDPG algorithm to obtain beamforming vectors in a computationally efficient manner. The main contributions of this paper are as follows:
\vspace{0.2cm}

\begin{itemize}
    \item To support reliable transmission of 360$^{\circ}$ video streaming in a THz wireless system, multiple multi-antenna APs are used for video transmission. To improve the users' QoE, we propose a 360$^{\circ}$ video streaming approach that includes (a) a viewport prediction framework to determine which video tiles to transmit and (b) two multi-agent DDPG algorithms to determine the bitrate selection of the video tiles and beamforming vectors at the APs. Fig.~\ref{our_approach} shows an illustration of our proposed 360$^{\circ}$ video streaming approach.

\begin{figure}[t]
\centering
  \includegraphics[scale=0.36]{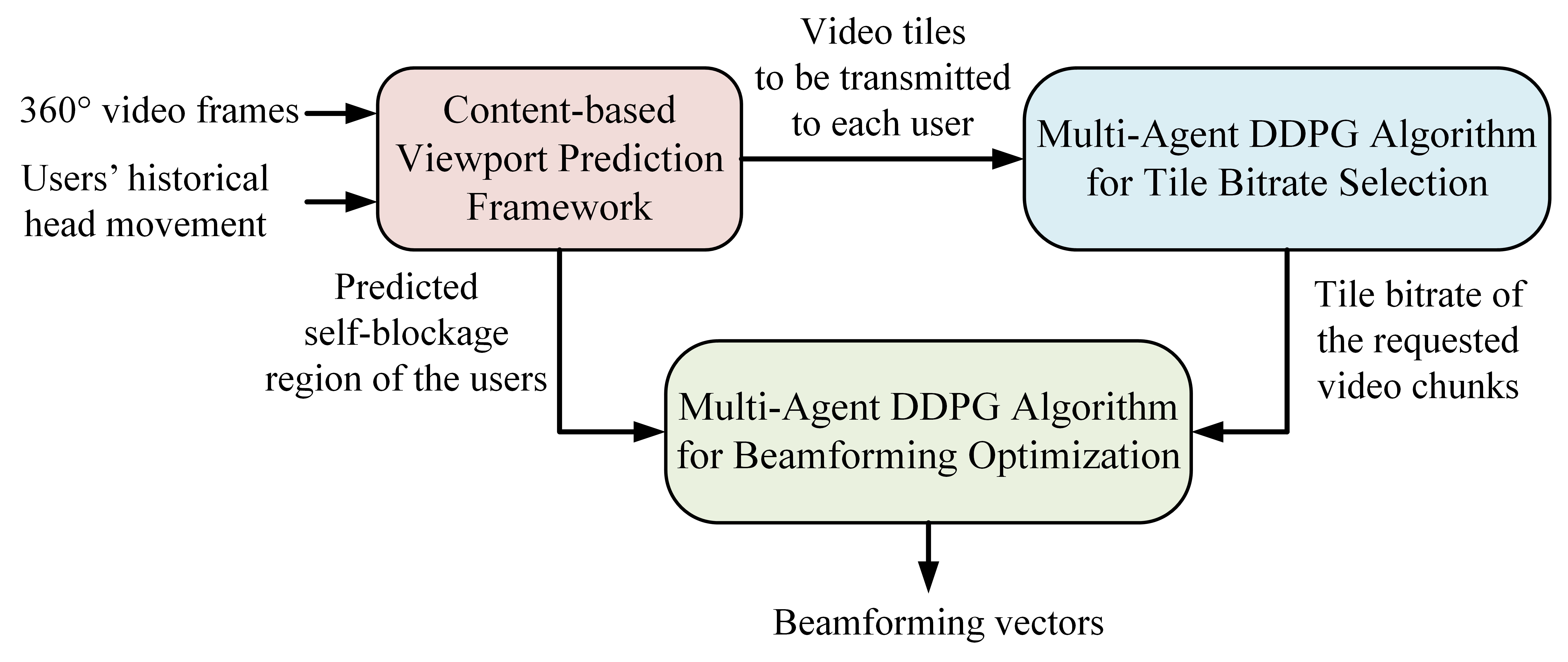}
  \caption{Our proposed 360$^{\circ}$ video streaming approach.}
  \label{our_approach}
  \vspace{0.2cm}
\end{figure}
    
   \item In particular, we propose a content-based viewport prediction framework that decouples the viewport prediction into two models. The first model focuses on saliency detection. The second model is for head movement prediction and is trained with a PFL algorithm. Due to the decoupling of these two models, any saliency detection model can be incorporated into this viewport prediction framework. With the use of PFL, our framework can address users' privacy concerns and mitigate the issue of data heterogeneity. To predict the viewport for each user, we integrate the outputs of the aforementioned models, namely the video saliency map and the user's head orientation map, through a fusion technique.
   
    \item We formulate a MacDec-POMDP problem to determine the policies for bitrate selection and beamforming optimization. Taking into account the asynchronous requests from users for new video chunks, we propose a multi-agent DDPG algorithm using a macro-action-based independent actor with individual centralized critic~(Mac-IAICC) approach. This algorithm can effectively obtain the policy for determining the bitrate of video tiles. Furthermore, we propose a multi-agent DDPG algorithm using a primitive-action-based centralized actor-critic~(Prim-CAC) approach to obtain the policy for beamforming design at the APs. To accommodate the hierarchical structure of the problem, we develop a hierarchical learning framework that facilitates training of the actor and critic networks in these algorithms.
    \item We evaluate the performance of our proposed video streaming approach using a public 360$^{\circ}$ video dataset~\cite{zhang2018saliency}. Simulation results show that  our proposed video streaming approach outperforms several benchmark algorithms in terms of the average QoE. In particular, when there are twelve users, our proposed approach provides an average QoE which is $23.77\%$, $62.7\%$, and $116.57\%$ higher than that of the video streaming with WMMSE beamforming algorithm, the combined field-of-view (FoV) tile-based adaptive streaming algorithm proposed in~\cite{yaqoob2021combined}, and the video streaming with viewport prediction proposed in~\cite{liu2021learning1}, respectively.
\end{itemize}
This paper is organized as follows. The related work is discussed in Section \ref{relW}. The system model is presented in Section \ref{sysModel}. Section~\ref{probFormulation} introduces the MacDec-POMDP problem formulation. Our proposed viewport prediction framework is presented in Section~\ref{viewportPred}. In Section~\ref{DRL_Alg_Sec}, we present our proposed DRL algorithms. Simulation results are presented in Section~\ref{Eval_Sec}. Conclusion is given in Section~\ref{conclusin_Sec}.\par
\emph{Notations}: In this paper, we represent vectors by boldface lowercase letters (e.g., $\bm{x}$), matrices by boldface uppercase letters (e.g., $\bm{X}$), and sets by calligraphic letters (e.g., $\mathcal{X}$). The cardinality of set $\mathcal{X}$ is denoted by $\lvert \mathcal{X} \rvert$. The symbol $(\cdot)^{H}$ denotes conjugate transpose operator. $\bm{I}_{N}$ denotes an identity matrix of size $N$. $\|\cdot\|$ denotes the norm of a vector. $\mathds{1}{(z \in \mathcal{Z})}$ denotes the indicator function which is equal to $1$ if $z \in \mathcal{Z}$, and is equal to zero otherwise. We define $[z]^+ = \textrm{max}\{0,\,z\}$.

\section{Related Work}\label{relW}

\subsection{Viewport Prediction}

Recently, deep neural networks (DNNs) have been incorporated into viewport prediction models to improve their prediction accuracy. Chao \emph{et al.} in \cite{chao2021transformer} proposed a transformer-based architecture to predict users' viewports. Liu \emph{et al.} in~\cite{liu2021learning} proposed long short-term memory (LSTM) and gated recurrent unit (GRU) architectures to predict future viewports. DNNs require a large amount of data for training. Thus, each user may not be able to obtain a prediction model with high accuracy using only its local data. The prediction accuracy can be improved if all users collaboratively train a shared model. Federated learning (FL) facilitates distributed training while preserving users' privacy. FedAvg~\cite{mcmahan2017communication}, which is a popular FL algorithm, has been used in~\cite{liu2021learning1} for viewport prediction. Furthermore, the issue of data heterogeneity among users' local data can be tackled using a PFL algorithm. In PFL, different users collaboratively train a shared global model. Then, each user utilizes its local data samples to fine-tune the global model and obtain a customized model~\cite{oh2021fedbabu}. Zhang \emph{et al.} in~\cite{zhang2021buffer} used a PFL algorithm based on meta-learning for viewport prediction. The aforementioned works fall into the category of content-independent viewport prediction approaches.\par
In content-based approaches, incorporating video content in viewport prediction can improve prediction accuracy by identifying the parts of the video frames that are more interesting for users to watch. Nguyen \emph{et al.} in \cite{nguyen2018your} proposed an LSTM architecture to predict the user's viewport using saliency maps of the past video frames and the user's historical head movements. Li \emph{et al.} in \cite{li2022spherical} proposed a spherical convolution-empowered model, where the users' future viewports are predicted by combining the salient spatial-temporal features of video frames with the users' historical viewport information. Wu \emph{et al.} in \cite{wu2020spherical} proposed a preference-aware viewport prediction model that utilizes an attention mechanism to combine visual features from 360$^{\circ}$ video frames with users' viewing historical data. The aforementioned content-based viewport prediction models require centralized training. Specifically, since the server has access to the previous video frames, users must provide their historical head movements to the server for viewport prediction in those models. Decoupling the viewport prediction model into a saliency detection model and a head movement prediction model as proposed in this paper can offer some advantages. First, any state-of-the-art saliency detection model can be incorporated into the viewport prediction model. Second, users’ privacy concerns and data heterogeneity issues can be addressed by using PFL for training the head movement prediction model.

\vspace{-0.1cm}
\subsection{360$^{\circ}$ Video Streaming over Wireless systems}

Recently, streaming 360$^{\circ}$ videos over wireless systems has received considerable attention. There are two main threads in related work. The first line of research aims to improve users' QoE by flexibly transmitting video tiles through adaptive bitrate selection mechanisms. The second line of research involves using new technologies, such as mobile edge computing~(MEC), rate-splitting~(RS), and millimeter wave~(mmWave) band communication, in wireless system infrastructure for video streaming. For flexible video tile transmission, the main focus is on viewport prediction, selecting the tiles that should be sent to users, and the bitrate selection for those tiles. Kan \emph{et al.} in~\cite{kan2021rapt360} proposed a viewport identification method, a viewport-aware adaptive tiling scheme, and a DRL-based rate adaptation algorithm for 360$^{\circ}$ video streaming. Yaqoob \emph{et al.} in \cite{yaqoob2021combined} proposed a combined FoV prediction-assisted 360$^{\circ}$ video streaming algorithm and a priority-based bitrate adaptation algorithm. The capability of adaptive bitrate selection mechanisms remains limited in fulfilling the data rate requirements of high-resolution 360$^{\circ}$ videos. Thus, new technologies should be incorporated into wireless system infrastructure to enable the delivery of such 360$^{\circ}$ videos and new generation of virtual reality (VR) services~\cite{hu2020cellular}.\par

The main focus of the following works is on utilizing new technologies for video streaming over wireless systems. Zhao \emph{et al.} in~\cite{zhao2022optimization} proposed iterative algorithms to determine the beamforming vectors for maximizing the weighted sum rate in a multicast RS VR streaming system. Yang \emph{et al.} in~\cite{yang2022feeling} proposed a DRL-based algorithm to improve the users' visual experience in a mmWave-enabled VR streaming system. Huang \emph{et al.} in~\cite{huang2022rate} proposed a DRL algorithm to optimize the intelligent reflecting surface (IRS) phase shifts, RS parameters, beamforming vectors, and bitrate selection of video tiles in an IRS-aided RS VR streaming system. The authors in \cite{zhao2022optimization,yang2022feeling,huang2022rate} have considered that the users' viewports are known \emph{a priori} (i.e., perfect viewport prediction). However, combining viewport prediction within a resource allocation optimization problem for 360$^{\circ}$ video streaming over wireless systems is not a trivial task. In particular, viewport prediction accuracy has an impact on bandwidth efficiency and users' QoE. Zhang \emph{et al.} in~\cite{zhang2021buffer} used a PFL-based viewport prediction and proposed a DRL algorithm for resource allocation in a multi-user MEC-enabled VR streaming system. The proposed approach in~\cite{zhang2021buffer} is based on considering only two possible quality levels (i.e., high and low) for tiles. Perfecto \emph{et al.} in~\cite{perfecto2020taming} proposed a deep recurrent neural network architecture for viewport prediction and an algorithm based on matching theory for video frame scheduling using mmWave multicast transmission. \par

The aforementioned works consider either synchronized chunk requests or single-user video streaming. However, in practical multi-user video streaming systems, users can request and download new video chunks asynchronously based on their buffer status. Asynchronous video streaming can further improve bandwidth efficiency~\cite{zhang2020adaptive}. Moreover, it is envisioned that a data rate of 6.37$-$95.55 Gbps is required for high-resolution 360$^{\circ}$ video streaming and ultimate VR, a new generation of VR services with better video quality and a multisensory experience~\cite{chaccour2022can}. Such a data rate is far beyond the maximum achievable data rate in the fifth generation (5G) wireless systems using mmWave. Migration towards higher frequency bands, e.g., THz bands, can address this challenge.

\vspace{-0.2cm}
\subsection{DRL for 360$^{\circ}$ Video Streaming}

DRL can be applied in 360$^{\circ}$ video streaming over wireless systems to learn bitrate selection and resource allocation policies. Since DRL-based algorithms can adapt well to network dynamics and provide desired solutions in a timely manner, they have recently attracted great attention. To determine the bitrates of tiles in a 360$^{\circ}$ video streaming system, the authors in~\cite{kan2021rapt360, zhang2019drl360, xiao2019deepvr} proposed different DRL-based algorithms using asynchronous advantage actor-critic, LSTM-based actor-critic, and Rainbow, respectively. These DRL-based algorithms are not multi-agent DRL. Thus, the interaction among users in a multi-user 360$^{\circ}$ video streaming system has not been explored in these works. Moreover, the DRL-based algorithms proposed in these works are not used to obtain a resource allocation policy in the wireless system. Specifically, it is assumed that the allocated throughput or bandwidth to a user in the previous time slot is given to be considered as one of the system states in the current time slot. On the other hand, the authors in \cite{zhang2021buffer, liu2021learning1, yang2022feeling} proposed different DRL-based algorithms to determine the resource allocation policy in the wireless system. In particular, the proposed DRL-based algorithms in \cite{zhang2021buffer}, \cite{liu2021learning1}, and \cite{yang2022feeling} are used for resource block allocation, IRS reflection coefficient matrix optimization, and satisfying a predefined data rate threshold for users, respectively. These works do not use DRL to obtain the bitrate selection policy. The authors in~\cite{huang2022rate} have shown that DRL can be used to jointly optimize the degrees of freedom provided by the wireless system infrastructure and the bitrate selection of video tiles, thereby improving users' QoE. However, the multi-agent DDPG algorithm proposed in~\cite{huang2022rate} cannot be used when users asynchronously request video tiles.

Asynchronous video tile requests from users make multi-user 360$^{\circ}$ video streaming a challenging problem. First, the time slot that a decision should be made for bitrate selection of tiles requested by a user may not be aligned with other users. Second, each user cannot request a new video chunk in every time slot due to its buffer status, while the wireless system requires resource allocation in each time slot. To address the first challenge, we formulate this problem as a MacDec-POMDP, and define macro-actions and a shared extrinsic reward to efficiently capture the impact of asynchronous decisions made upon each user's chunk request. The second challenge is tackled by using a hierarchical learning framework, which can consider interactions among bitrate selection and resource allocation policies at different levels of temporal abstraction.

\section{System Model}\label{sysModel}
\begin{figure}[t]
\centering
  \includegraphics[scale=0.34]{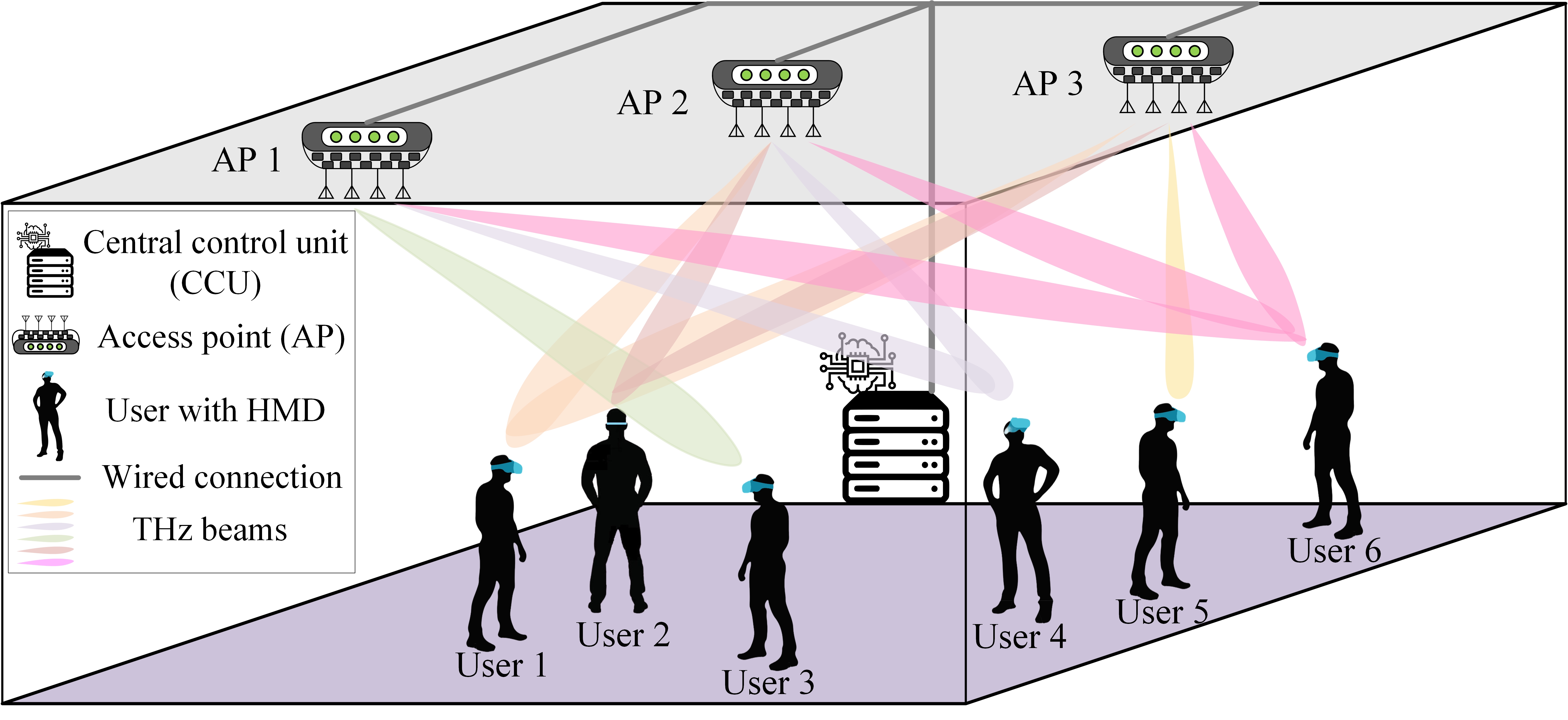}
  \caption{A THz-enabled 360$^{\circ}$ video streaming system. There are six users and three APs in the environment. 360$^{\circ}$ video streams are sent to each user by the APs which are not in the user's self-blockage region. The APs are connected to a CCU via a wired connection.}
  \vspace{0.05cm}
  \label{SysModel}
\end{figure}

Consider $U$ users who are watching 360$^{\circ}$ videos in an indoor environment, using THz wireless links as shown in Fig.~\ref{SysModel}. We denote the set of users by $\mathcal{U}=\{1,\ldots,U\}$. The users are stationary. However, they can turn around to watch different parts of the video. Each user is equipped with a wireless HMD operating at THz frequency band. Let $\bm{l}_u=\left(x_u,\,y_u,\,h_u\right)$ denote the location of the HMD that is worn by user $u \in \mathcal{U}$, where $x_u$, $y_u$, and $h_u$ denote the $x-$axis coordinate, the $y-$axis coordinate, and the height of user $u$'s HMD from the ground, respectively. In order to mitigate self-blockage of THz links, multiple APs are used to transmit the 360$^{\circ}$ video streams to the users. Let $\mathcal{A}=\{1,\ldots,N_{\text{AP}}\}$ denote the set of ceiling-mounted APs. We denote the location of AP~$a \in \mathcal{A}$ by $\bm{l}_a=\left(x_a,\,y_a,\,h^\textrm{AP}\right)$, where $x_a$ and $y_a$ denote the coordinate of AP $a$ on the $x-$ and $y-$axes, respectively, and $h^\textrm{AP}$ is the height of the ceiling from the ground. Each AP and each user's HMD are equipped with a uniform linear array~(ULA) of $N_{\textrm{t}}$ and $N_{\textrm{r}}$ antenna elements, respectively. Let $f_{\textrm{c}}$ denote the carrier frequency of the transmitted signals by the APs' antennas. The spacing between adjacent antenna elements is chosen to be $d=\frac{\lambda_{\textrm{c}}}{2}$, where $\lambda_{\textrm{c}}$ is the wavelength of carrier frequency $f_{\textrm{c}}$. \par

\subsection{THz Channel and Downlink Transmission Model}
THz band communications suffer from attenuation due to molecular absorption. Let $\kappa(f)$~(in m$^{-1}$) denote the molecular absorption coefficient at frequency $f$. $\kappa(f)$ can be predicted for a given transmission medium using the high resolution transmission molecular absorption database~(HITRAN)~\cite{jornet2011channel}. We consider that $\kappa(f)$ remains relatively constant for the frequencies within the transmission bandwidth of the APs~\cite{shafie2021spectrum,shafie2021coverage}. Moreover, THz band communications are highly directional, especially when high gain antennas are used by the APs and users. Thus, we consider only the LoS path to obtain the THz channel gain between each AP and user~\cite{mehrabian2024joint}. Let $\gamma_{u,a}$ denote the LoS path gain between AP~$a \in \mathcal{A}$ and user $u \in \mathcal{U}$. $\gamma_{u,a}$ is composed of the spreading and molecular absorption losses for the LoS path \cite{shafie2021spectrum,jornet2011channel}. We have $\gamma_{u,a} = \frac{c_{\textrm{0}}}{4\pi f_{\textrm{c}} \| \bm{l}_a -  \bm{l}_u \|}e^{-\frac{1}{2}\kappa(f_{\textrm{c}})\|\bm{l}_a -  \bm{l}_u \|}$, where $c_{\textrm{0}}$ is the speed of light. Let $\textbf{a}_a(\psi_{u,a}^\textrm{AoD}) \in \mathbb{C}^{N_{\textrm{t}}}$ and $\textbf{a}_u(\psi_{u,a}^\textrm{AoA}) \in \mathbb{C}^{N_{\textrm{r}}}$ denote the array steering vectors for ULA at AP $a$ and user $u$, respectively. $\psi_{u,a}^\textrm{AoD}$ and $\psi_{u,a}^\textrm{AoA}$ represent the angle-of-departure~(AoD) and the angle-of-arrival~(AoA) of the THz beam transmitted from AP $a$ to user $u$, respectively. We have $\textbf{a}_a(\psi_{u,a}^\textrm{AoD})=\big(1,e^{j\frac{2 \pi d}{\lambda_{\textrm{c}}} \sin(\psi_{u,a}^\textrm{AoD})},\ldots,e^{j\frac{2 \pi d}{\lambda_{\textrm{c}}}{(N_{\textrm{t}}-1)} \sin(\psi_{u,a}^\textrm{AoD})}\big)$ and $\textbf{a}_u(\psi_{u,a}^\textrm{AoA})=\big(1,e^{j\frac{2 \pi d}{\lambda_{\textrm{c}}} \sin(\psi_{u,a}^\textrm{AoA})},\ldots,e^{j\frac{2 \pi d}{\lambda_{\textrm{c}}}{(N_{\textrm{r}}-1)} \sin(\psi_{u,a}^\textrm{AoA})}\big)$. Let $\bm{G}_{u,a} \in \mathbb{C}^{N_{\textrm{t}} \times N_{\textrm{r}}}$ denote the channel gain matrix between AP~$a$ and user $u$. We have $\bm{G}_{u,a}= \sqrt{g_a g_u} \gamma_{u,a}  \textbf{a}_a(\psi_{u,a}^\textrm{AoD}) \textbf{a}^{H}_u(\psi_{u,a}^\textrm{AoA})$, where $g_a$ and $g_u$ are the antenna gains (in dBi) at AP $a$ and user $u$, respectively.\par
The APs are connected to a central control unit (CCU) via a wired connection. Given the THz LoS link availability between the APs and users, as well as the channel gain matrices, the beamforming vectors can be determined by the CCU in each time slot. Let $\mathcal{T} = \{1,2,\ldots\}$ denote the set of time slots. Each time slot has a duration of $T^\textrm{slot}$~(in millisecond~(ms)~\cite{perfecto2020taming}). Let $\bm{b}_{u,a}(t) \in \mathbb{C}^{N_{\textrm{t}}}$ denote the beamforming vector from AP~$a \in \mathcal{A}$ to user $u \in \mathcal{U}$ in time slot $t \in \mathcal{T}$. Considering $P^{\textrm{max}}$ as the maximum transmit power of an AP, we have
\vspace{0.05cm}
\begin{align}\label{pow_cons}
\sum_{u \in \mathcal{U}}{\|\bm{b}_{u,a}(t)\|^2} \leq P^{\textrm{max}}, \,\, a\in \mathcal{A},\,\, t\in \mathcal{T}.
\end{align}\par
\begin{figure}[t]
    \centering
    \includegraphics[scale=0.58]{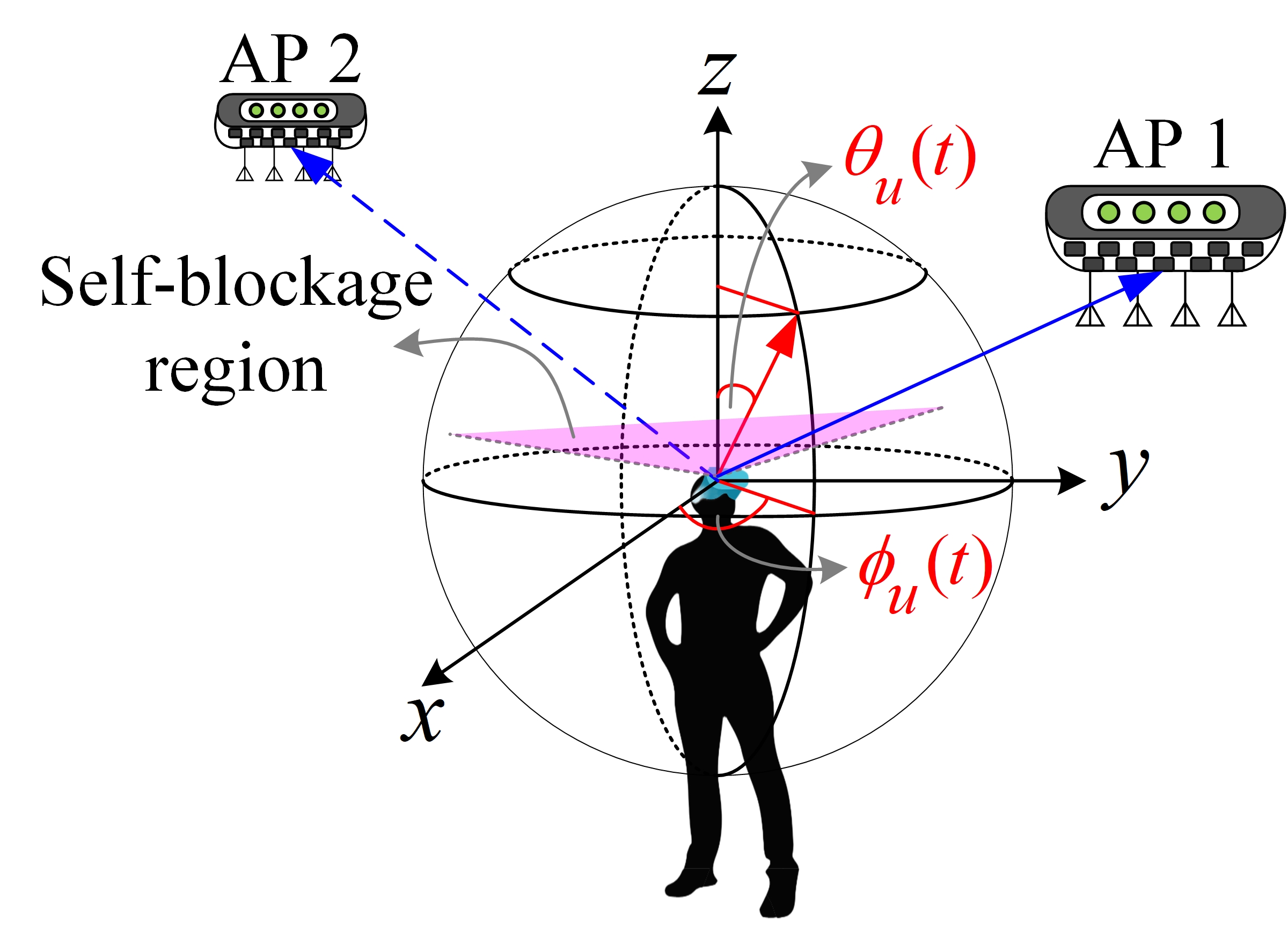}
    \caption{Illustration of the latitude and longitude angles of a user’s head, as well as its self-blockage region. AP 2 is located within the self-blockage region, whereas AP 1 is not.}
    \label{angles}
\end{figure}

\vspace{0.05cm}
An HMD has internal sensors (e.g., gyroscope, accelerometer) that enable it to track the head movement of its user~\cite{hu2021virtual}. For each user $u \in \mathcal{U}$, let $\theta_u(t)$ and $\phi_u(t)$, respectively, denote the latitude and longitude angles of its head orientation in time slot $t \in \mathcal{T}$ in the local spherical coordinate system\footnote{The local spherical coordinate system of each user $u \in \mathcal{U}$ has its origin at point $\bm{l}_u$ and its axes are aligned with the three-dimensional~(3D) Cartesian coordinate system.}. The latitude angle $\theta_u(t)$, where $0\leq\theta_u(t) \leq \pi$, is the angle measured from the $z-$axis. The longitude angle $\phi_u(t)$, where $0\leq \phi_u(t) \leq 2\pi$, is the angle measured from the $x-$axis after projection onto the $x-y$ plane. We define a self-blockage angle $\phi^\textrm{blocked}$ to characterize the self-blockage region of the users with respect to the locations of the APs~\cite{chaccour2022can, shafie2021coverage}. Fig.~\ref{angles} shows an illustration of a user's self-blockage region. Let $\mathcal{A}_u^{\textrm{nb}}(t)$ denote the set of APs which are not in the self-blockage region of user $u \in \mathcal{U}$ in time slot $t \in \mathcal{T}$. We have  
\begin{align}\label{non-block_reg}
\mathcal{A}_u^{\textrm{nb}}(t) = \left\{a \,\Big |\, a\in \mathcal{A},\, \left \lvert \phi_u(t) - \phi_{u,a} -\pi \right \rvert \geq \frac{\phi^\textrm{blocked}}{2} \right\},
\end{align}
where $\phi_{u,a}=\bmod(\arctantwo({y_a-y_u},{x_a-x_u}),2\pi)$ denotes the longitude angle of the LoS link between user $u$ and AP~$a$. $\arctantwo(\cdot,\cdot)$ returns an angle, ranging from $-\pi$ to $\pi$, that represents the angle between the positive $x-$axis and a given vector. $\bmod(\cdot,2\pi)$ is the modulo operator, which is used to obtain a value between $0$ and $2\pi$ for $\phi_{u,a}$.\par

 Let $r_{u}(t)$ denote the data rate of user $u \in \mathcal{U}$ in time slot $t \in \mathcal{T}$. For the sake of brevity, we define vector $\bm{d}_{u,u'}(t)=\sum_{a \in \mathcal{A}_{u'}^{\textrm{nb}}(t) \cap \mathcal{A}_u^{\textrm{nb}}(t)}{\bm{G}_{u,a}^{H} \bm{b}_{u',a}(t)}$. We have \cite{shi2011iteratively}:
\vspace{-0.2cm}
\begin{align}\label{Data_Rate}
r_{u}(t) =B \log_{2} \left( 1 +  \bm{d}^{H}_{u,u}(t) \bm{\Gamma}_{u}^{-1}(t) \bm{d}_{u,u}(t)   \right),
\end{align}
where $B$ is the transmission bandwidth and $\bm{\Gamma}_{u}(t)$ is the interference-plus-noise covariance matrix at user $u$. We have $\bm{\Gamma}_{u}(t)= \sum_{u' \in \mathcal{U}\backslash\{u\}}\bm{d}_{u,u'}(t) \bm{d}_{u,u'}^{H}(t) + \sigma^2 \bm{I}_{N_{\textrm{r}}}$, where $\sigma^2$ is the variance of the additive white Gaussian noise.

\begin{figure}[t]
    \centering
    \includegraphics[scale=0.524]{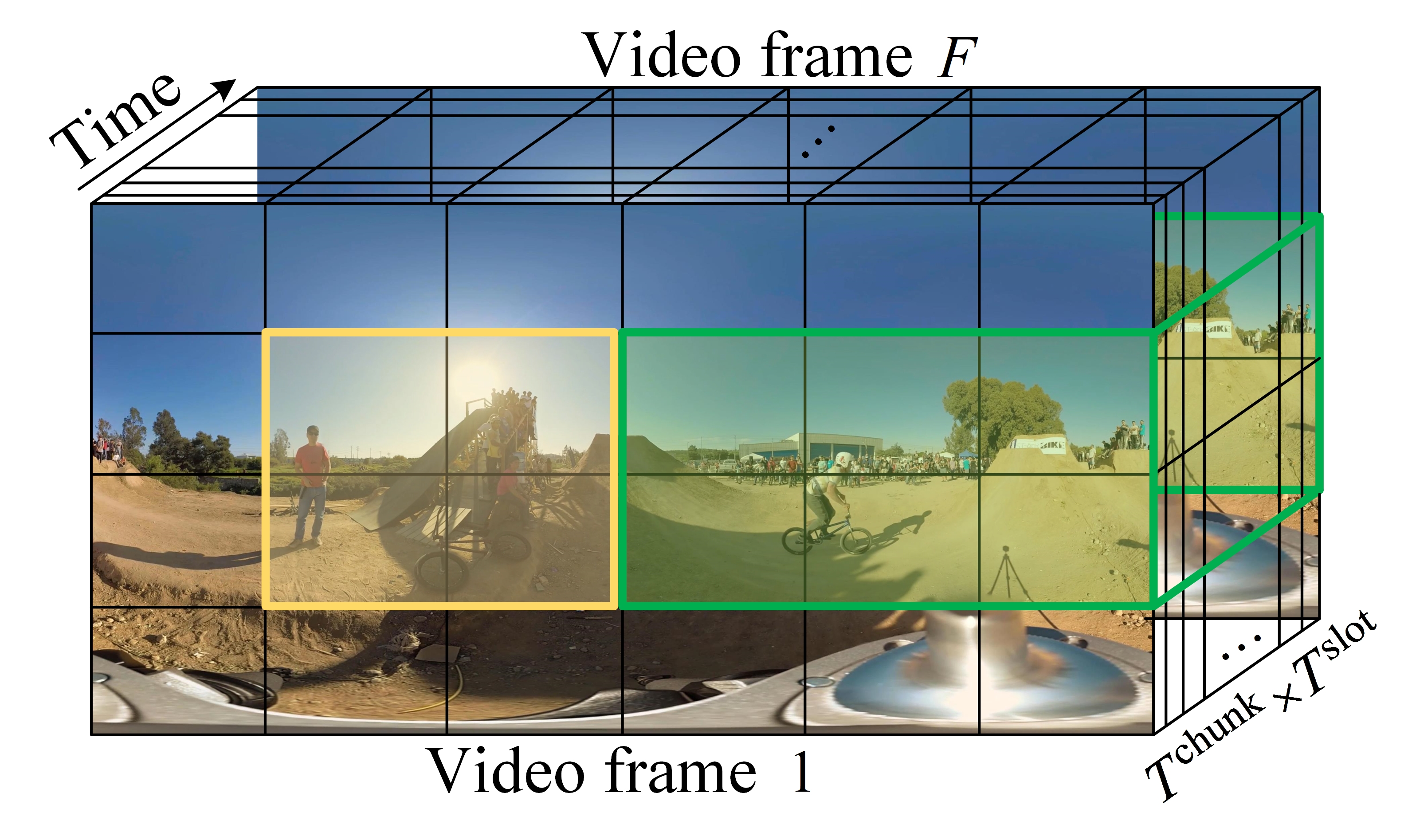}
    \caption{Illustration of a video chunk. There are $F$ video frames within each chunk. Each video frame is divided into $24$ tiles using $6 \times 4$ tiling pattern. The tiles in the viewport are shown in green, while those in the marginal region are shown in yellow. Other tiles are considered in the invisible region.}
    \label{chunks}
\end{figure}

\subsection{Tile Request and Buffer Model}\label{tiling}
To enhance bandwidth efficiency in streaming 360$^{\circ}$ videos, we leverage a tile-based approach. Let $\mathcal{V}=\{1,\ldots,V\}$ denote the set of available 360$^{\circ}$ videos. In the time domain, each video $v \in \mathcal{V}$ is segmented into $C_v$ chunks. Let $\mathcal{C}_v = \{1,\ldots,C_v\}$ denote the chunk indices for video $v$. As shown in Fig. \ref{chunks}, we consider that each video chunk has a fixed duration of $T^\textrm{chunk}$~(in time slot) and contains $F$ video frames. Let $\mathcal{F}=\{1,\ldots,F\}$ denote the set of indices of the video frames within a chunk. In the spatial domain, each video frame is divided into $N$ tiles. Let $\mathcal{N}=\{1,\ldots,N\}$ denote the set of indices corresponding to the tiles of each video frame.\par

360$^{\circ}$ videos are streamed to users as chunks. We consider a prefetching scheme in which each user downloads one chunk at a time and subsequently requests the next chunk based on its buffer status~\cite{zhang2021joint}. When a user requests a video chunk, a viewport prediction framework, to be presented in Section~\ref{viewportPred}, is utilized to predict the tiles that may be viewed by the user for that chunk. The viewport prediction framework facilitates dividing each video frame into three regions: viewport, marginal, and invisible regions. To save wireless system bandwidth, the tiles corresponding to the invisible region are not transmitted to the users. To provide a high QoE, the tiles covering both the viewport and marginal regions are transmitted to the users.  For user $u \in \mathcal{U}$ requesting chunk $c \in \mathcal{C}_{v}$ of video $v \in \mathcal{V}$, let $\mathcal{N}_{u,f,c,v}^{\textrm{view}}$ and $\mathcal{N}_{u, f, c, v}^{\textrm{marg}} \subset \mathcal{N}$ denote the set of tile indices corresponding to the viewport and marginal regions of video frame~$f \in \mathcal{F}$, respectively. Let $\mathcal{N}_{u,c,v}^\textrm{pred}$ denote the set of tile indices that are predicted to be transmitted to user $u$ upon its request for chunk $c$ of video $v$. We have $\mathcal{N}_{u,c,v}^\textrm{pred}=\cup_{f\in \mathcal{F}}\big({\mathcal{N}_{u,f,c,v}^{\textrm{view}}} \cup {\mathcal{N}_{u, f, c, v}^{\textrm{marg}}} \big)$.\par

In our proposed tile-based 360$^{\circ}$ video streaming approach, after receiving the chunk request from a user, the CCU needs to determine the quality level of the tiles. When a tile is encoded at a better quality level, it requires a higher bitrate for transmission accordingly. Let $\mathcal{M}=\{1,\ldots,M\}$ denote the set of quality levels, where the lowest quality is represented by $1$. We consider that the tiles with the same quality level have identical bitrate~\cite{huang2022rate,long2020optimal}. Let $\nu_{m}$ (in bits/s) denote the bitrate required to encode a tile at quality level $m \in \mathcal{M}$. We use the binary decision variable $\beta_{u,n,m}$ to indicate whether quality level $m \in \mathcal{M}$ is selected for tile $n \in \mathcal{N}_{u,c,v}^\textrm{pred}$ when user $u \in \mathcal{U}$ requests chunk $c \in \mathcal{C}_{v}$ of video $v \in \mathcal{V}$ ($\beta_{u,n,m}=1$) or not ($\beta_{u,n,m}=0$). We have
\begin{align}\label{quality-level_assign}
\sum_{m \in \mathcal{M}}{\beta_{u,n,m}} \leq 1, \,\, u \in \mathcal{U},\,\, n\in \mathcal{N}_{u,c,v}^\textrm{pred}, \,\, c \in \mathcal{C}_{v},\,\, v \in \mathcal{V}.
\end{align}

 The time when a user requests for a new video chunk depends on its buffer status (i.e., the playback time of the previously downloaded video chunks in its buffer). Let $\tau_{u,c,v}^\textrm{REQ}$~(in time slot) denote the time when the tiles of video chunk~$c \in \mathcal{C}_{v}$ are requested by user $u \in \mathcal{U}$ who is watching video $v \in \mathcal{V}$.  Since the chance of playback stalling depends on the video delivery time, we employ a time slot-based definition for the buffer status of the users. Let $B_{u}\left(\tau_{u,c,v}^\textrm{REQ}\right)$ (in time slot) denote the buffer status of user $u$ in time slot $\tau_{u,c,v}^\textrm{REQ}$. To prevent buffer overflow, each user only requests a new video chunk when its buffer status is below a certain threshold \cite{kan2021rapt360}. Let $B_u^\textrm{THR}$~(in time slot) denote the buffer size threshold for user $u$. When the buffer status of user $u$ is above $B_u^\textrm{THR}$, the user will wait for a period of time before requesting the next video chunk to avoid buffer overflow. Let $\tau_{u,c,v}^\textrm{WT}$~(in time slot) denote the waiting time for user $u$ after receiving chunk $c$ of video~$v$. We have
\begin{align}\label{wait_time}
\tau_{u,c,v}^\textrm{WT} = \left[\left[B_{u}\left(\tau_{u,c,v}^\textrm{REQ}\right) - \tau_{u,c,v}^\textrm{TD}\right]^{+} + T^\textrm{chunk}-B_u^\textrm{THR}\right]^{+},
\end{align}
where $\tau_{u,c,v}^\textrm{TD}$ (in time slot) denotes the time it takes for chunk~$c$ of video~$v$ to be transmitted from the APs to user $u$. It is given by $\tau_{u,c,v}^\textrm{TD} =  \min \big\{t' \in \mathcal{T} \,\big|\, \sum_{t=\tau_{u,c,v}^\textrm{REQ}}^{t'} {r_{u}(t) } \geq T^\textrm{chunk} \sum_{n \in \mathcal{N}_{u,c,v}^\textrm{pred}}\sum_{m \in \mathcal{M}}{\beta_{u,n,m} \nu_{m}} \big\} - \tau_{u,c,v}^\textrm{REQ} + 1$.\par

Considering the waiting time $\tau_{u,c,v}^\textrm{WT}$, the next chunk of video~$v \in \mathcal{V}$ is requested by user $u \in \mathcal{U}$ in the following time slot:
\begin{align}\label{next_chunk_time}
&\tau_{u,c+1,v}^\textrm{REQ} = \tau_{u,c,v}^\textrm{REQ} + \tau_{u,c,v}^\textrm{TD} +\tau_{u,c,v}^\textrm{WT},\nonumber\\
&\hspace{3.82cm}u \in \mathcal{U},\,\, c \in \mathcal{C}_{v}\backslash \{C_v\},\,\, v \in \mathcal{V}.
\end{align}
The buffer status of user $u$ at time slot $\tau_{u,c+1,v}^\textrm{REQ}$ is given by $B_u(\tau_{u,c+1,v}^\textrm{REQ})=\big[[B_{u}(\tau_{u,c,v}^\textrm{REQ}) - \tau_{u,c,v}^\textrm{TD}]^{+} + T^\textrm{chunk} -\tau_{u,c,v}^\textrm{WT}\big]^{+}$.

\subsection{Quality of Experience Model}\label{QoE_Model}
The CCU aims to maximize the users' QoE. Let $\Upsilon_{u,c,v}$ denote the QoE of user $u \in \mathcal{U}$ for chunk $c \in \mathcal{C}_v$ of video~$v \in \mathcal{V}$. We consider that $\Upsilon_{u,c,v}$ depends on four factors: the average quality of the tiles in the viewport~$\Bar{\ell}_{u,c,v}^\textrm{view}$, the spatial quality smoothness of the tiles in the viewport~${\ell}_{u,c,v}^\textrm{spatial}$, the temporal quality smoothness of the tiles in the viewport~${\ell}_{u,c,v}^\textrm{temp}$, and the rebuffering delay $\tau_{u,c,v}^\textrm{RD}$~\cite{yaqoob2021combined}. In the following, we describe how CCU obtains each of these QoE factors.\par

Let $\mathcal{N}_{u, c, v}^{\textrm{actual}}$ denote the set of tile indices that user $u \in \mathcal{U}$ has actually viewed for chunk $c \in \mathcal{C}_v$ of video $v \in \mathcal{V}$. $\Bar{\ell}_{u,c,v}^\textrm{view}$ is obtained by averaging the quality of the tiles in set $\mathcal{N}_{u, c, v}^{\textrm{actual}}$. We have $\Bar{\ell}_{u,c,v}^\textrm{view} = \frac{1}{\left\lvert \mathcal{N}_{u, c, v}^{\textrm{actual}} \right\rvert}{\sum_{n\in \mathcal{N}_{u, c, v}^{\textrm{actual}}}\sum_{m \in \mathcal{M}}{\beta_{u,n,m} m}}$.
\par

The spatial quality smoothness factor measures the intra-chunk quality switch. In particular, the variance of the quality level of the tiles in the viewport, i.e., ${\ell}_{u,c,v}^\textrm{spatial}$, may lead to viewing irritation, cybersickness, and other physiological effects including fatigue, nausea, and aversion \cite{yaqoob2021combined}. ${\ell}_{u,c,v}^\textrm{spatial}$ is obtained as ${\ell}_{u,c,v}^\textrm{spatial} = \frac{1}{\left\lvert \mathcal{N}_{u, c, v}^{\textrm{actual}} \right\rvert}{\sum_{n\in \mathcal{N}_{u, c, v}^{\textrm{actual}}}\left(\sum_{m \in \mathcal{M}}{\beta_{u,n,m} m}-\Bar{\ell}_{u,c,v}^\textrm{view} \right)^2}$.\par

The temporal quality smoothness factor measures the inter-chunk quality switch. The user's QoE degrades when the average quality level of the tiles in the viewport differs between two consecutive chunks \cite{kan2021rapt360}. For $c=1$, we set ${\ell}_{u,c,v}^\textrm{temp}=0$. For $c>1$, we have ${\ell}_{u,c,v}^\textrm{temp} = \left\lvert \Bar{\ell}_{u,c,v}^\textrm{view} - \Bar{\ell}_{u,c-1,v}^\textrm{view} \right\rvert$. \par

The rebuffering delay $\tau_{u,c,v}^\textrm{RD}$ captures video stalling during playback \cite{kan2021rapt360}. A video is stalled when the downloading time of chunk $c$ exceeds the user's buffer status at chunk $c$'s request time. We have $\tau_{u,c,v}^\textrm{RD} = \left[ \tau_{u,c,v}^\textrm{TD} - B_{u}\left(\tau_{u,c,v}^\textrm{REQ}\right) \right]^{+}$. \par

The QoE of user $u\in \mathcal{U}$ for chunk $c \in \mathcal{C}_v$ of video $v \in \mathcal{V}$ is the weighted sum of the mentioned factors. We have
\begin{align}\label{chunk_QoE}
&\Upsilon_{u,c,v} = \Bar{\ell}_{u,c,v}^\textrm{view} \hspace{-0.25mm}- \lambda^{\textrm{spatial}} {\ell}_{u,c,v}^\textrm{spatial} \hspace{-0.25mm}- \lambda^{\textrm{temp}} {\ell}_{u,c,v}^\textrm{temp} \hspace{-0.25mm}- \lambda^{\textrm{RD}} \tau_{u,c,v}^\textrm{RD}, \nonumber\\
&\hspace{4.77cm}u \in \mathcal{U},\,\, c \in \mathcal{C}_{v},\,\, v \in \mathcal{V},
\end{align}
where $\lambda^{\textrm{spatial}} $, $\lambda^{\textrm{temp}}$, and $\lambda^{\textrm{RD}}$ are the non-negative weighting coefficients which penalize user~$u$'s QoE due to the nonzero  intra-chunk quality switch,  inter-chunk quality switch, and rebuffering delay, respectively.   
\section{Optimizing Tile Bitrate Selection and Beamforming Design}\label{probFormulation}
Each user asynchronously requests a new video chunk based on its buffer status (see eqn.~(\ref{next_chunk_time})). Upon receiving a chunk request, the CCU chooses an action (i.e., tile bitrate selection) given the observed system state information. The CCU aims to maximize the users' expected long-term QoE. To this end, the CCU should make \emph{asynchronous} decisions on the quality level of the tiles requested by the users. Furthermore, the design of beamforming vectors in each time slot affects the rebuffering delay. At the beginning of each time slot, the CCU obtains an observation from the system and chooses an action (i.e., beamforming design) based on the selected bitrate for tiles. Given the asynchronous decision-making and hierarchical structure of this problem, we formulate it as a MacDec-POMDP~\cite{xiao2022asynchronous,lyu2022multi} with $T^\textrm{max}$ decision epochs. In particular, we consider that for each user, an agent in the CCU is responsible to make decision on behalf of that user\footnote{Considering one agent for each user makes our approach scalable. It also facilitates compatibility with existing adaptive bitrate streaming techniques.}. The agents cooperatively determine the bitrate of tiles by taking asynchronous macro-actions. The agents also cooperatively design the beamforming vectors through synchronized primitive-actions. Next, we describe the observation, action, and reward of each agent.

\vspace{-0.2cm}
\subsection{Observation}\label{observation}
The global system state comprises the set of indices of the tiles that are actually viewed by the users~(i.e., $\mathcal{N}_{u, c, v}^{\textrm{actual}}$). It also contains the set of APs which are not in the users' self-blockage region in each time slot~(i.e., $\mathcal{A}_u^{\textrm{nb}}(t)$). However, the CCU does not have access to the global system state. Instead, it obtains a partial observation
of the underlying system state. Among the $N$ available tiles, let binary vector $\bm{\upsilon}_{u,c,v} \in \{0,1\}^{N}$ indicate the tile indices predicted by the viewport prediction framework to be transmitted to user~$u \in \mathcal{U}$ upon its request for chunk~$c \in \mathcal{C}_v$ of video $v \in \mathcal{V}$. The $n$-th element of vector $\bm{\upsilon}_{u,c,v}$ is equal to $\mathds{1}{(n \in \mathcal{N}_{u,c,v}^\textrm{pred})}$. As we will discuss in Section \ref{Integration_technique}, our proposed viewport prediction framework provides a fused feature map for each video frame by integrating the video saliency map and the user's head orientation map. We segment the fused feature map into $N$ tiles. A higher feature value assigned to a tile on the map indicates that it is more likely for the user to view that tile in the corresponding video frame. We normalize the feature value of the tiles in the fused feature map to be within $[0,1]$. Let vector $\bm{\Bar{\chi}}_{u,c,v} \in [0,1]^{N}$ denote the average feature value of tiles when chunk $c$ of video $v$ is requested by user $u$. The $n$-th element of vector $\bm{\bar{\chi}}_{u,c,v}$ is equal to the average feature value of the $n$-th tile across all video frames in chunk $c$. We also denote the average quality level of the tiles transmitted to user $u$ upon its request for chunk $c$ of video $v$ by $\Bar{\ell}_{u,c,v}^\textrm{trans}$. We have $\Bar{\ell}_{u,c,v}^\textrm{trans}=\frac{1}{\left\lvert \mathcal{N}_{u,c,v}^\textrm{pred} \right\rvert}{\sum_{n\in \mathcal{N}_{u,c,v}^\textrm{pred}}\sum_{m \in \mathcal{M}}{\beta_{u,n,m} m}}$.\par

For the sake of brevity, we refer to the agent responsible to make decision on behalf of user $u \in \mathcal{U}$ as agent~$u$. Let $\bm{o}_u^{\textrm{m}}(t)$ denote the macro-observation vector of agent $u$ at the beginning of time slot~$t \in \mathcal{T}$. When chunk $c \in \mathcal{C}_v$ of video $v \in \mathcal{V}$ is requested by user $u$, the macro-observation vector of agent $u$ contains $\bm{\upsilon}_{u,c,v}$ and $\bm{\bar{\chi}}_{u,c,v}$. $\bm{o}_u^{\textrm{m}}(t)$ also contains the average quality level of the previously transmitted video chunk to user $u$ (i.e., $\Bar{\ell}_{u,c-1,v}^\textrm{trans}$) and the buffer status of user $u$ in time slot $\tau_{u,c,v}$~(i.e., $B_{u}\left(\tau_{u,c,v}^\textrm{REQ}\right)$). For $t \in \{\tau_{u,c,v}^\textrm{REQ},\tau_{u,c,v}^\textrm{REQ}+1.\ldots,\tau_{u,c+1,v}^\textrm{REQ}-1\}$, we have $\bm{o}_u^{\textrm{m}}(t)=\left(\bm{\upsilon}_{u,c,v},\, \bm{\bar{\chi}}_{u,c,v},\,\Bar{\ell}_{u,c-1,v}^\textrm{trans},\, B_{u}\left(\tau_{u,c,v}^\textrm{REQ}\right) \right)$. Let $\mathcal{O}_u^{\textrm{m}}$ denote the macro-observation space over agent~$u$. We have $\mathcal{O}_u^{\textrm{m}} = \{0,1\}^{N} \times [0,1]^{N} \times [1,M] \times \{0,\ldots, B_u^\textrm{THR}\}$. \par

Each agent $u \in \mathcal{U}$ selects the bitrate of the tiles in set~$\mathcal{N}_{u,c,v}^\textrm{pred}$ based on its macro-observation vector $\bm{o}_u^{\textrm{m}}(t) \in \mathcal{O}_u^{\textrm{m}}$. At the beginning of each time slot~$t \in \mathcal{T}$, the agents design the beamforming vectors based on their primitive-observation vector. We denote agent $u$'s primitive-observation vector at the beginning of time slot~$t$ by $\bm{o}_u^{\textrm{p}}(t)$. At time slot $t$, the CCU does not have access to ${\phi}_u(t)$. Instead, using our proposed viewport prediction framework, the CCU obtains the predicted longitude angle of user~$u$'s head orientation (i.e. $\hat{\phi}_u(t)$). Thus, the CCU can \emph{proactively} determine the self-blockage region of user $u$. We denote the set of APs which are predicted to be not in user~$u$'s self-blockage region in time slot $t$ by $\hat{\mathcal{A}}_u^{\textrm{nb}}(t)$. Let binary vector $\bm{\varrho}_{u}(t) \in \{0,1\}^{N_{\text{AP}}}$ indicate the APs which are not in the self-blockage region of user $u$ in time slot $t$. The $a$-th element of vector $\bm{\varrho}_{u}(t)$ is equal to $\mathds{1}{(a \in \hat{\mathcal{A}}_u^{\textrm{nb}}(t))}$. \par

Let $\Delta_u^\textrm{rem}(t)$ denote the remaining bits of the requested chunk available for transmission to user~$u \in \mathcal{U}$ in time slot $t \in \mathcal{T}$. At time slot~$t=\tau_{u,c,v}^\textrm{REQ}$, $\Delta_u^\textrm{rem}(t)$ is initialized to be $T^\textrm{chunk} T^\textrm{slot} \sum_{n \in \mathcal{N}_{u,c,v}^\textrm{pred}}{\sum_{m \in \mathcal{M}}{\beta_{u,n,m} \nu_{m}}}$. For $t \in \{\tau_{u,c,v}^\textrm{REQ}+1,\tau_{u,c,v}^\textrm{REQ}+2,\ldots,\tau_{u,c+1,v}^\textrm{REQ}-1 \}$, $\Delta_u^\textrm{rem}(t)$ is updated as $
\Delta_u^\textrm{rem}(t) = \left[ \Delta_u^\textrm{rem}(t-1) - T^\textrm{slot} {r_{u}(t-1) } \right]^{+}$.
We also define parameter $\Delta_u^\textrm{time}(t)$ as the remaining time that user $u$ can obtain its requested chunk without experiencing video stalling during playback. We initialize $\Delta_u^\textrm{time}(t)=B_{u}\left(\tau_{u,c,v}^\textrm{REQ}\right)$ at time slot $t=\tau_{u,c,v}^\textrm{REQ}$. For $t \in \{\tau_{u,c,v}^\textrm{REQ}+1,\tau_{u,c,v}^\textrm{REQ}+2,\ldots,\tau_{u,c+1,v}^\textrm{REQ}-1 \}$, $\Delta_u^\textrm{time}(t)$ is updated as $\Delta_u^\textrm{time}(t) = \Delta_u^\textrm{time}(t-1) - 1$  when $\Delta_u^\textrm{rem}(t)>0$, and is set to $\left[B_{u}\left(\tau_{u,c,v}^\textrm{REQ}\right) - \tau_{u,c,v}^\textrm{TD}\right]^{+} + T^\textrm{chunk}$ when $\Delta_u^\textrm{rem}(t)=0$. \par

At the beginning of time slot $t \in \mathcal{T}$, we have $\bm{o}_u^{\textrm{p}}(t) = \big(\bm{\varrho}_{u}(t),\, \Delta_u^\textrm{rem}(t),\,\Delta_u^\textrm{time}(t)\big)$ for each agent $u \in \mathcal{U}$. We denote the primitive-observation space over agent $u$ by $\mathcal{O}_u^{\textrm{p}} = \{0,1\}^{N_{\text{AP}}} \times \mathbb{R}^{+} \cup \{0\} \times \mathbb{Z}^{-} \cup \{0,\ldots, B_u^\textrm{THR} + T^\textrm{chunk} - 1 \}$. Agent~$u$ takes a primitive-action in time slot $t$ based on $\bm{o}_u^{\textrm{p}}(t) \in \mathcal{O}_u^{\textrm{p}}$.
\subsection{Action}\label{action}
When user $u \in \mathcal{U}$ requests chunk $c \in \mathcal{C}_v$ of video $v \in \mathcal{V}$ at time slot $t=\tau_{u,c,v}^\textrm{REQ}$, agent $u$ will take a macro-action which remains unchanged for $t \in \{\tau_{u,c,v}^\textrm{REQ},\tau_{u,c,v}^\textrm{REQ}+1,\ldots,\tau_{u,c+1,v}^\textrm{REQ}-1\}$. Let $\bm{a}_u^\textrm{m}(t)$ denote the macro-action which is taken by agent $u$ at time slot $t$. From Section \ref{tiling}, recall that the quality level of the tiles is selected from set $\mathcal{M}$. Let $\nu_{u,n}$ denote the bitrate of tile $n \in \mathcal{N}_{u,c,v}^\textrm{pred}$ when user $u$ requests chunk $c$ of video $v$. We set $\nu_{u,n}=0$ for $n \notin \mathcal{N}_{u,c,v}^\textrm{pred}$. Agent $u$ selects the bitrate of tile $n \in \mathcal{N}_{u,c,v}^\textrm{pred}$ using the following relaxed constraint:
\begin{align}\label{relaxed_bitrate}
\nu_{1} \leq \nu_{u,n} \leq \nu_{M}, \,\, u \in \mathcal{U},\,\, n\in \mathcal{N}_{u,c,v}^\textrm{pred}, \,\, c \in \mathcal{C}_{v},\,\, v \in \mathcal{V}.
\end{align}
Thus, we have $\bm{a}_u^\textrm{m}(t)= \left(\nu_{u,n},\, n \in \mathcal{N}\right)$ for  $u\in \mathcal{U}$ and $t \in \{\tau_{u,c,v}^\textrm{REQ},\tau_{u,c,v}^\textrm{REQ}+1,\ldots,\tau_{u,c+1,v}^\textrm{REQ}-1\}$. After determining $\nu_{u,n}$, agent $u$ rounds it down to the nearest possible bitrate based on the quality levels available in set $\mathcal{M}$. Note that the variables $\beta_{u,n,m}$, $u \in \mathcal{U}$, $n \in \mathcal{N}$, $m \in \mathcal{M}$ can be determined using the obtained bitrate for the tiles. We denote the macro-action space over agent $u$ by $\mathcal{A}_u^\textrm{m}=\{\{0\} \cup [\nu_{1}, \nu_{M}]\}^{N}$. Each agent ends its previous macro-action upon receiving a new video chunk request. \par

At the beginning of each time slot $t \in \mathcal{T}$, agent $u \in \mathcal{U}$ will take a primitive-action to determine the beamforming vectors. Let $\bm{a}_u^\textrm{p}(t)$ denote the primitive-action which is taken by agent $u$ at the beginning of time slot $t$. We have $\bm{a}_u^\textrm{p}(t) = \left(\bm{b}_{u,a}(t),\, a \in \mathcal{A}\right)$, $u \in \mathcal{U}$, $t \in \mathcal{T}$. We denote the primitive-action space over agent $u$ by $\mathcal{A}_{u}^\textrm{p}=\mathbb{C}^{N_{\textrm{t}} \times N_{\text{AP}}}$. 
\subsection{Reward}\label{reward}

The agents aim to cooperatively maximize the QoE of the users. In each time slot~$t \in \mathcal{T}$, we consider a \emph{shared} extrinsic reward over agents denoted by $R^\textrm{extr}(t)$. Given $\Upsilon_{u,c,v}$ from eqn. (\ref{chunk_QoE}) as the QoE of user $u \in \mathcal{U}$ for chunk $c \in \mathcal{C}_v$ of video $v \in \mathcal{V}$, we have 
\begin{align}\label{high-level-reward}
R^\textrm{extr}(t) = \sum_{u\in \mathcal{U}} \sum_{v\in \mathcal{V}} \sum_{c \in \mathcal{C}_v} \Upsilon_{u,c,v}\mathds{1}{\left( \tau_{u,c,v}^\textrm{REQ} + \tau_{u,c,v}^\textrm{TD} = t \right)}.
\end{align}
Based on the defined $R^\textrm{extr}(t)$ in (\ref{high-level-reward}), the agents receive a non-zero reward in time slot $t$ when at least one of the users has completely downloaded its requested video chunk\footnote{In this work, we use a 360$^{\circ}$ video dataset that includes the users' viewport during video playback. Thus, the CCU can obtain $\Upsilon_{u,c,v}$ once user $u$ has downloaded chunk $c$ of video $v$. However, without such a dataset, the CCU can obtain $\Upsilon_{u,c,v}$ only after user $u$ has watched chunk $c$ of video $v$ from its playback buffer. Hence, obtaining a policy without using a recorded 360$^{\circ}$ video dataset may lead to a delayed reward environment \cite{han2022off}. While dealing with delayed rewards is not within the scope of this work, this issue can be addressed by redefining the reward function and the observation-action trajectories in our proposed algorithms.}.  

By performing macro-action $\bm{a}_u^\textrm{m}(t) \in \mathcal{A}_u^\textrm{m}$ in time slot~$t=\tau_{u,c,v}$, the number of bits required for transmitting chunk~$c \in \mathcal{C}_v$ of video $v \in \mathcal{V}$ to user $u$ can be determined. The agents aim to maximize the system's sum-rate in each time slot while satisfying the selected bitrate of the requested video tiles for users through a sequence of primitive actions. The agents receive an intrinsic reward, denoted by $R^\textrm{intr}(t)$, for performing primitive-actions $\bm{a}_u^\textrm{p}(t) \in \mathcal{A}_{u}^\textrm{p}$, $u \in \mathcal{U}$ in time slot $t \in \mathcal{T}$. $R^\textrm{intr}(t)$ is shared over the agents and is obtained as follows: 
\begin{align}\label{low-level-reward}
\hspace{-0.15cm}R^\textrm{intr}(t) = \sum_{u\in \mathcal{U}}{ {r_{u}(t) } } -\lambda^{\textrm{intr}} \sum_{u\in \mathcal{U}}{ \left[ \Delta_u^\textrm{rem}(t) - T^\textrm{slot}{r_{u}(t) } \right]^{+}},
\end{align}
where $\lambda^{\textrm{intr}}$ is a positive scaling factor.
\subsection{Problem Formulation}\label{problem}
In a POMDP, an agent does not have direct access to the underlying system state. However, a history of observations and actions provides sufficient statistics for the agent to make decisions~\cite{yu2023asynchronous,lillicrap2015continuous}. Let ${\bm{h}}_u^{\textrm{m}}(t)$ and ${\bm{h}}_u^{\textrm{p}}(t)$, respectively, denote the macro- and primitive-observation-action history of agent~$u$ in time slot $t \in \mathcal{T}$. Both macro- and primitive-actions are high-dimensional continuous control variables. Each agent $u$ selects a macro-action $\bm{a}_u^\textrm{m}(t) = {\mu}_u\left({\bm{h}}_u^{\textrm{m}}(t)\right)$ using a deterministic policy ${\mu}_u: {\mathcal{H}}_u^{\textrm{m}} \rightarrow \mathcal{A}_u^\textrm{m}$ based on its macro-observation-action history ${\bm{h}}_u^{\textrm{m}}(t) \in {\mathcal{H}}_u^{\textrm{m}}$ in time slot $t$. Agent $u$ also uses a deterministic policy ${\pi}_u: {\mathcal{H}}_u^{\textrm{p}} \rightarrow \mathcal{A}_u^\textrm{p}$ to select a primitive-action $\bm{a}_u^\textrm{p}(t) ={\pi}_u\left({\bm{h}}_u^{\textrm{p}}(t)\right)$ given ${\bm{h}}_u^{\textrm{p}}(t) \in {\mathcal{H}}_u^{\textrm{p}}$ in time slot~$t$. Let $\bm{{\mu}} = \left({\mu}_u,\, u\in \mathcal{U} \right)$ and $\bm{{\pi}}=\left({\pi}_u,\, u\in \mathcal{U}\right)$ denote the agents' joint policies for bitrate selection and beamforming design, respectively. We also denote the joint macro- and primitive-actions performed by all agents in time slot $t$ by ${\bm{a}}^\textrm{m}(t) = \left(\bm{a}_u^\textrm{m}(t),\, u\in \mathcal{U} \right)$ and ${\bm{a}}^\textrm{p}(t)=\left(\bm{a}_u^\textrm{p}(t),\, u\in \mathcal{U}\right)$, respectively. \par

By employing a hierarchical learning framework \cite{kulkarni2016hierarchical},
both $\bm{{\mu}}$ and $\bm{{\pi}}$ are learned in order to maximize the expected discounted extrinsic and intrinsic rewards, respectively. Let $Q^{*}\left({{\bm{h}}}^{\textrm{m}},{\bm{a}}^\textrm{m}\right)$ denote the maximum action-value function when agents choose the joint macro-action ${\bm{a}}^\textrm{m}$ given the joint macro-observation-action history ${{\bm{h}}}^{\textrm{m}}$. For tile bitrate selection, we formulate the following optimization problem:
\begin{align}
&\mathcal{P}^\textrm{m}:\,\,\,Q^{*}\left({{\bm{h}}}^{\textrm{m}},{\bm{a}}^\textrm{m}\right) = \underset{\bm{{\mu}}}{\textrm{maximize}}\,\, \mathbb{E}_{\bm{{\mu}}} \Bigg\{\sum_{t' = t}^{T^\textrm{max}}\gamma^{t'-t}R^\textrm{extr}(t') \Bigm|\nonumber\\
&{{\bm{h}}}^{\textrm{m}}(t) = {{\bm{h}}}^{\textrm{m}},\,{\bm{a}}^\textrm{m}(t)={\bm{a}}^\textrm{m} \Bigg\}\hspace{1.2cm}\textrm{subject to constraint (\ref{relaxed_bitrate})}, \nonumber
\end{align}
where $\gamma$ is the discount factor. Problem $\mathcal{P}^\textrm{m}$ aims to learn a joint policy that maximizes the expected discounted extrinsic reward over a sequence of macro-actions when starting from ${{\bm{h}}}^{\textrm{m}}(t) = {{\bm{h}}}^{\textrm{m}} \in \prod_{u \in \mathcal{U}}{\mathcal{H}}_u^{\textrm{m}}$, having the agents take ${\bm{a}}^\textrm{m}(t)={\bm{a}}^\textrm{m} \in \prod_{u \in \mathcal{U}}{\mathcal{A}_u^\textrm{m}}$, and thereafter following policy $\bm{{\mu}}$.\par

Let $Q^{*}\left({\bm{h}}^{\textrm{p}},{\bm{a}}^\textrm{p};{\bm{a}}^\textrm{m}\right)$ denote the maximum action-value function when the agents choose the joint primitive-action ${\bm{a}}^\textrm{p}$ given the joint primitive-observation-action history ${{\bm{h}}}^{\textrm{p}}$ and the specified joint macro-action ${\bm{a}}^\textrm{m}$. The beamforming problem can be formulated as follows:
\begin{align}
&\mathcal{P}^\textrm{p}:\,\,\, Q^{*}\left({\bm{h}}^{\textrm{p}},{\bm{a}}^\textrm{p};{\bm{a}}^\textrm{m}\right)=\underset{\bm{{\pi}}}{\textrm{maximize}}\,\, \mathbb{E}_{\bm{{\pi}}} \Bigg\{\sum_{t' = t}^{T^\textrm{max}}\gamma^{t'-t}R^\textrm{intr}(t') \Bigm| \nonumber\\
&\hspace{3.21cm} {\bm{h}}^{\textrm{p}}(t) = {\bm{h}}^{\textrm{p}},\, {\bm{a}}^\textrm{p}(t)={\bm{a}}^\textrm{p},\, {\bm{a}}^\textrm{m}(t)={\bm{a}}^\textrm{m} \Bigg\} \nonumber \\
&\hspace{3.57cm}\textrm{subject to constraint (\ref{pow_cons})}.\nonumber 
\end{align}
 Problem $\mathcal{P}^\textrm{p}$ aims to learn a joint policy that leads to the maximum expected discounted intrinsic reward when starting from ${{\bm{h}}}^{\textrm{p}}(t) = {{\bm{h}}}^{\textrm{p}}  \in \prod_{u \in \mathcal{U}}{\mathcal{H}}_u^{\textrm{p}}$, given the joint macro-action ${\bm{a}}^\textrm{m} \in \prod_{u \in \mathcal{U}}{\mathcal{A}_u^\textrm{m}}$ executed by the agents in time slot~$t$. In problem $\mathcal{P}^\textrm{p}$, the expected discounted intrinsic reward is maximized over a sequence of primitive-actions, when the agents take ${\bm{a}}^\textrm{p}(t)={\bm{a}}^\textrm{p} \in \prod_{u \in \mathcal{U}}{\mathcal{A}_u^\textrm{p}}$, and thereafter follow policy $\bm{{\pi}}$. Problems $\mathcal{P}^\textrm{m}$ and $\mathcal{P}^\textrm{p}$ are finite-horizon stochastic optimal control problems. These problems are difficult to solve due to their asynchronous decision-making and hierarchical structure. Moreover, the dynamics of users' viewports and their connections to APs are not known in advance. To address these challenges, we first propose a viewport prediction framework in Section \ref{viewportPred} to predict the tiles in set $\mathcal{N}_{u,c,v}^\textrm{pred}$ as well as the head orientation of each user $u$. Then, in Section~\ref{DRL_Alg_Sec}, we present two DRL algorithms based on multi-agent actor-critic methods to enable efficient macro- and primitive-action selection for the agents. These two algorithms are developed within a hierarchical learning framework to learn the joint policies $\bm{{\mu}}$ and $\bm{{\pi}}$ for problems $\mathcal{P}^\textrm{m}$ and $\mathcal{P}^\textrm{p}$, respectively.

\section{Viewport Prediction Framework}\label{viewportPred}
In this section, we follow the idea proposed in our previous work \cite{setayesh2023PredFramework} to develop a content-based viewport prediction framework. In particular, our proposed viewport prediction framework consists of three main components: a saliency detection model, a head movement prediction model, and an integration mechanism using fusion techniques. In the following subsections, we describe each of these three components.

\vspace{-0.155cm}
\subsection{Saliency Detection Model} \label{SalModel}
A saliency detection model determines the saliency map for each video frame. This map helps identify those parts of a video frame that are more interesting to users. In this work, we use the PAVER model, proposed in \cite{yun2022panoramic}, for saliency detection. The PAVER model can be trained without requiring explicit supervision. In particular, when using this model, it is not required that the saliency map of the video frames be part of the training dataset. This is important because the availability of saliency datasets is limited, and collecting saliency maps for 360$^\circ$ videos through human supervision is expensive~\cite{nguyen2018your}. Another advantage of PAVER is its ability to leverage the pretrained weights of a vision transformer (ViT) model, which has been trained on 2D images or videos. By using these pretrained weights, the saliency detection performance of the PAVER model on 360$^{\circ}$ video frames can be improved. \par

\begin{figure}[t]
    \centering
    \includegraphics[scale=0.396]{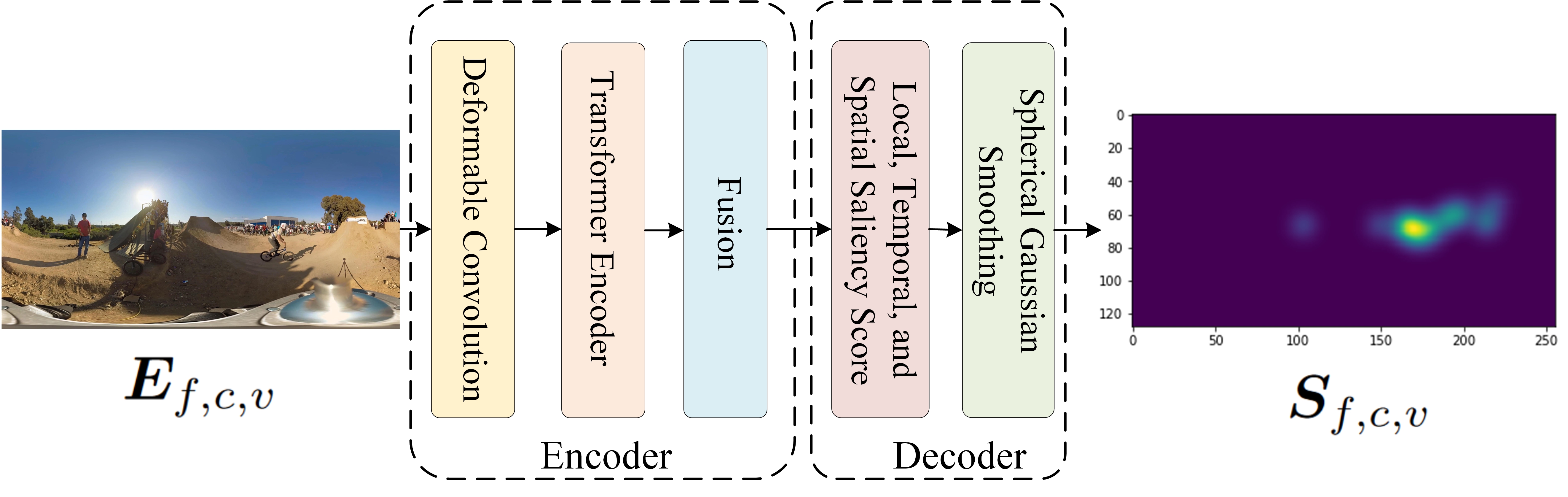}
    \caption{PAVER model for saliency detection.}
    \label{Paver}
\end{figure}

 Let $\bm{E}_{f,c,v} \in \mathbb{R}^{3 \times W \times H}$ and $\bm{S}_{f,c,v} \in \mathbb{R}^{W \times H}$ denote frame~$f \in \mathcal{F}$ of chunk $c \in \mathcal{C}_v$ of 360$^{\circ}$ video $v \in \mathcal{V}$ in equirectangular projection (ERP) format and its corresponding saliency map, respectively. $W \times H$ denotes the video frame resolution (i.e., the  number of pixels across the width and height of the video frame). As shown in Fig. \ref{Paver}, the PAVER model takes an ERP video frame as input. The output is the predicted saliency map. The input is divided into $I = \frac{W}{L} \times \frac{H}{L}$ patches,  each having a resolution of $L \times L$. These patches pass through an encoder and a decoder module. The encoder module consists of a deformable convolution layer, a transformer encoder layer, and a fusion layer. The encoder module captures the local and global context of the input video frame. \par

The decoder module computes a saliency score for each patch. Local, temporal, and spatial saliencies are considered in determining the saliency score for each patch. All the saliency scores corresponding to the patches of the video frame $\bm{E}_{f,c,v}$ are collected in the saliency matrix $\hat{\bm{S}}_{f,c,v} \in \mathbb{R}^{\frac{W}{L} \times \frac{H}{L}}$. The final saliency map $\bm{S}_{f,c,v}$ is obtained using spherical Gaussian smoothing \cite{yun2022panoramic} to upsample from $ \mathbb{R}^{\frac{W}{L} \times \frac{H}{L}}$ to $\mathbb{R}^{ W \times H}$. \par

All 360$^{\circ}$ video frames are stored at the edge server. Thus, the saliency detection model can be trained entirely at the edge server. The weights of the deformable convolution and transformer encoder layers can be transferred from the pretrained ViT models without fine-tuning. Without any ground truth labels, the other layers in the PAVER model are trained by minimizing a weighted sum of the spatio-temporal consistency losses and the global context loss using Adam optimizer~\cite{kingma2014adam}.

\subsection{Head Movement Prediction Model} \label{headPredModel}
User $u \in \mathcal{U}$ is watching frame $\Tilde{f} \in \mathcal{F}$ of chunk $\Tilde{c}  \in \mathcal{C}_v$ from its playback buffer, when it requests chunk $c$ of video~$v \in \mathcal{V}$ at time slot $\tau_{u,c,v}^\textrm{REQ}$. The head orientation angles of user $u$ at time slot $\tau_{u,c,v}^\textrm{REQ}$, i.e., $\big( \theta_u(\tau_{u,c,v}^\textrm{REQ}), \phi_u(\tau_{u,c,v}^\textrm{REQ}) \big)$, can be equivalently represented by $\big( \theta_{u,\Tilde{f},\Tilde{c},v}, \phi_{u,\Tilde{f},\Tilde{c},v} \big)$. To predict the head orientation angles of user $u$, a head movement prediction model takes a sequence with length $Q^\textrm{hist}$ of the current and previous head movements~(i.e., $\bm{s}_{u,\Tilde{f},\Tilde{c},v}^\textrm{hist}$) as input, and returns a sequence with length $Q^\textrm{pred}$ which contains the predicted head movements for the frames of chunk $c$~(i.e., $\bm{s}_{u,\Tilde{f},\Tilde{c},v}^\textrm{pred}$). To train the same model for all users, we set $Q^\textrm{pred}=F + \big \lfloor {F \max_{u\in \mathcal{U}}{B_u^\textrm{THR}}}/{T^\textrm{chunk}} \big \rfloor$. 

\begin{figure}[t]
    \centering
    \includegraphics[scale=0.33]{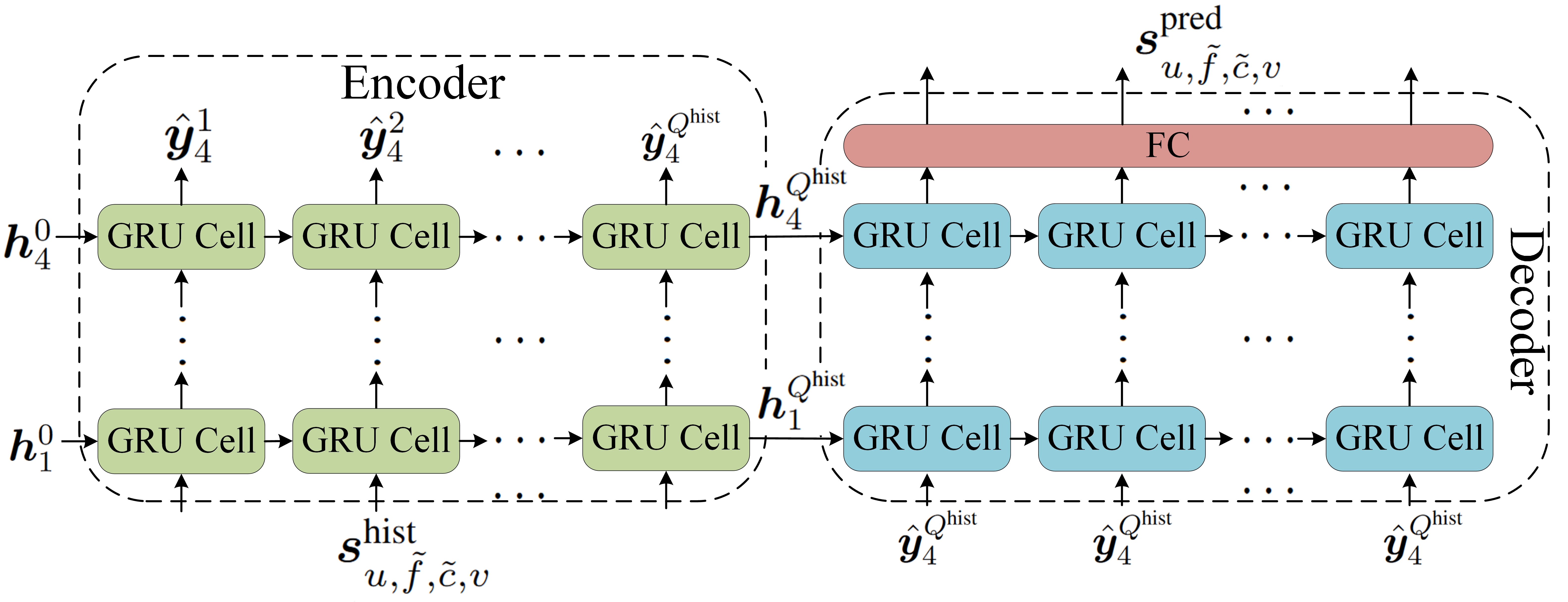}
    \caption{GRU-based head movement prediction model.}
    \label{GRU_model}
\end{figure}

\noindent \textbf{Model}: As shown in Fig. \ref{GRU_model}, the head movement prediction model comprises an encoder and a decoder, each containing four GRU layers. Each GRU layer has $Q^\textrm{hist}$ and $Q^\textrm{pred}$ GRU cells in the encoder and decoder modules, respectively. Let $d^{\textrm{GRU}}$ denote the number of features in the hidden state of each GRU cell. $\bm{h}^0_i \in \mathbb{R}^{d^{\textrm{GRU}}}$ is the initial hidden state of layer~$i \in \{1,\ldots,4\}$. $\bm{h}^0_i$ is initialized to zero for the first input sequence of user $u$'s head movements corresponding to each video. The input of the first GRU layer in the encoder module is the sequence $\bm{s}_{u,\Tilde{f},\Tilde{c},v}^\textrm{hist}$. The output of the last GRU layer in the decoder module passes through a fully connected (FC) layer. The output of the FC layer is the sequence~$\bm{s}_{u,\Tilde{f},\Tilde{c},v}^\textrm{pred}$.

\noindent \textbf{Loss Function}: Let $\mathcal{V}^{\textrm{head-tr}}_{u}$ denote the set of videos for which user $u \in \mathcal{U}$ has its head movement measurements in the local training dataset. Let $\bm{w}_{u}$ denote the learning parameters of the head movement prediction model for user $u$. Each user~$u$ aims to determine $\bm{w}_{u}$ by minimizing the following loss function based on its local historical head~movement:
\begin{align}\label{viewerLoss}
&\mathcal{L}^{\textrm{head}}_{u} = \frac{1}{\left\lvert \mathcal{V}^{\textrm{head-tr}}_{u} \right\rvert}  \sum_{v \in \mathcal{V}^{\textrm{head-tr}}_{u}} \frac{1}{C_v} \sum_{\Tilde{c} \in \mathcal{C}_{v}\backslash \{C_v\}} \frac{1}{F}  \sum_{\Tilde{f} \in \mathcal{F}} \epsilon_{u,\Tilde{f},\Tilde{c},v}^2,
\end{align}
where $\epsilon_{u,\Tilde{f},\Tilde{c},v}=\big\|\bm{s}_{u,\Tilde{f},\Tilde{c},v}^\textrm{pred} - \bm{s}_{u,\Tilde{f},\Tilde{c},v}^\textrm{actual} \big\|$, and $\bm{s}_{u,\Tilde{f},\Tilde{c},v}^\textrm{actual}$ is the sequence of user $u$'s actual head movements in its dataset.

\noindent \textbf{Training}: We propose a PFL algorithm for training the head movement prediction model to address the data heterogeneity issue and preserve the users' privacy. Model decomposition has recently emerged as a promising method for PFL. For model decomposition, we use an approach similar to FedBABU \cite{oh2021fedbabu} and PerFedMask~\cite{setayesh2023perfedmask}. For each user~$u \in \mathcal{U}$, we decompose the learning model $\bm{w}_{u}$ into a global learning model $\bm{w}^\textrm{GRU}$ and a local head model $\bm{w}_{u}^{\textrm{FC}}$. We have $\bm{w}_{u} = \{\bm{w}^\textrm{GRU}, \bm{w}_{u}^{\textrm{FC}}\}$. $\bm{w}^\textrm{GRU}$ and $\bm{w}_{u}^{\textrm{FC}}$ contain the learning parameters of the GRU layers and the FC layer, respectively. All the local head models $\bm{w}_{u}^{\textrm{FC}}$, $u \in \mathcal{U}$, are initialized with the same random weights $\bm{w}_{0}^{\textrm{FC}}$ and are kept fixed during training of the global model $\bm{w}^\textrm{GRU}$. After convergence to a global model, each user~$u$ obtains its personalized model by fine-tuning the learning model $\bm{w}_{u}$ based on its local historical data.\par   

\begin{algorithm}[t]\footnotesize 
 \caption{PFL-based Training Algorithm}
 \begin{algorithmic}[1] \label{TrAlg}
 \STATE Set the number of communication rounds $R$, the number of local update iterations~$\rho$, and the learning rate~$\eta^{\textrm{head}}$.
 \STATE Randomly initialize $\bm{w}^{\textrm{GRU}}_{0}$ and $\bm{w}^{\textrm{FC}}_{0}$; set $\bm{w}_{u}^{\textrm{FC}} := \bm{w}^{\textrm{FC}}_{0}$, $u \in \mathcal{U}$.
 \STATE \textbf{for} each communication round $r \in \mathcal{R}= \{1, \ldots, R \}$ \textbf{do}
 \begin{ALC@g}
  \STATE \textbf{for} each user $u \in \mathcal{U}$ in parallel \textbf{do}
 \begin{ALC@g}
\STATE Given $\bm{w}^{\textrm{GRU}}_{r-1}$ and $\mathcal{L}^{\textrm{head}}_{u}$ in (\ref{viewerLoss}), perform $\rho$ local update iterations to obtain the updated $\bm{w}_{u,r}^{\textrm{GRU}}$.
\end{ALC@g}
 \STATE \textbf{end for}
 \STATE $\bm{w}^{\textrm{GRU}}_{r} := \sum_{u \in \mathcal{U}} \alpha_{u} \bm{w}_{u,r}^{\textrm{GRU}}$.
\end{ALC@g}
 \STATE \textbf{end for} 
   \STATE \textbf{for} each user $u \in \mathcal{U}$ in parallel \textbf{do}
 \begin{ALC@g}
 \STATE Fine-tune the learning model $\bm{w}_{u} := \{\bm{w}^{\textrm{GRU}}_{R}, \bm{w}_{u}^{\textrm{FC}}\}$ using the local historical data.
 \end{ALC@g}
 \STATE \textbf{end for}
 \end{algorithmic} 
\end{algorithm}

The global model $\bm{w}^\textrm{GRU}$ is trained through communication rounds. Let $\mathcal{R}=\{1,\ldots, R \}$ denote the set of communication rounds. At the beginning of each communication round $r \in \mathcal{R}$, the users download the latest global model $\bm{w}^{\textrm{GRU}}_{r-1}$ from the server. $\bm{w}^{\textrm{GRU}}_{0}$ is initialized randomly. Each user~$u \in \mathcal{U}$ initializes its global model $\bm{w}_{u,r}^{\textrm{GRU}}$ by the downloaded global model. We have $\bm{w}_{u,r}^{\textrm{GRU}} = \bm{w}^{\textrm{GRU}}_{r-1}$, $u \in \mathcal{U}$. Then, each user performs $\rho$ local update iterations to update the global model using its local historical data. For each local update iteration, we have $\bm{w}_{u,r}^{\textrm{GRU}} \leftarrow \bm{w}_{u,r}^{\textrm{GRU}} -\eta^{\textrm{head}} \nabla_{\bm{w}^\textrm{GRU}} \mathcal{L}^{\textrm{head}}_{u}\big|_{ \bm{w}^\textrm{GRU}= \bm{w}_{u,r}^{\textrm{GRU}}}$, where $\eta^{\textrm{head}}$ is the learning rate. After completing the local update iterations, each user uploads its updated global model $\bm{w}_{u,r}^{\textrm{GRU}}$ to the server. At the end of each communication round $r$, the server computes a new global model $\bm{w}_{r}^{\textrm{GRU}}$ by aggregating the updated global models it has received from the users. We have $\bm{w}^{\textrm{GRU}}_{r} = \sum_{u \in \mathcal{U}} \alpha_{u} \bm{w}_{u,r}^{\textrm{GRU}}$, where $\alpha_{u} = \frac{\sum_{v \in \mathcal{V}^{\textrm{head-tr}}_{u}} {C_v}}{\sum_{u' \in \mathcal{U}} \sum_{v' \in \mathcal{V}^{\textrm{head-tr}}_{u'}} {C_{v'}}}$ is the weight of user $u$ in model aggregation. Algorithm \ref{TrAlg} summarizes our PFL-based training.\par

\noindent \textbf{Head Orientation Map}: User $u \in \mathcal{U}$ uses its pre-trained head movement prediction model~$\bm{w}_{u}$ to predict its head orientation for frame $f \in \mathcal{F}$ of chunk $c \in \mathcal{C}_v$ of video~$v \in \mathcal{V}$. When user~$u$ requests video chunk $c$ at time slot~$\tau_{u,c,v}^\textrm{REQ}$, it also provides the server with its predicted head orientations $\big( \hat{\theta}_{u,f,c,v}, \hat{\phi}_{u,f,c,v} \big)$, $f\in \mathcal{F}$. Then, the server employs a Gaussian kernel \cite{nguyen2018your} to obtain the head orientation maps of user $u$. We denote the head orientation map corresponding to user $u$'s predicted head orientation, i.e., $\big( \hat{\theta}_{u,f,c,v}, \hat{\phi}_{u,f,c,v} \big)$, by ${\bm{M}}_{u,f,c,v} \in \mathbb{R}^{ W \times H}$.

\subsection{Integration of Saliency and Head Orientation Maps}\label{Integration_technique}
When the server receives a request from user $u \in \mathcal{U}$ for chunk $c \in \mathcal{C}_v$ of video $v \in \mathcal{V}$, it aims to obtain the set of tile indices corresponding to viewport and marginal regions of each video frame~$f \in \mathcal{F}$, i.e., $\mathcal{N}_{u,f,c,v}^{\textrm{view}}$ and $\mathcal{N}_{u, f, c, v}^{\textrm{marg}} \subset \mathcal{N}$, respectively. The saliency and the user's head orientation maps for video frame $f$ are first divided into $N$ tiles. Then, a fused feature map is obtained by integrating those maps using the \emph{regional} fusion technique~\cite{setayesh2023PredFramework}. Finally, the tiles covering the viewport and marginal regions of a video frame are determined based on the feature value of the tiles in the fused feature map.

\begin{figure}[t]
    \centering
    \includegraphics[scale=0.227]{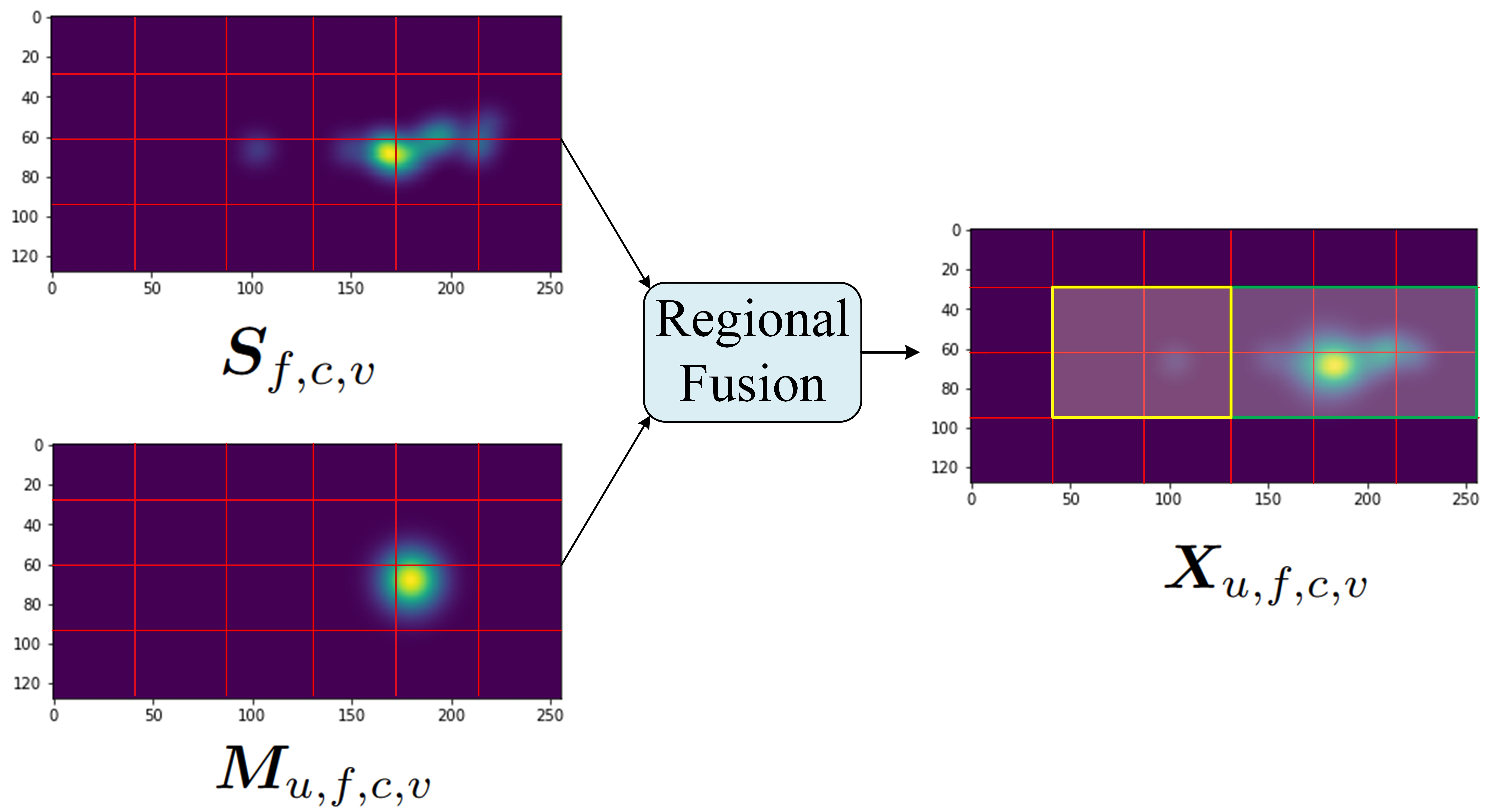}
    \caption{Regional fusion for obtaining the fused feature map.}
    \label{fusion}
\end{figure}

\noindent \textbf{Regional Fusion Technique:} In this fusion technique, we first normalize each feature in the maps $\bm{S}_{f,c,v}$ and ${\bm{M}}_{u,f,c,v}$ to be within~$[0,1]$. Let $\tilde{\bm{S}}_{f,c,v}$ and $\tilde{{\bm{M}}}_{u,f,c,v} \in \mathbb{R}^{ W \times H}$ denote the normalized versions of $\bm{S}_{f,c,v}$ and ${\bm{M}}_{u,f,c,v}$, respectively. We segment $\tilde{\bm{S}}_{f,c,v}$ and $\tilde{{\bm{M}}}_{u,f,c,v}$ into $N$ tiles. We obtain the maximum saliency score of the pixels in each tile of $\tilde{\bm{S}}_{f,c,v}$ and $\tilde{{\bm{M}}}_{u,f,c,v}$, and collect them in vectors ${\bm{s}}_{f,c,v}^{\textrm{max}}$ and ${\bm{m}}_{u,f,c,v}^{\textrm{max}} \in \mathbb{R}^{N}$, respectively. By integrating $\tilde{\bm{S}}_{f,c,v}$ and $\tilde{{\bm{M}}}_{u,f,c,v}$, the fused feature map $\bm{X}_{u,f,c,v}$ is obtained as shown in Fig. \ref{fusion}. We have $\bm{X}_{u,f,c,v} = \big(\max_{n\in \mathcal{N}}{{\bm{s}}_{f,c,v}^{\textrm{max}}[n]}-\frac{1}{N}\sum_{n\in \mathcal{N}}{{{\bm{s}}_{f,c,v}^{\textrm{max}}[n]}} \big)^2 \tilde{\bm{S}}_{f,c,v} + \big(\max_{n\in \mathcal{N}}{{\bm{m}}_{u,f,c,v}^{\textrm{max}}[n]} - \frac{1}{N}\sum_{n\in \mathcal{N}}{{\bm{m}}_{u,f,c,v}^{\textrm{max}} } \big)^2 \tilde{{\bm{M}}}_{u,f,c,v}$.\par

\noindent \textbf{Tile Selection:} The number of tiles that cover the viewport region depends on the FoV of the user's HMD. Thus, the size of FoV specifies $\big\lvert \mathcal{N}_{u,f,c,v}^{\textrm{view}} \big\rvert$. The server selects the tiles in the viewport region (i.e., $\mathcal{N}_{u,f,c,v}^{\textrm{view}}$) based on the adjacent tiles with the maximum feature values in the fused feature map~$\bm{X}_{u,f,c,v}$. On the other hand, the number of tiles that cover the marginal region depends on the prediction accuracy of the user $u$'s predicted head orientation \cite{teng2021qoe}. The difference between the requested chunk index $c$ and the last watched chunk index $\tilde{c}$ by user $u$ at time slot $\tau_{u,c,v}^\textrm{REQ}$ has a direct impact on the prediction accuracy. Thus, we consider that the number of tiles in the marginal region is obtained as $ N_{u, f, c, v}^{\textrm{marg}} = \big[\big \lfloor (\alpha^\textrm{marg} (c-\tilde{c})-1)\lvert \mathcal{N}_{u,f,c,v}^{\textrm{view}} \rvert  \big \rfloor \big]^{+}$, where $\alpha^\textrm{marg}$ is a scaling factor that can be empirically determined. To obtain the tiles in set $\mathcal{N}_{u, f, c, v}^{\textrm{marg}}$, the server initializes $\mathcal{N}_{u, f, c, v}^{\textrm{pred}}$ to be $\mathcal{N}_{u,f,c,v}^{\textrm{view}}$ and invokes the following step  $N_{u, f, c, v}^{\textrm{marg}}$ times: the server selects a tile $n \notin \mathcal{N}_{u, f, c, v}^{\textrm{pred}}$ which is adjacent to set $\mathcal{N}_{u, f, c, v}^{\textrm{pred}}$ and has the maximum feature value in the fused feature map to be added to set $\mathcal{N}_{u, f, c, v}^{\textrm{pred}}$. The tiles in the marginal region are determined as $\mathcal{N}_{u, f, c, v}^{\textrm{marg}} = \mathcal{N}_{u, f, c, v}^{\textrm{pred}} \backslash \mathcal{N}_{u,f,c,v}^{\textrm{view}}$.

\section{DRL Algorithm Design}\label{DRL_Alg_Sec}
\begin{figure}[t]
    \centering
    \includegraphics[scale=0.312]{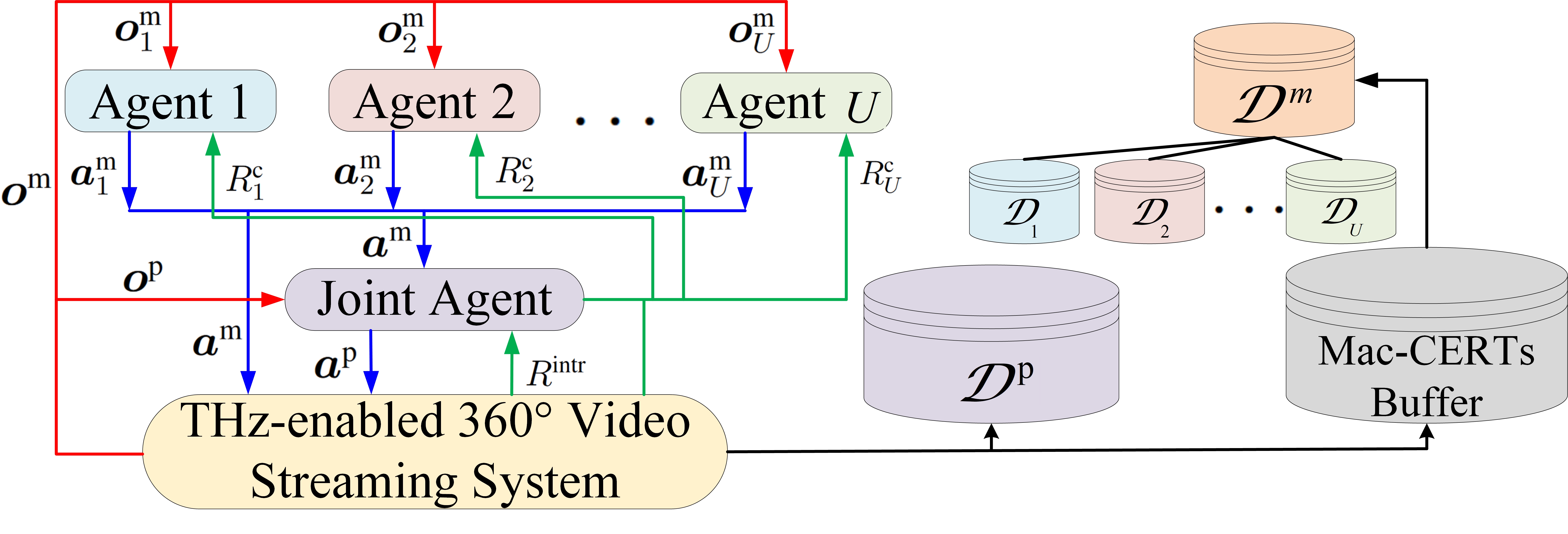}
    \caption{Mac-IAICC and Prim-CAC are used within the agents and the joint agent, respectively. Each agent executes a macro-action based on its macro-observation. The joint agent executes the primitive-action based on the primitive-observation. At each time slot, the transition tuples of the agents and the joint agent are stored in the Mac-CERTs buffer and $\mathcal{D}^\textrm{p}$, respectively.}
    \label{HDDPG}
\end{figure}

In this section, we develop a hierarchical learning framework using two DRL algorithms based on multi-agent cooperative actor-critic methods to solve problems $\mathcal{P}^\textrm{m}$ and $\mathcal{P}^\textrm{p}$. For each problem, we propose a DDPG algorithm to learn the agents' policy and action-value functions using actor and critic networks, respectively. Note that independent actor and critic networks for each agent may lead to a non-stationary learning environment in a multi-agent setting~\cite{lowe2017multi}. In particular, if each agent independently performs  policy updating and exploring, it may not be possible for the agents to obtain high-quality cooperative policies. To address this issue, we leverage the centralized training with decentralized execution~(CTDE) approach \cite{xiao2022asynchronous} to solve problem $\mathcal{P}^\textrm{m}$. By using this approach, we can learn a Mac-IAICC for each agent. Mac-IAICC facilitates offline training using centralized information and online execution in a decentralized manner for the agents. Thus, it is a suitable approach for selecting the bitrate of tiles requested asynchronously by users. To solve problem $\mathcal{P}^\textrm{p}$, which requires the agents to work in a synchronized manner, we use the Prim-CAC approach. This approach enables the agents to cooperatively design the beamforming vectors in each time slot. Fig. \ref{HDDPG} shows an illustration of the interaction among the agents and the environment. Next, we present two algorithms for solving $\mathcal{P}^\textrm{m}$ and $\mathcal{P}^\textrm{p}$ using Mac-IAICC and Prim-CAC, respectively. Then, we propose a hierarchical learning framework to train the actor and critic networks.

\vspace{-0.15cm}
\subsection{Solving Problem $\mathcal{P}^\textrm{m}$ Using Mac-IAICC}\label{Mac_IAICC_Sec}
To solve problem $\mathcal{P}^\textrm{m}$, we develop a DDPG algorithm to learn an independent actor and an individual centralized critic for each agent. Let $\bm{\omega}_u$ and $\bm{\vartheta}_u$ denote the learnable parameters of the neural networks corresponding to the actor and critic of agent~$u \in \mathcal{U}$, respectively. Agent $u$'s actor network specifies the policy of that agent for tile bitrate selection. We have $\bm{a}_u^\textrm{m}(t) = {\mu}_{\bm{\omega}_u}\left({\bm{h}}_u^{\textrm{m}}(t)\right)$. The centralized critic network of agent~$u$ learns the action-value function $Q_{\bm{\vartheta}_u}\left({{\bm{h}}}^{\textrm{m}}(t),{\bm{a}}^\textrm{m}(t)\right)$. Note that ${\bm{h}}_u^{\textrm{m}}(t)$ can be generated implicitly for the actor and critic networks by leveraging an LSTM layer within their neural network architectures \cite{xiao2022asynchronous}.\par

\begin{algorithm}[t]\footnotesize 
 \caption{Parameters Update in Mac-IAICC with DDPG}
 \begin{algorithmic}[1] \label{mac-update-params}
 \STATE \textbf{function} MacUpdateParams$(B^{\textrm{m}},\,\mathcal{D}^\textrm{m},\,\mathcal{D}_u,\,\bm{\vartheta}_u,\,\bm{\omega}_u)$
 \begin{ALC@g}
  \STATE Sample a random minibatch of $B^{\textrm{m}}$ consecutive transition tuples from the sets $\mathcal{D}^\textrm{m}$ and $\mathcal{D}_u$.
 \STATE Obtain the gradient of $\mathcal{L}^{\textrm{TD}}_{u}(\bm{\vartheta}_u)$ in (\ref{TD-error-macro}) and update $\bm{\vartheta}_u$ using Adam optimizer.
 \STATE Determine the gradient in (\ref{policy-grad}) and update $\bm{\omega}_u$ using Adam optimizer.
 \STATE \textbf{Return} $\bm{\vartheta}_u$ and $\bm{\omega}_u$
\end{ALC@g}
 \STATE \textbf{end function}
 \end{algorithmic} 
\end{algorithm}

DDPG is an off-policy algorithm that is able to learn the policy and action-value functions in a stable and robust manner using a replay buffer and separate target networks~\cite{lillicrap2015continuous}. Since agents in a MacDec-POMDP asynchronously start and complete their macro-actions, a new replay buffer needs to be designed. Thus, we collect the macro-observation, macro-action, and extrinsic reward of the agents into a buffer called macro-action concurrent experience replay trajectories~(Mac-CERTs) in each time slot. The Mac-CERTs buffer has been used in \cite{xiao2022asynchronous} to solve a MacDec-POMDP problem using policy gradient algorithm. In this work, we utilize Mac-CERTs buffer to train the actor and critic networks of each agent using a DDPG algorithm. For $t \in \{\tau_{u,c,v}^\textrm{REQ},\tau_{u,c,v}^\textrm{REQ}+1,\ldots,\tau_{u,c+1,v}^\textrm{REQ}-1\}$, agent $u \in \mathcal{U}$ receives a cumulative reward for the macro-action at time slot $\tau_{u,c,v}^\textrm{REQ}$ as $R^\textrm{c}_{u}(t) = \sum_{t' = \tau_{u,c,v}^\textrm{REQ}}^{t}\gamma^{t'-\tau_{u,c,v}^\textrm{REQ}}R^\textrm{extr}(t')$. Then, at the end of each time slot $t$, agent $u$ stores its transition experience in the Mac-CERTs buffer as a tuple $\left(\bm{o}_u^{\textrm{m}}(t), \bm{a}_u^\textrm{m}(t), \bm{o}_u^{\textrm{m}}(t+1), R^\textrm{c}_{u}(t) \right)$. Note that $\bm{o}_u^{\textrm{m}}(t)$, $\bm{a}_u^\textrm{m}(t)$, and $\bm{o}_u^{\textrm{m}}(t+1)$ remain unchanged until agent~$u$ completes its current macro-action at the end of time slot $t = \tau_{u,c+1,v}^\textrm{REQ}-1$. Thus, a macro action $\bm{a}_u^\textrm{m}(t)$ takes $\Delta_\tau(\bm{a}_u^\textrm{m}(t))=\tau_{u,c+1,v}^\textrm{REQ}-\tau_{u,c,v}^\textrm{REQ}$ time slots to complete. When training the actor network of agent $u$, we only consider the tuples in the Mac-CERTs buffer that correspond to agent $u$. We filter those tuples by selecting the ones that agent $u$ completes its macro-action. However, training each agent's critic network requires all the joint macro-observation-action information. To train the critic networks, we filter the tuples in the Mac-CERTs buffer by selecting the ones that an agent has completed its macro-action. Let $\mathcal{D}_u$ and $\mathcal{D}^\textrm{m}$ denote the set of tuples which are used for training agent $u$'s actor and critic networks, respectively. \par

Each agent $u$ updates the learnable parameters of its critic network (i.e., $\bm{\vartheta}_u$) by minimizing the temporal difference~(TD) error over the tuples sampled from set~$\mathcal{D}^\textrm{m}$. We create a copy of the actor and critic networks to be used as target networks. Let $\bm{\omega^-}_u$ and $\bm{\vartheta^-}_u$, respectively, denote the weights of agent $u$'s target actor and critic networks. The weights of agent $u$'s target networks are updated by slowly tracking the learned parameters of its actor and critic networks. The TD error is obtained as follows:
\begin{align}\label{TD-error-macro}
\mathcal{L}^{\textrm{TD}}_{u}(\bm{\vartheta}_u) = \mathbb{E}_{\Lambda^{\textrm{m}}_{u} \sim \mathcal{D}^\textrm{m}}\Big\{ \big(Q_{\bm{\vartheta}_u}\big({{\bm{h}}}^{\textrm{m}},{\bm{a}}^\textrm{m}\big) - \hat{Q}_{\bm{\vartheta^-}_u}\left({{\bm{h}}}^{\textrm{m}},{\bm{a}}^\textrm{m}\right)\big)^2\Big\},
\end{align}
where $\Lambda^{\textrm{m}}_{u} = \left({{\bm{o}}}^{\textrm{m}}, {\bm{a}}^\textrm{m}, {{\bm{o}}'}^{\textrm{m}}, R^\textrm{c}_{u} \right)$ and $\hat{Q}_{\bm{\vartheta^-}_u}\big({{\bm{h}}}^{\textrm{m}},{\bm{a}}^\textrm{m}\big) = R^\textrm{c}_{u} + \gamma^{\Delta_\tau(\bm{a}_u^\textrm{m})} Q_{\bm{\vartheta^-}_u}\big({{\bm{h}}'}^{\textrm{m}},$ ${{\bm{a}}'}^{\textrm{m}}\big)$. We obtain ${{\bm{a}}'}^{\textrm{m}}$ using the target actor networks of the agents. We have ${{\bm{a}}'}^{\textrm{m}} = \big({\mu}_{\bm{\omega^-}_u}\left({\bm{h}'}_u^{\textrm{m}}\right),\, u\in \mathcal{U} \big)$.\par

Each agent $u \in \mathcal{U}$ updates the learnable parameters of its actor network (i.e., $\bm{\omega}_u$) using the gradient of the action-value function. The policy gradient is obtained as follows \cite{lillicrap2015continuous}:

\begin{align}\label{policy-grad}
&\nabla_{\bm{\omega}_u} \mathcal{L}^{\textrm{PG}}_{u} = \mathbb{E}_{\bm{o}_u^{\textrm{m}} \sim \mathcal{D}_{u}^{\bm{o}}}\Big\{ \nabla_{\bm{a}_u^\textrm{m}}Q_{\bm{\vartheta}_u}\left({{\bm{h}}}^{\textrm{m}},{\bm{a}}^\textrm{m}\right)\big|_{\bm{a}_u^\textrm{m}= {\mu}_{\bm{\omega}_u}\left({\bm{h}}_u^{\textrm{m}}\right)}\nonumber\\
&\hspace{5.05cm} \times \nabla_{\bm{\omega}_u} {\mu}_{\bm{\omega}_u}\left({\bm{h}}_u^{\textrm{m}}\right) \Big\},
\end{align}
where $\mathcal{D}_{u}^{\bm{o}} = \big\{{\bm{o}_u^{\textrm{m}} \bigm|\left(\bm{o}_u^{\textrm{m}}, \bm{a}_u^\textrm{m}, {\bm{o}'}_u^{\textrm{m}}, R^\textrm{c}_{u} \right) \in \mathcal{D}_{u}} \big\}$. Algorithm \ref{mac-update-params} describes the MacUpdateParams function, which updates the actor and critic network parameters for each agent $u$.

\begin{algorithm}[t]\footnotesize  
 \caption{Parameters Update in Prim-CAC with DDPG}
 \begin{algorithmic}[1] \label{primi-update-params}
 \STATE \textbf{function} PrimUpdateParams$(B^{\textrm{p}},\,\mathcal{D}^\textrm{p},\,\bm{\vartheta},\,\bm{\omega})$
 \begin{ALC@g}
  \STATE Sample a random minibatch of $B^{\textrm{p}}$ consecutive transition tuples from the set $\mathcal{D}^\textrm{p}$.
 \STATE Obtain the gradient of $\mathcal{L}^{\textrm{TD}}(\bm{\vartheta})$ in (\ref{TD-error-primitive}) and update $\bm{\vartheta}$ using Adam optimizer.
 \STATE Determine the gradient in (\ref{policy-grad-prim}) and update $\bm{\omega}$ using Adam optimizer.
 \STATE \textbf{Return} $\bm{\vartheta}$ and $\bm{\omega}$
\end{ALC@g}
 \STATE \textbf{end function}
 \end{algorithmic} 
\end{algorithm}

\vspace{-0.1cm}
\subsection{Solving Problem $\mathcal{P}^\textrm{p}$ Using Prim-CAC}\label{Prim-CAC_Sec}
The agents aim to cooperatively solve problem $\mathcal{P}^\textrm{p}$ in a synchronized manner. This problem can be efficiently solved by treating all the agents as a joint agent. As a result, the learning environment turns into a fully centralized learning setting, where the joint agent trains both the centralized actor and critic networks to obtain $\bm{\pi}_{\bm{\omega}}\big({\bm{h}}^{\textrm{p}}\big)$ and $Q_{\bm{\vartheta}}\big({\bm{h}}^{\textrm{p}},{\bm{a}}^\textrm{p};{\bm{a}}^\textrm{m}\big)$, respectively. To this end, the joint agent collects the joint primitive-observations and actions, as well as the intrinsic rewards into a replay buffer. Recall from Section \ref{reward} that $R^\textrm{intr}(t)$ represents the agents' intrinsic reward in time slot $t \in \mathcal{T}$. At the end of each time slot~$t$, the tuple $\big({\bm{o}}^{\textrm{p}}(t), {\bm{a}}^\textrm{p}(t), {\bm{o}}^{\textrm{p}}(t+1), R^\textrm{intr}(t) \big)$ is stored in the replay buffer. Let  $\mathcal{D}^\textrm{p}$ denote the set of tuples in the replay buffer. The learnable parameters of the centralized critic network (i.e., $\bm{\vartheta}$) are updated by minimizing the following TD error:
\begin{align}\label{TD-error-primitive}
\mathcal{L}^{\textrm{TD}}(\bm{\vartheta}) \!=\!\mathbb{E}_{\Lambda^{\textrm{p}} \sim \mathcal{D}^\textrm{p}}\Big\{ \!\big(Q_{\bm{\vartheta}}\left({\bm{h}}^{\textrm{p}},{\bm{a}}^\textrm{p};{\bm{a}}^\textrm{m}\right)\!-\! \hat{Q}_{\bm{\vartheta^-}}\left({\bm{h}}^{\textrm{p}},{\bm{a}}^\textrm{p};{\bm{a}}^\textrm{m}\right)\big)^2\!\Big\}, 
\end{align}
where $\Lambda^{\textrm{p}} = \left({{\bm{o}}}^{\textrm{p}}, {\bm{a}}^\textrm{p}, {{\bm{o}}'}^{\textrm{p}}, R^\textrm{intr} \right)$ and $\hat{Q}_{\bm{\vartheta^-}}\big({\bm{h}}^{\textrm{p}},{\bm{a}}^\textrm{p};{\bm{a}}^\textrm{m}\big) = R^\textrm{intr}$ $+\, \gamma Q_{\bm{\vartheta^-}}\big({{\bm{h}}'}^{\textrm{p}},$ $\bm{\pi}_{\bm{\omega^-}}\big({{\bm{h}}'}^{\textrm{p}}\big);{{\bm{a}}'}^{\textrm{m}}\big)$. We update the centralized actor network's parameters using the following policy gradient:
\begin{align}\label{policy-grad-prim}
&\nabla_{\bm{\omega}} \mathcal{L}^{\textrm{PG}} = \mathbb{E}_{{{\bm{o}}}^{\textrm{p}} \sim \mathcal{D}^{\bm{o}}}\Big\{ \!\nabla_{{\bm{a}}^\textrm{p}}Q_{\bm{\vartheta}}\left({\bm{h}}^{\textrm{p}},{\bm{a}}^\textrm{p};{\bm{a}}^\textrm{m}\right)\big|_{{\bm{a}}^\textrm{p}= \bm{\pi}_{\bm{\omega}}\left({\bm{h}}^{\textrm{p}}\right)} \nonumber \\
& \hspace{5.4cm}\times \nabla_{\bm{\omega}} \bm{\pi}_{\bm{\omega}}\left({\bm{h}}^{\textrm{p}}\right) \Big\},
\end{align}
where $\mathcal{D}^{\bm{o}} = \big\{{{{\bm{o}}}^{\textrm{p}} \bigm|{\Lambda^{\textrm{p}} \in \mathcal{D}^\textrm{p}}} \big\}$. Algorithm~\ref{primi-update-params} describes the PrimUpdateParams function, which updates the actor and critic network parameters for the joint agent.

\subsection{Hierarchical Learning Framework}
The parameters of the Mac-IAICC for each agent and the parameters of the Prim-CAC for the considered joint agent are learned at different time scales. In particular, the joint agent's transition experiences are collected at each time slot. However, to train the Mac-IAICC for each agent, we use the agents' transition experiences when their macro-actions are completed. Furthermore, the agents' macro-actions will change the primitive-observation vector for the joint agent, while the joint agent's action affects the extrinsic rewards obtained by the agents. To effectively learn both policies $\bm{{\mu}}$ and $\bm{{\pi}}$, we propose a hierarchical learning framework which considers the interaction among the agents' policies at different levels of temporal abstraction and trains their actor-critic networks accordingly. Algorithm~\ref{h-drl} describes our proposed hierarchical learning framework. \par

In Algorithm~\ref{h-drl}, each agent $u \in \mathcal{U}$ and the joint agent randomly explore the environment for $T^\textrm{warm-up}$ time slots and initialize the sets $\mathcal{D}_u$, $\mathcal{D}^\textrm{m}$, and  $\mathcal{D}^\textrm{p}$. Then, we train the actor and critic networks of the agents and the joint agent for $E^\textrm{max}$ episodes, each with $T^\textrm{max}$ time slots. To encourage effective exploration and learning, we add an exploration noise $\bm{\varrho}^\textrm{m}$ to the actions determined by the actor network of each agent. We also add an exploration noise $\bm{\varrho}^\textrm{p}$ to the actions determined by the actor network of the joint agent. Both $\bm{\varrho}^\textrm{m}$ and $\bm{\varrho}^\textrm{p}$ follow the Ornstein-Uhlenbeck process with parameters $(\theta_{\bm{\varrho}^\textrm{m}}, \sigma_{\bm{\varrho}^\textrm{m}})$ and $(\theta_{\bm{\varrho}^\textrm{p}}, \sigma_{\bm{\varrho}^\textrm{p}})$, respectively. To stabilize the training process, we update the target networks at the end of each time slot using the soft update technique with constant $\varepsilon$, where $0<\varepsilon<1$.

\begin{algorithm}[t]\footnotesize 
 \caption{Our Proposed Hierarchical Learning Framework}
 \begin{algorithmic}[1] \label{h-drl}
  \STATE Set the maximum number of episodes $E^\textrm{max}$, the minibatch sizes $B^{\textrm{m}}$ and $B^{\textrm{p}}$, the learning rate for the actor network $\eta^{\textrm{a}}$, the learning rate for the critic network $\eta^{\textrm{c}}$, and the soft target network update constant $\varepsilon \in (0,1)$.
 \STATE Randomly initialize the learnable parameters $\bm{\vartheta}_u$ and $\bm{\omega}_u$ for each agent~$u \in \mathcal{U}$, and the learnable parameters $\bm{\vartheta}$ and $\bm{\omega}$ for the joint agent.
 \STATE Set the target network weights $\bm{\vartheta^-}_u := \bm{\vartheta}_u$ and $\bm{\omega^-}_u :=\bm{\omega}_u$ for each agent~$u \in \mathcal{U}$, as well as $\bm{\vartheta^-} := \bm{\vartheta}$ and $\bm{\omega^-} := \bm{\omega}$ for the joint agent.
 \STATE Initialize the set $\mathcal{D}_u$ for each agent $u \in \mathcal{U}$, as well as the sets $\mathcal{D}^\textrm{m}$ and $\mathcal{D}^\textrm{p}$ by performing random exploration for $T^\textrm{warm-up}$ time slots.
 \STATE \textbf{for} each episode \textbf{do}
 \begin{ALC@g}
 \STATE Initialize the macro-observation vector $\bm{o}_u^{\textrm{m}}$ for each agent $u \in \mathcal{U}$.
 \STATE Determine the macro-action $\bm{a}_u^\textrm{m} \leftarrow {\mu}_{\bm{\omega}_u}\left({\bm{h}}_u^{\textrm{m}}\right) + \bm{\varrho}^\textrm{m}$ for each agent $u$.
 \STATE Obtain the joint primitive-observation ${{\bm{o}}}^{\textrm{p}}$.
  \STATE \textbf{for} each time slot $t \in \{1,\ldots,T^\textrm{max}\}$ \textbf{do}
 \begin{ALC@g}
\STATE Determine the joint primitive-action ${\bm{a}}^\textrm{p}= \bm{\pi}_{\bm{\omega}}\big({\bm{h}}^{\textrm{p}}\big)+\bm{\varrho}^\textrm{p}$.
\STATE Obtain the next macro-observation vector ${\bm{o}'}_u^{\textrm{m}}$ and the cumulative reward $R^\textrm{c}_{u}$ for each agent $u$.
\STATE \textbf{for} each agent $u \in \mathcal{U}$ \textbf{do}
\begin{ALC@g}
\STATE Store the tuple $\big(\bm{o}_u^{\textrm{m}}, \bm{a}_u^\textrm{m}, {\bm{o}'}_u^{\textrm{m}}, R^\textrm{c}_{u} \big)$ in the Mac-CERTs buffer.
\STATE \textbf{if} macro-action $\bm{a}_u^\textrm{m}$ is completed \textbf{then}
\begin{ALC@g}
\STATE $\bm{o}_u^{\textrm{m}} \leftarrow {\bm{o}'}_u^{\textrm{m}}$ and $\bm{a}_u^\textrm{m} \leftarrow {\mu}_{\bm{\omega}_u}\left({\bm{h}}_u^{\textrm{m}}\right) + \bm{\varrho}^\textrm{m}$.
\end{ALC@g}
\STATE \textbf{end if}
\end{ALC@g}
\STATE \textbf{end for}
\STATE Obtain the next observation ${{\bm{o}}'}^{\textrm{p}}$ and the intrinsic reward $R^\textrm{intr}$.
\STATE Store the tuple $\big({{\bm{o}}}^{\textrm{p}}, {\bm{a}}^\textrm{p}, {{\bm{o}}'}^{\textrm{p}}, R^\textrm{intr} \big)$ in set $\mathcal{D}^\textrm{p}$.
\STATE Update $\bm{\vartheta}_u$ and $\bm{\omega}_u$ using Algorithm \ref{mac-update-params} for each agent $u \in \mathcal{U}$.
\STATE Update $\bm{\vartheta}$ and $\bm{\omega}$ using Algorithm \ref{primi-update-params} for the joint agent.
\STATE Update the target network weights for each agent $u \in \mathcal{U}$ as follows: $\bm{\vartheta^-}_u \leftarrow \varepsilon \bm{\vartheta}_u + \left(1-\varepsilon\right)\bm{\vartheta^-}_u$ and $\bm{\omega^-}_u \leftarrow \varepsilon \bm{\omega}_u + \left(1-\varepsilon\right)\bm{\omega^-}_u$.
\STATE Update the target network weights for the joint agent as follows: $\bm{\vartheta^-} \leftarrow \varepsilon \bm{\vartheta} + \left(1-\varepsilon\right)\bm{\vartheta^-}$ and $\bm{\omega^-} \leftarrow \varepsilon \bm{\omega} + \left(1-\varepsilon\right)\bm{\omega^-}$.
\STATE ${{\bm{o}}}^{\textrm{p}} \leftarrow {{\bm{o}}'}^{\textrm{p}}$.
\end{ALC@g}
 \STATE \textbf{end for}
\end{ALC@g}
 \STATE \textbf{end for} 
   \STATE Outputs are $\bm{\omega}_u$ for each agent $u \in \mathcal{U}$, and $\bm{\omega}$ for the joint agent.
 \end{algorithmic} 
\end{algorithm}

\subsection{Computational Complexity}
For the actor and critic networks, we consider neural networks comprising one LSTM layer and two FC layers. There is a leaky rectified linear unit (leaky ReLU) activation layer between the LSTM and the first FC layer, as well as between the two FC layers. A hyperbolic tangent (tanh) activation layer is utilized after the second FC layer in the actor networks. Let $d_\textrm{h}^\textrm{A}$ and $d_\textrm{h}^\textrm{J}$ denote the hidden sizes of each agent's LSTM layer and the joint agent's LSTM layer, respectively. The output sizes of the first and second FC layers corresponding to the neural networks of each agent are denoted by $d^\textrm{A}_\textrm{fc}$ and $d^\textrm{A}_\textrm{out}$, respectively. We denote the output sizes of the first and second FC layers corresponding to the neural networks of the joint agent by $d^\textrm{J}_\textrm{fc}$ and $d^\textrm{J}_\textrm{out}$, respectively.

After training, the computational complexity of the online tile bitrate selection
using the pre-trained actor network obtained by Algorithms~\ref{mac-update-params} and \ref{h-drl} for each agent is $O(d^\textrm{A}_\textrm{in}d^\textrm{A}_\textrm{h}+(d^\textrm{A}_\textrm{h})^2+d^\textrm{A}_\textrm{h}d^\textrm{A}_\textrm{fc}+d^\textrm{A}_\textrm{fc}d^\textrm{A}_\textrm{out})$, where $d^\textrm{A}_\textrm{in}$ is the size of the input layer for the agent's actor network. Similarly, after training, the computational complexity of the online design of beamforming vectors 
using the pre-trained actor network obtained by Algorithms~\ref{primi-update-params} and \ref{h-drl} for the joint agent is $O(d^\textrm{J}_\textrm{in}d^\textrm{J}_\textrm{h}+(d^\textrm{J}_\textrm{h})^2+d^\textrm{J}_\textrm{h}d^\textrm{J}_\textrm{fc}+d^\textrm{J}_\textrm{fc}d^\textrm{J}_\textrm{out})$, where $d^\textrm{J}_\textrm{in}$ is the size of the input layer for the joint agent's actor network.\par

\section{Performance Evaluation}\label{Eval_Sec}
\subsection{Experimental Setup}
\noindent \textbf{Wireless Environment:} We consider a $10$ m $\times$ $10$ m $\times$ $4$~m indoor environment and three APs as shown in Fig. \ref{SysModel}. The APs are located at $\bm{l}_1=\left(9,1,4 \right)$, $\bm{l}_2=\left(5,5,4 \right)$, and $\bm{l}_3=\left(1,9,4 \right)$. We consider the APs to be operating at a carrier frequency of $f_{\textrm{c}} = 1.05$ THz. The molecular absorption coefficient is determined as $\kappa(f_{\textrm{c}})=0.07512$ m$^{-1}$. Unless stated otherwise, we consider $U = 6$ users, each with a height of $h_u = 1.6$ m. The floor of the indoor environment is divided into $U$ equal areas. Each user is positioned at the center of an area. We set the self-blockage angle of the users to $\phi^\textrm{blocked} = \pi$. We set the number of antenna elements for each AP $N_{\textrm{t}}$ and each user's HMD $N_{\textrm{r}}$ to be $6$ and $2$, respectively.

\noindent \textbf{Dataset:}  We conduct our experiments using a public 360$^{\circ}$ video dataset \cite{zhang2018saliency}. The dataset consists of $104$ videos including five sports events: basketball, parkour, BMX, skateboarding, and dance. There are $27$ viewers in this dataset. Each video has been watched by at least $18$ viewers. For each viewer, the eye gaze points are recorded when watching the videos.

\noindent \textbf{Simulation Setting:} A frame rate of $30$ frames per second and a chunk duration of one second are considered for the videos. Each video frame is divided into $24$ tiles using $6 \times 4$ tiling pattern as shown in Fig. \ref{chunks}. The time slot duration $T^\textrm{slot}$ is set to $100$ ms. The FoV of each user's HMD is set to $90^{\circ} \times 135^{\circ}$. We consider five quality levels. We select the bitrate value for encoding the tiles from set $\{28, 33, 38, 43, 48 \}$ Mbps. 

\begin{figure}
     \begin{subfigure}[b]{0.238\textwidth}
         \includegraphics[scale=0.248]{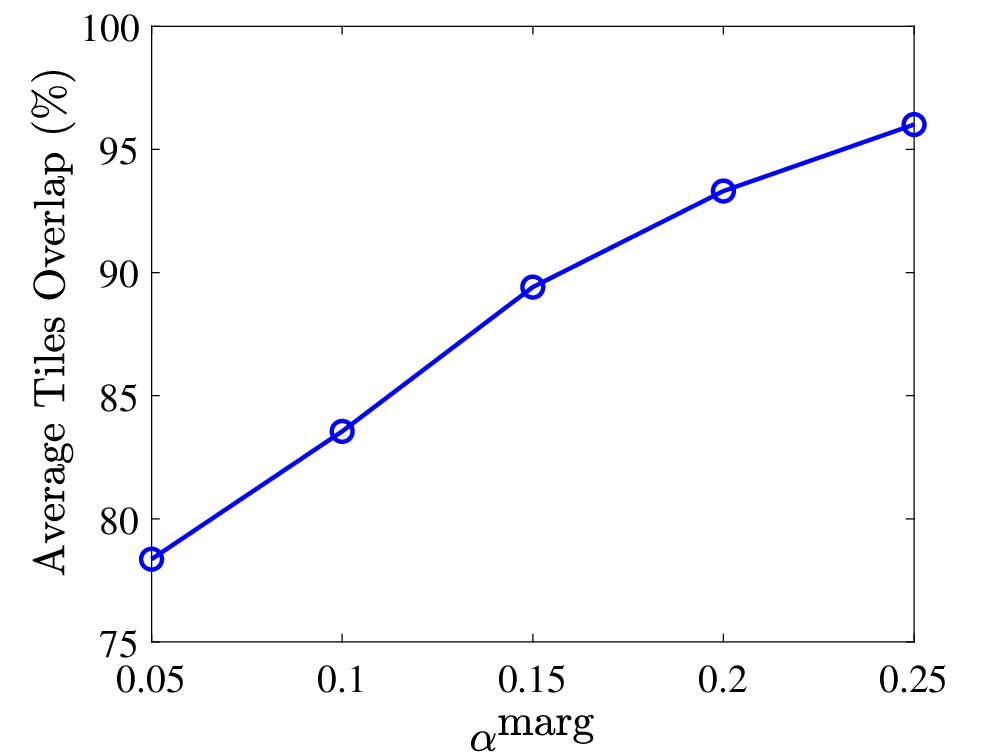}
         \vspace{-1mm}
         \caption{}
         \label{ThrvsArivalRate}
     \end{subfigure}\,\,
     \begin{subfigure}[b]{0.238\textwidth} 
         \includegraphics[scale=0.248]{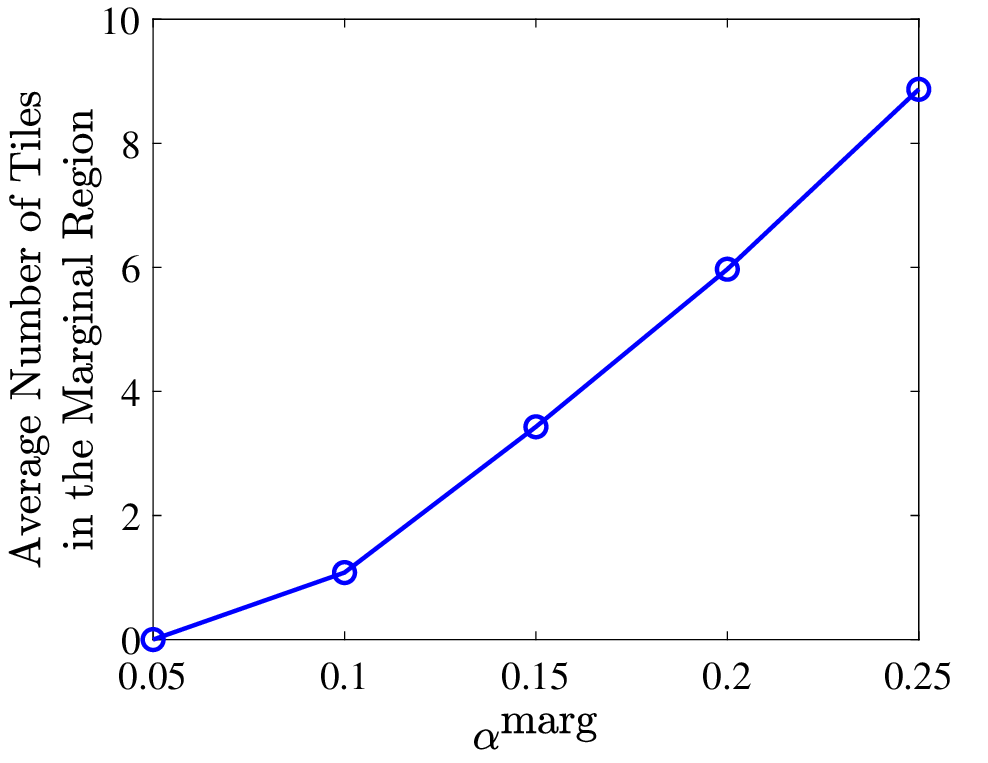}
         \vspace{-1mm}
         \caption{}
         \label{SsrvsArivalRate}
     \end{subfigure}
\caption{(a) Average tiles overlap and (b) average number of tiles in the marginal region versus the scale factor $\alpha^\textrm{marg}$.}
\label{saleFactorEff}
\end{figure}

\begin{figure*}
     \begin{subfigure}[b]{0.32\textwidth}
         \includegraphics[trim=0 0cm 17.6cm 0,clip,scale=0.165]{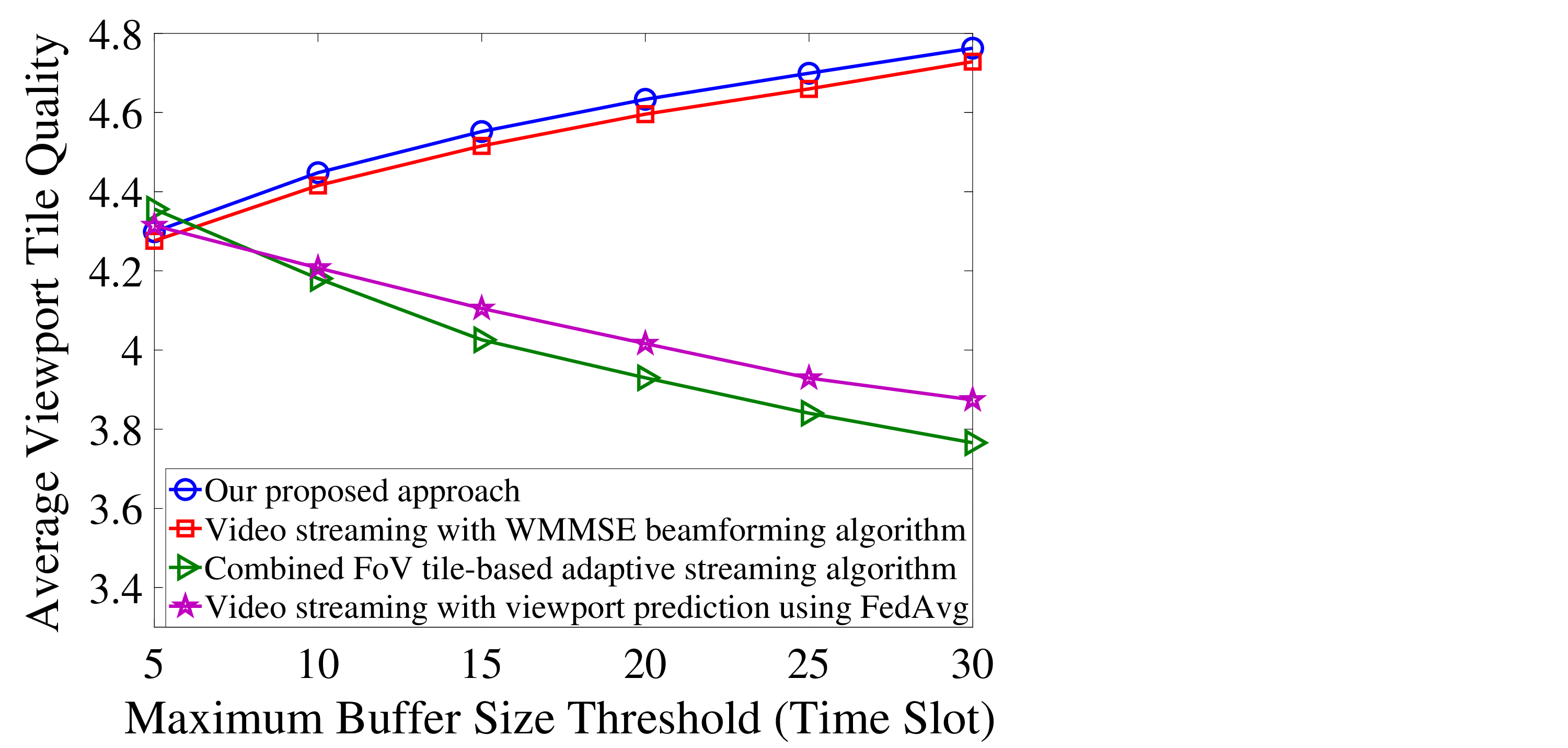}
         \vspace{-1mm}
         \caption{}
         \label{viewTileQ}
     \end{subfigure}\quad
     \begin{subfigure}[b]{0.32\textwidth} 
         \includegraphics[trim=0 0cm 17.6cm 0,clip,scale=0.165]{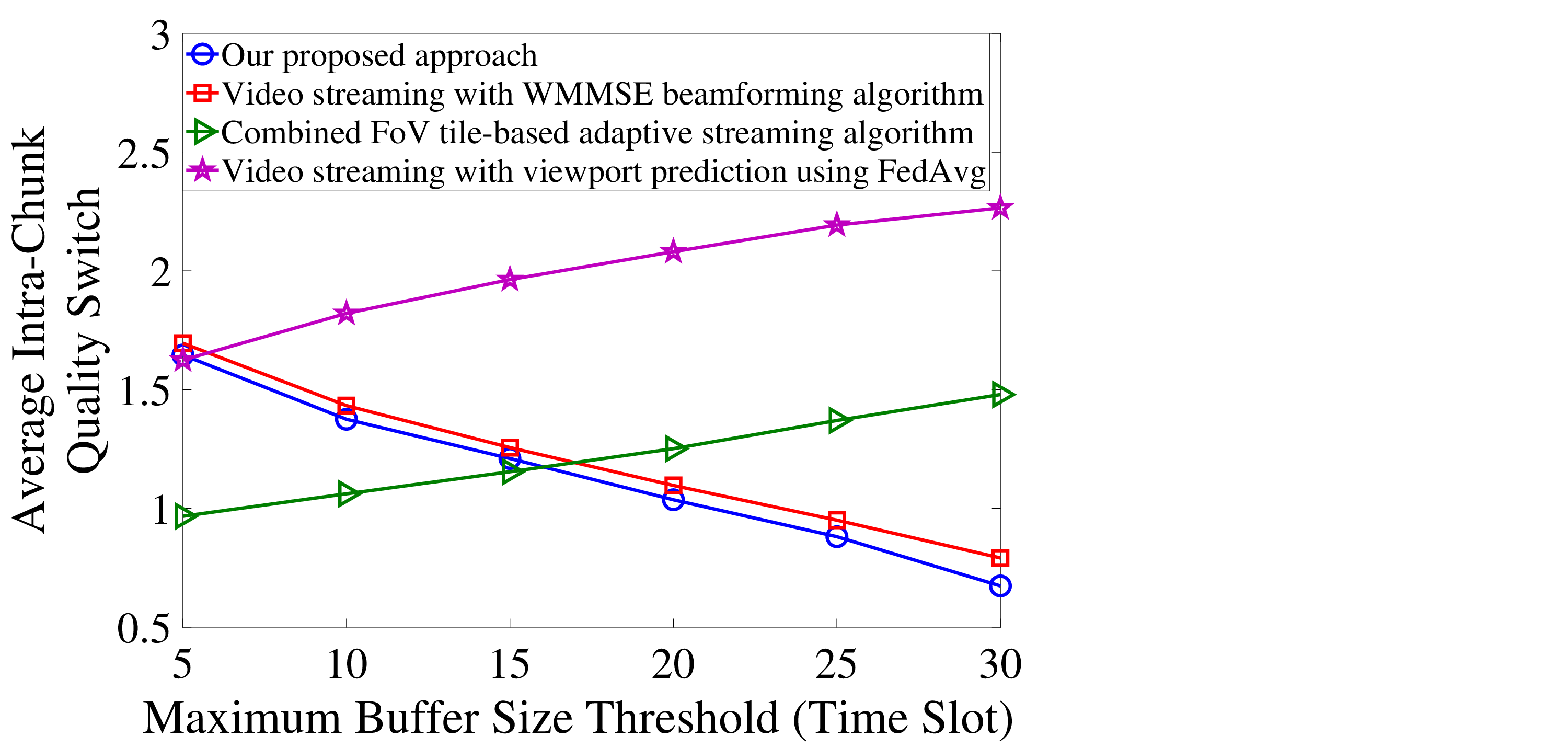}
         \vspace{-1mm}
         \caption{}
         \label{SpaQ}
     \end{subfigure}\quad
     \begin{subfigure}[b]{0.32\textwidth} 
         \includegraphics[trim=0 0cm 17.6cm 0,clip,scale=0.165]{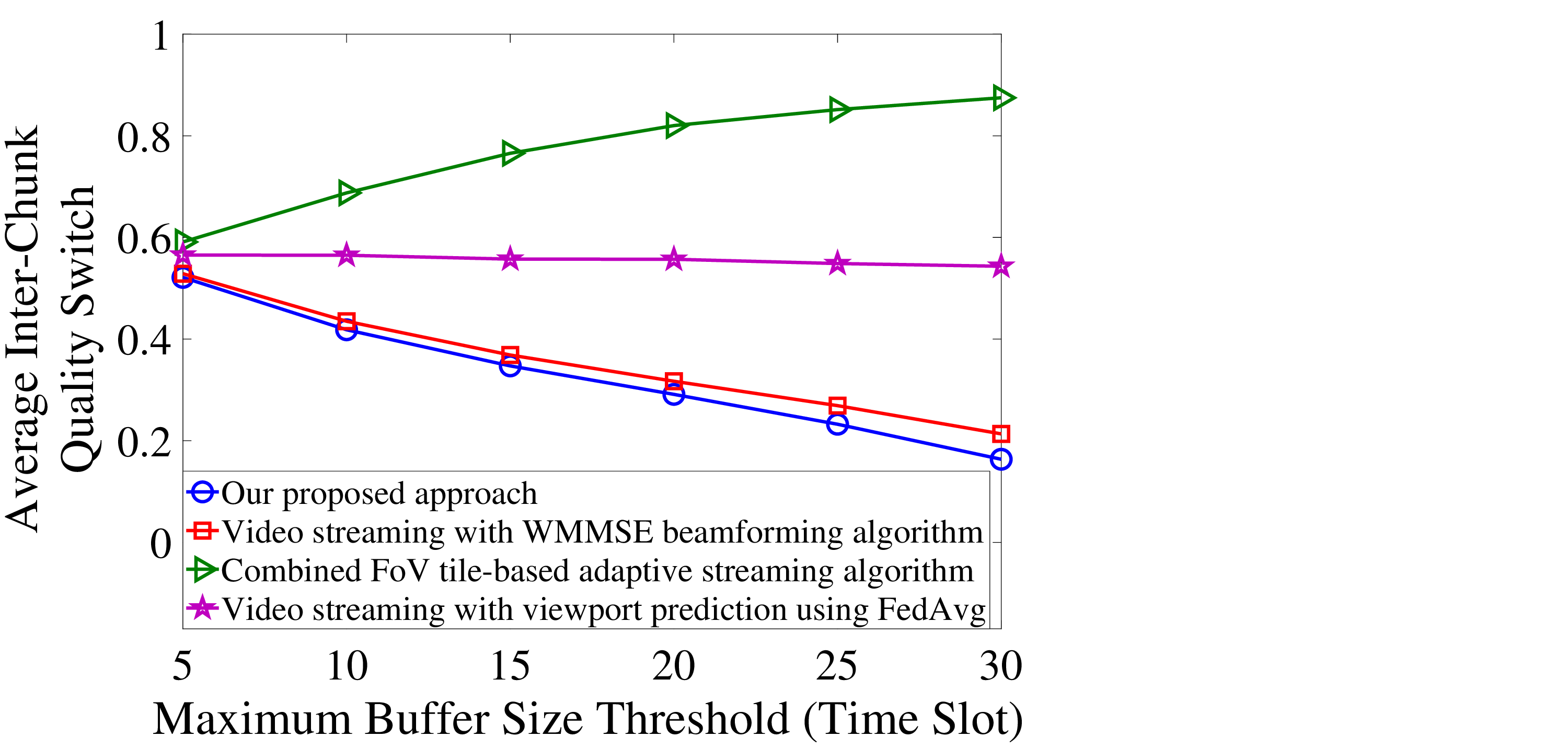}
         \vspace{-1mm}
         \caption{}
         \label{TempQ}
     \end{subfigure}\quad
     \begin{subfigure}[b]{0.32\textwidth} 
         \includegraphics[trim=0 0cm 17.6cm 0,clip,scale=0.165]{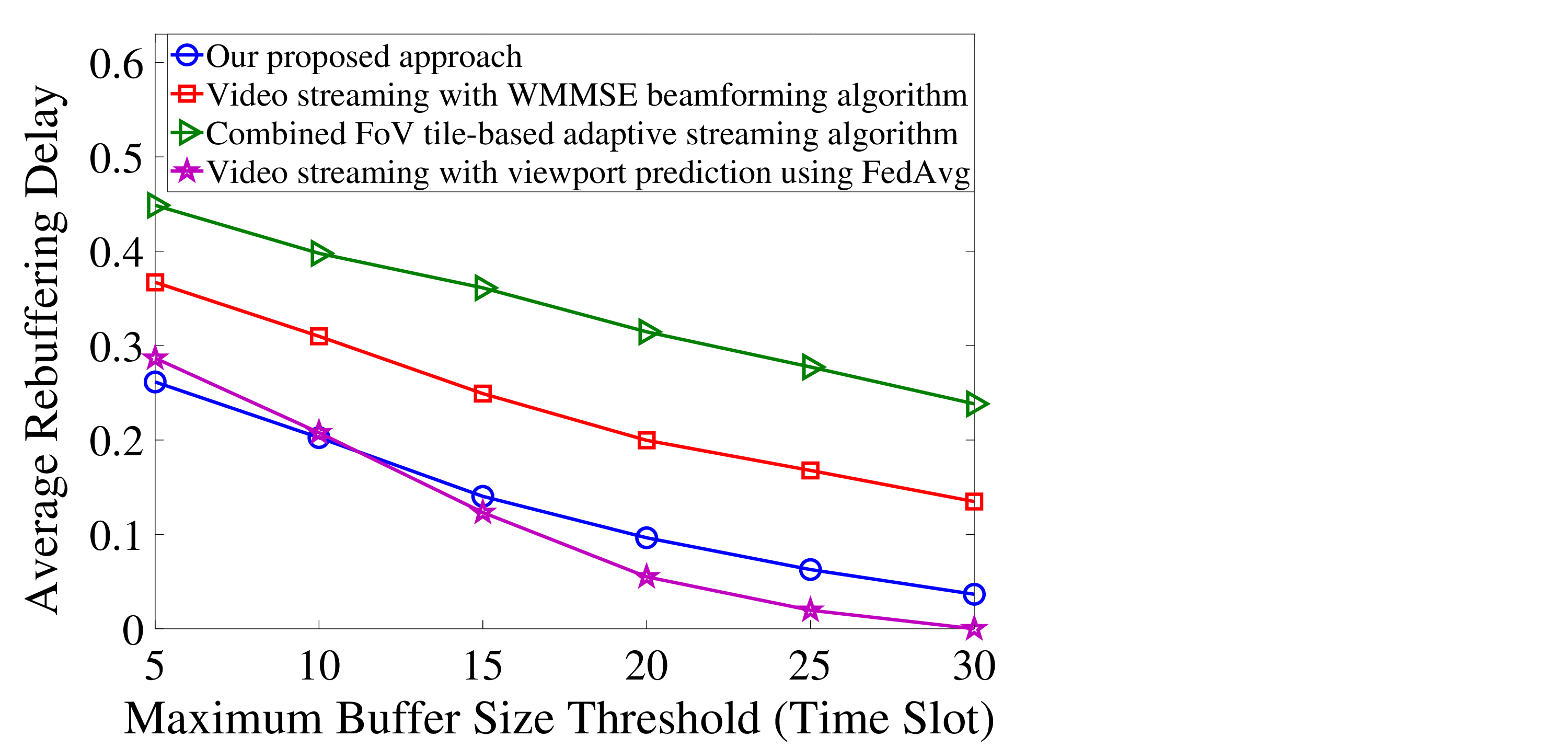}
         \vspace{-1mm}
         \caption{}
         \label{RebuffD}
     \end{subfigure}
     \quad
     \begin{subfigure}[b]{0.32\textwidth} 
         \includegraphics[trim=0 0cm 17.6cm 0,clip,scale=0.165]{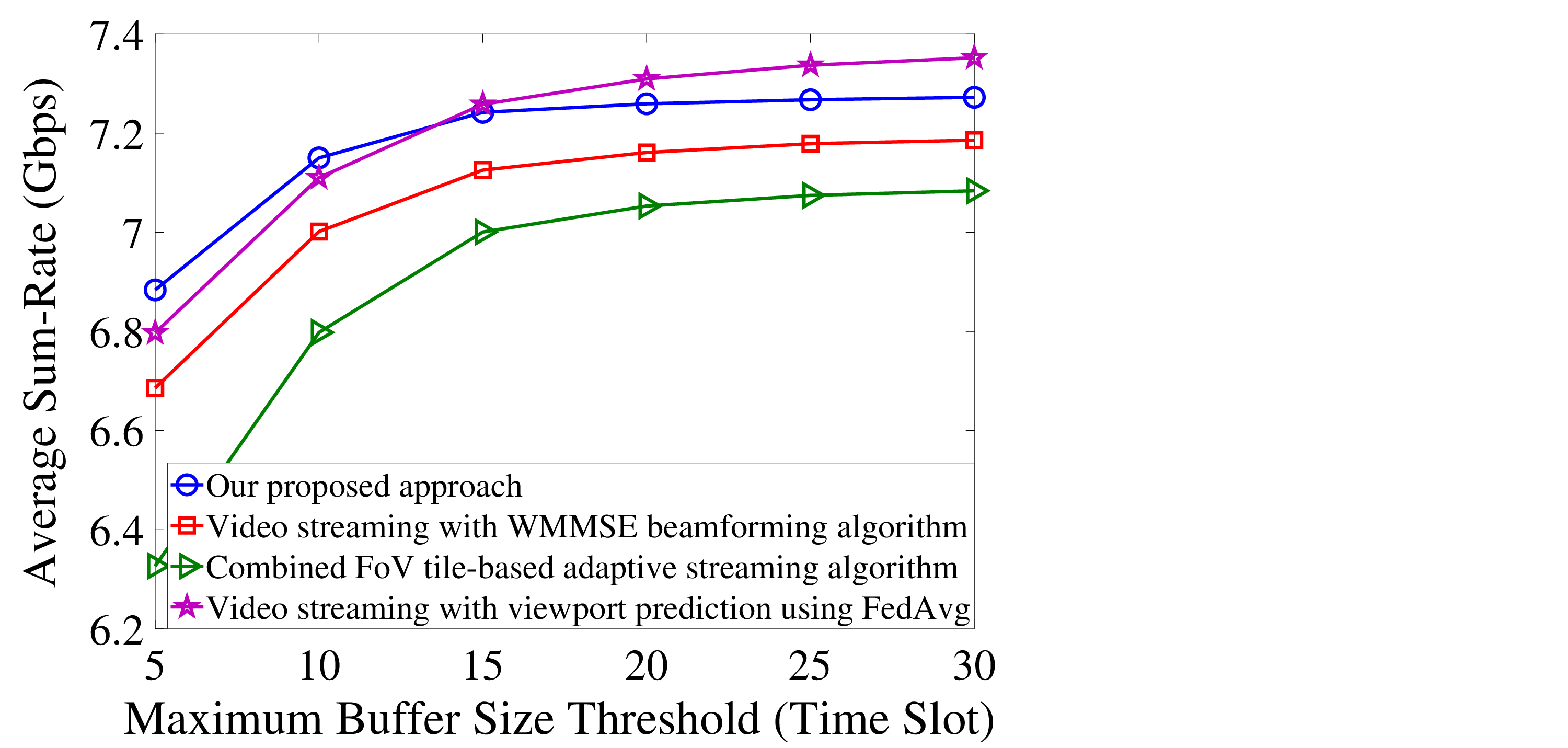}
         \vspace{-1mm}
         \caption{}
         \label{SumR}
     \end{subfigure}\quad
     \begin{subfigure}[b]{0.32\textwidth} 
         \includegraphics[trim=0 0cm 17.6cm 0,clip,scale=0.165]{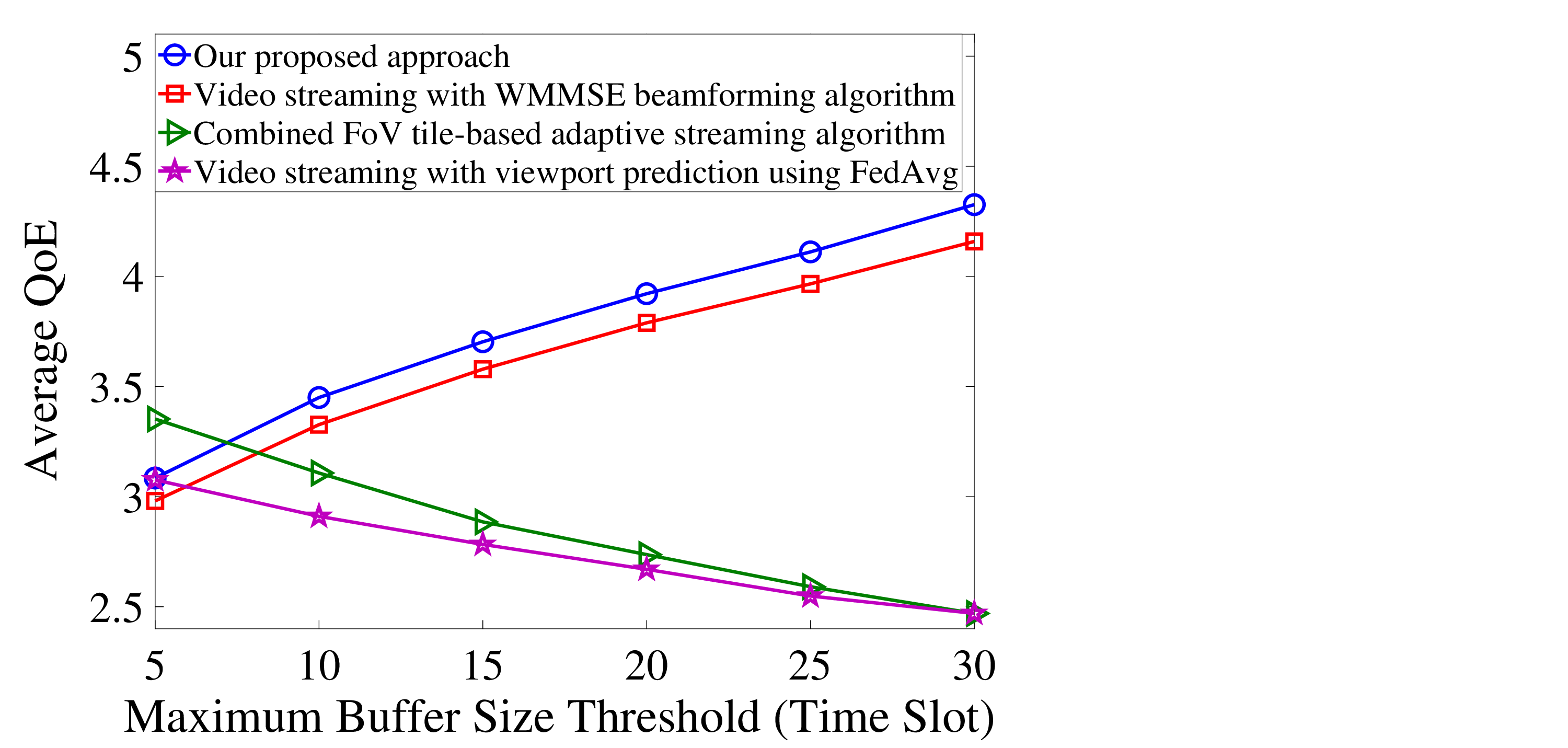}
         \vspace{-1mm}
         \caption{}
         \label{AveQ}
     \end{subfigure}

\caption{(a) Average viewport tile quality, (b) average intra-chunk quality switch, (c) average inter-chunk quality switch, (d) average rebuffering delay, (e)~average sum-rate, and (f) average QoE over users versus the maximum buffer size threshold. We set $B=0.5$ GHz.}
\label{BuffSizeEff}
\end{figure*}

\noindent \textbf{Algorithms Parameters Setting:} For the head movement prediction model, we set $Q^\textrm{hist}=90$ and $d^{\textrm{GRU}}=64$. In Algorithm \ref{TrAlg}, we set $R=50$, $\rho=3$, and $\eta^{\textrm{head}}=0.01$. For each user, we consider $80$ and $24$ videos in the training dataset $\mathcal{V}^{\textrm{head-tr}}_{u}$ and test dataset $\mathcal{V}^{\textrm{head-tst}}_{u}$, respectively. Since all the video frames are available at the server and our proposed viewport prediction framework decouples the saliency detection and head movement prediction models, the saliency map corresponding to each video frame can be obtained and stored at the server in advance. In particular, each user sends its predicted head orientations to the server when requesting a new video chunk. Upon receiving a new video chunk request, the server can retrieve the saliency maps of the video frames corresponding to the requested video chunk, as well as the user's head orientation maps corresponding to those video frames. Given the saliency maps and the user's head orientation maps, the tiles covering the viewport and marginal regions of the requested video frames can be determined using the integration mechanism described in Section~\ref{Integration_technique}, which has a computational complexity of $O(N)$. We empirically determine the value of $\alpha^\textrm{marg}$ by comparing the average tiles overlap and the average number of tiles in the marginal region for various values of $\alpha^\textrm{marg}$. The tiles overlap is obtained as ${\big\lvert \mathcal{N}_{u, c, v}^{\textrm{pred}} \cap \mathcal{N}_{u, c, v}^{\textrm{actual}}\big\lvert}/{\left\lvert \mathcal{N}_{u, c, v}^{\textrm{actual}} \right\rvert}$ for tiles that are predicted to be transmitted for user $u \in \mathcal{U}$ upon its request for chunk~$c \in \mathcal{C}_{v}$ of video $v \in \mathcal{V}^{\textrm{head-tst}}_{u}$. Fig. \ref{saleFactorEff} shows that increasing $\alpha^\textrm{marg}$ leads to a higher average tiles overlap and a higher average number of tiles in the marginal region. To achieve an average tiles overlap above $90 \%$ without consuming excessive network bandwidth on the transmission of additional tiles, we set $\alpha^\textrm{marg}=0.15$. For the agents, we set $d^\textrm{A}_\textrm{h} = 512$ and $d^\textrm{A}_\textrm{fc}=d^\textrm{A}_\textrm{out}=256$. For the joint agent, we set $d^\textrm{J}_\textrm{h} = 1024$ and $d^\textrm{J}_\textrm{fc}=d^\textrm{J}_\textrm{out}=512$. Other simulation parameters are summarized in Table~\ref{sim_par}. 

\begin{table}[t]
    \caption{Simulation Parameters}
    \label{sim_par}
    \centering
    {
    \centering
    \scalebox{0.746}{
    \begin{tabular}{ |c|c||c|c||c|c||c|c| }
      \hline
      Parameter & Value & Parameter & Value & Parameter & Value & Parameter & Value\\ \hhline{|=|=||=|=||=|=||=|=|}
      $\eta^{\textrm{a}}$, $\eta^{\textrm{c}}$            & $10^{-4}$ & $\sigma_{\bm{\varrho}^\textrm{m}}$, $\sigma_{\bm{\varrho}^\textrm{p}}$           & $0.15$      & $E^\textrm{max}$     & $5000$ & $P^{\textrm{max}}$ &  $5$ dBm  \\\hline
      $\varepsilon$        & $10^{-2}$    & $\lambda^{\textrm{spatial}} $, $\lambda^{\textrm{temp}}$                  & $0.5$ & $T^\textrm{max}$   & $153$    & $g_a$ &  $25$ dBi  \\\hline
      $B^\textrm{m}$, $B^\textrm{p}$        & $512$       & $\lambda^{\textrm{RD}}$                     & $0.5$     & $T^\textrm{warm-up}$  & $5200$  & $g_u$ & $15$ dBi  \\\hline
      $\theta_{\bm{\varrho}^\textrm{m}}$, $\theta_{\bm{\varrho}^\textrm{p}}$       & $0.1$          & $\lambda^{\textrm{intr}}$            & $2$ & $\gamma$   &  $0.99$  & $\sigma^2$ &  $-77$ dBm   \\\hline
    \end{tabular}}}
\end{table}

\noindent \textbf{Benchmarks:} We compare the performance of our proposed algorithms with the following algorithms as benchmarks:
\begin{itemize}
    \item \textbf{Video streaming with viewport prediction using FedAvg}: In this algorithm, instead of using our proposed viewport prediction framework, we use FedAvg to train the head movement prediction model as proposed in \cite{liu2021learning1}.
    \item \textbf{Combined FoV tile-based adaptive streaming algorithm \cite{yaqoob2021combined}}: In this algorithm, the viewport is predicted by combining the current chunk's viewport and another viewport obtained by spherical walk. A priority-based bitrate adaptation algorithm is used to select the bitrate of tiles. Each user's data rate is predicted by dividing the number of bits in the previous video chunk by its transmission delay. Since users' head movements are not predicted, we consider that the APs in the self-blockage region of users are determined only after detecting beam failure, a process that takes $300$~ms.
    \item \textbf{Video streaming with WMMSE beamforming algorithm}: In this algorithm, the joint agent employs a WMMSE algorithm \cite{huang2022rate, shi2011iteratively} to obtain the beamforming vectors instead of utilizing our actor-critic algorithm.
\end{itemize}

\vspace{-0.35cm}
\subsection{Experimental Results}
In Fig. \ref{BuffSizeEff}, we study the impact of increasing the maximum buffer size threshold on the 360$^{\circ}$ video streaming system performance. The users can prefetch more video chunks when the buffer size threshold increases. The results in Fig.~\ref{BuffSizeEff}(a) illustrate that as the maximum buffer size threshold increases, the average viewport tile quality decreases for the combined FoV tile-based adaptive streaming algorithm and video streaming with viewport prediction using FedAvg. This is because the increase in the maximum buffer size threshold leads to a degradation in viewport prediction performance for these algorithms. Moreover, such degradation in viewport prediction performance results in lower average intra- and inter-chunk quality switch as shown in Figs.~\ref{BuffSizeEff}(b) and \ref{BuffSizeEff}(c), respectively. However, by transmitting more video tiles based on the value of $\alpha^\textrm{marg}$ and by using the Mac-IAICC approach for tile bitrate selection, our proposed video streaming approach can achieve a higher average viewport tile quality and a lower average intra- and inter-chunk quality switch. By increasing the maximum buffer size threshold, users have more time to prefetch their requested chunks without experiencing video stalling. Thus, as shown in Fig.~\ref{BuffSizeEff}(d), the average rebuffering delay decreases for all the algorithms. In video streaming with viewport prediction using FedAvg, only the predicted tiles for viewport are sent to the users. Thus, this algorithm can achieve a lower average rebuffering delay and a higher average sum-rate as shown in Figs.~\ref{BuffSizeEff}(d) and \ref{BuffSizeEff}(e), respectively. The combined FoV tile-based adaptive streaming algorithm uses a reactive THz beam failure detection. Thus, it has the highest average rebuffering delay and the lowest average sum-rate. Moreover, the results in Figs.~\ref{BuffSizeEff}(d) and \ref{BuffSizeEff}(e) illustrate that the Prim-CAC approach employed by the joint agent in our proposed video streaming approach effectively reduces the average rebuffering delay and increases the average sum-rate compared with using the WMMSE beamforming algorithm. Overall, as shown in Fig.~\ref{BuffSizeEff}(f), our proposed approach provides a higher average QoE compared with the considered benchmarks when the maximum buffer size threshold increases.

\begin{figure}[t]
    \centering
    \includegraphics[trim=0 0cm 15.8cm 0,clip,scale=0.21]{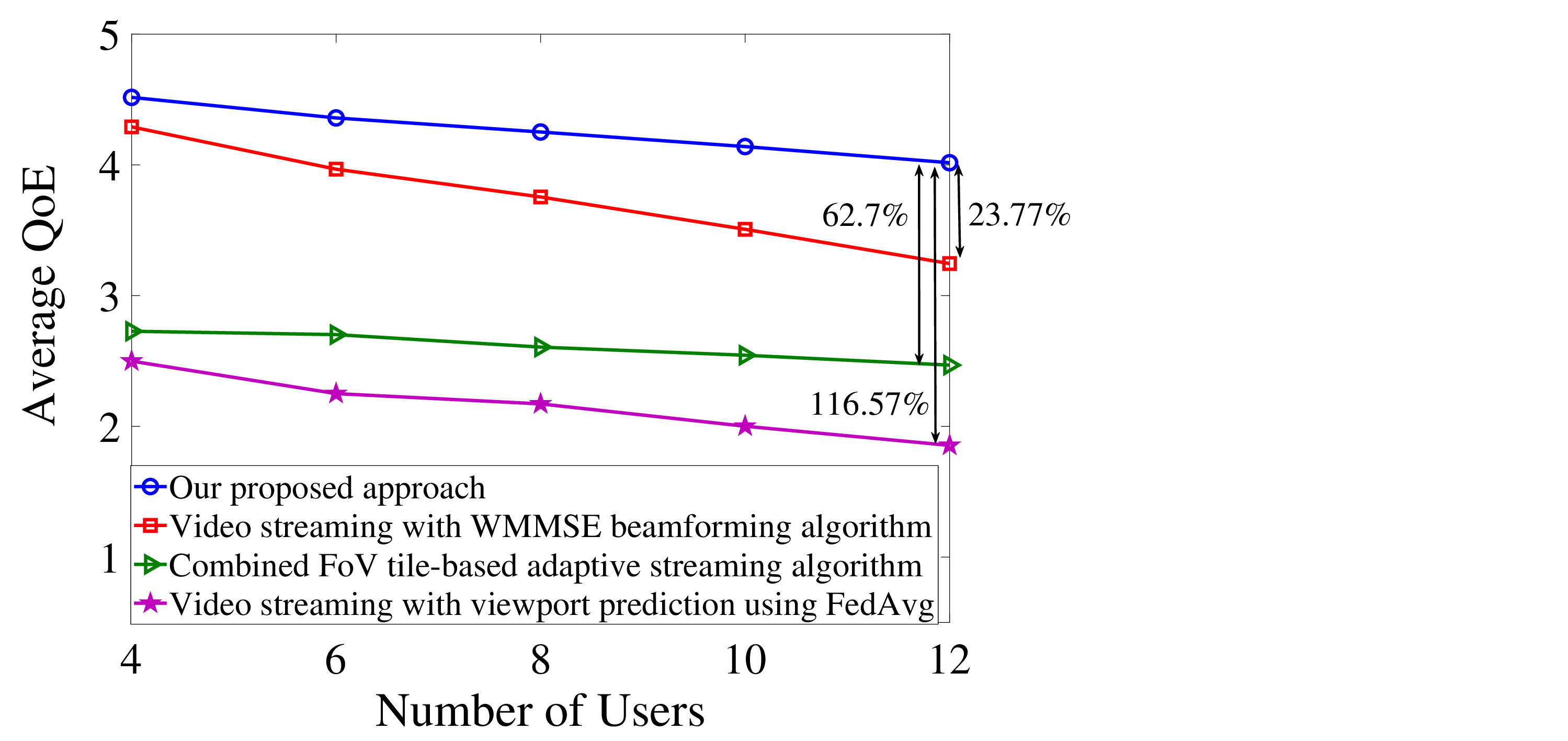}
    \caption{Average QoE over users when the number of users is increased. To support more users, we set $B=5$ GHz and $\max_{u\in \mathcal{U}}{B_u^\textrm{THR}} = 30$ time slots.}
    \label{UserNumResult}
    \vspace{0.13cm}
\end{figure}

In Fig. \ref{UserNumResult}, we investigate the impact of increasing the number of users on average QoE. With more users, designing the beamforming vectors becomes more challenging due to the limited bandwidth of the APs and the self-blockage region of the users. The results in Fig.~\ref{UserNumResult} illustrate that when the number of users is equal to $12$, our proposed video streaming approach can achieve an average QoE which is $23.77\%$, $62.7\%$, and $116.57\%$ higher than that~of the video streaming with WMMSE beamforming algorithm, combined FoV tile-based adaptive streaming algorithm, and video streaming with viewport prediction using FedAvg, respectively. 

\begin{figure}
     \begin{subfigure}[b]{0.498\textwidth}
     \centering
         \includegraphics[trim=0 0cm 16.2cm 0,clip,scale=0.182]{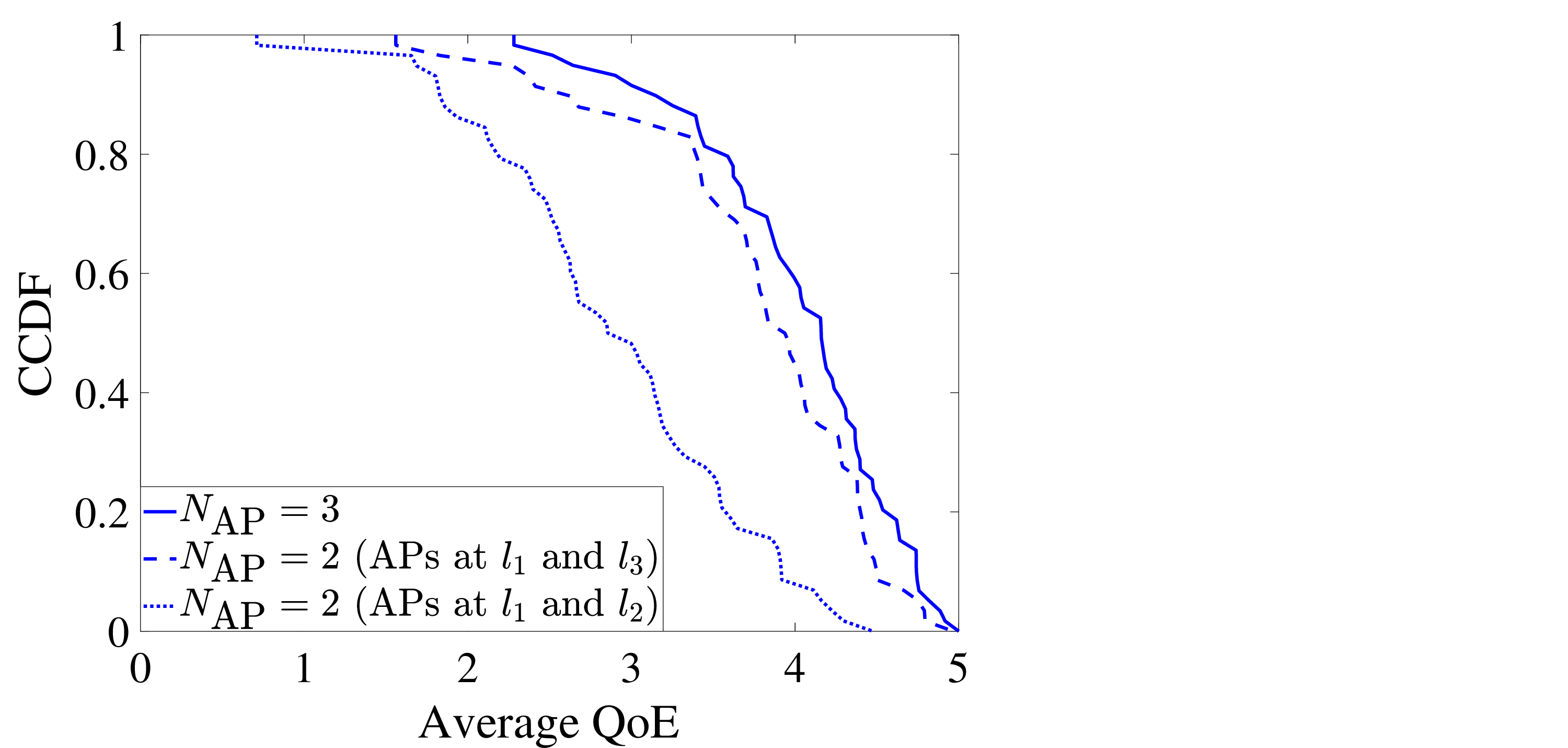}
         \vspace{-1mm}
         \caption{}
         \label{APsQoE}
     \end{subfigure}\vspace{0.1cm}
     \begin{subfigure}[b]{0.498\textwidth} 
     \centering
         \includegraphics[trim=0 0cm 16.2cm 0,clip,scale=0.182]{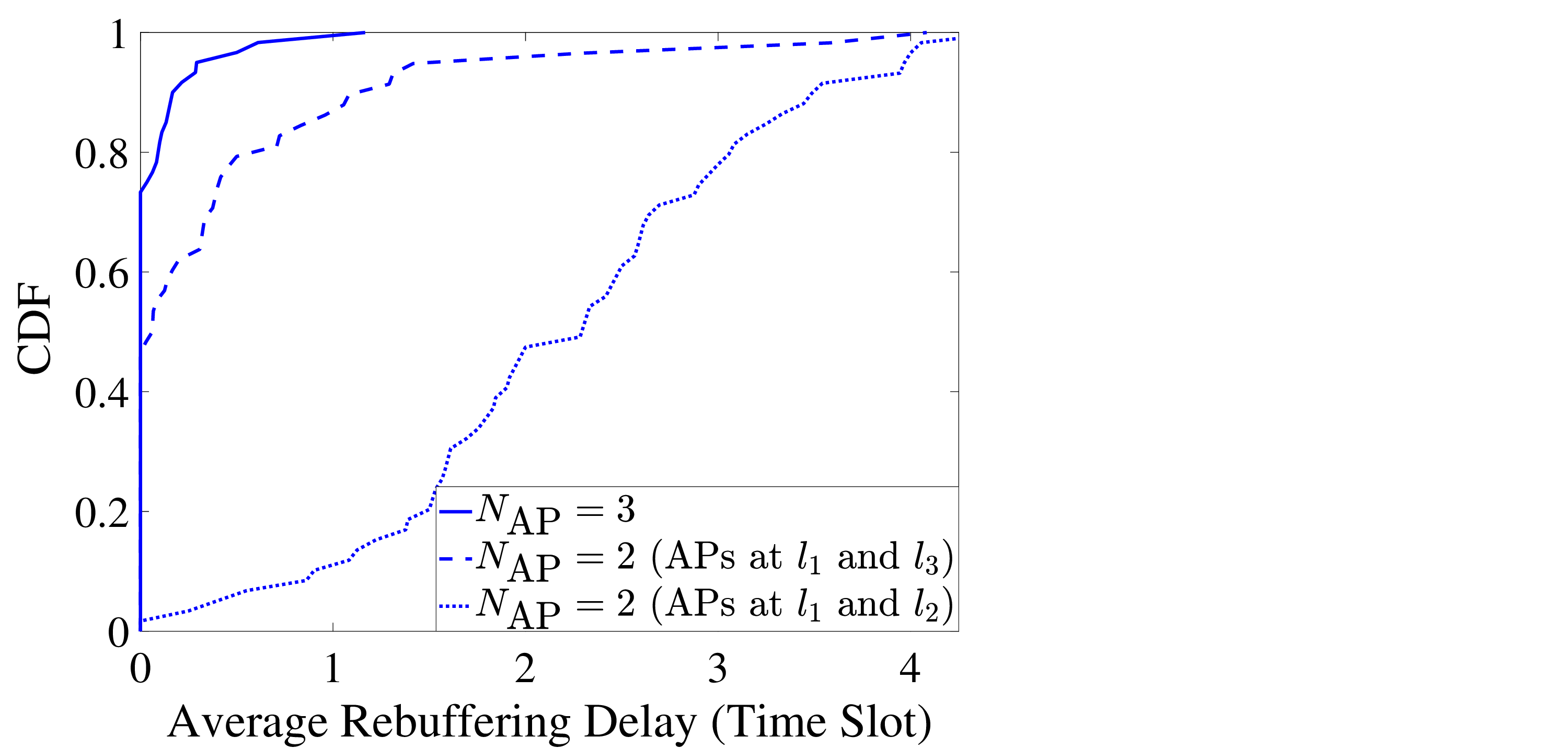}
         \vspace{-1mm}
         \caption{}
         \label{APsBuff}
     \end{subfigure}
\caption{(a) CCDF of the average QoE and (b) CDF of the average rebuffering delay. We set $B=0.5$ GHz and $\max_{u\in \mathcal{U}}{B_u^\textrm{THR}} = 30$ time slots.}
\label{APsNum}
\end{figure}

In Fig. \ref{APsNum}, we show the impact of the number and location of APs on the performance of the considered THz-enabled 360$^{\circ}$ video streaming system. Due to self-blockage, the system requires more than one AP. Also, as shown in Fig.~\ref{SysModel}, there is a symmetry in the APs' locations. Thus, we consider the following three scenarios: all three APs are available; only two APs located at $\bm{l}_1=\left(9,1,4 \right)$ and $\bm{l}_3=\left(1,9,4 \right)$ are available; and only two APs located at $\bm{l}_1=\left(9,1,4 \right)$ and $\bm{l}_2=\left(5,5,4 \right)$ are available to transmit the video tiles to the users. Fig.~\ref{APsNum}(a) illustrates the complementary cumulative distribution function (CCDF) of the average QoE. A higher CCDF indicates a higher probability of obtaining an average QoE above a given threshold value. Fig.~\ref{APsNum}(b) presents the cumulative distribution function (CDF) of the average rebuffering delay for each of the considered scenarios. A higher CDF indicates a higher probability of obtaining an average rebuffering delay below a given threshold value. The results in Fig.~\ref{APsNum} show that with three APs, we can consistently achieve a higher average QoE ($> 2.28$) and a lower average rebuffering delay ($<1.67$ time slots) with a higher probability compared to scenarios with only two APs. Furthermore, Fig.~\ref{APsNum} illustrates that the APs' locations have a significant impact on the system performance when only two APs are available for video tile transmission to the users. In particular, locating two APs at $\bm{l}_1$ and $\bm{l}_3$ can better cope with self-blockage. For example, as shown in Fig.~\ref{APsNum}(b), the probability of having an average rebuffering delay below $2$~time slots is approximately $0.96$ when APs are located at $\bm{l}_1$ and $\bm{l}_3$, while this probability is close to $0.47$ when APs are located at $\bm{l}_1$ and $\bm{l}_2$. Thus, a more reliable transmission of 360$^\circ$ videos over THz links is provided if the two APs are located at $\bm{l}_1$ and $\bm{l}_3$.

\begin{figure}[t]
    \centering
    \includegraphics[trim=0 0cm 24.5cm 0,clip,scale=0.16]{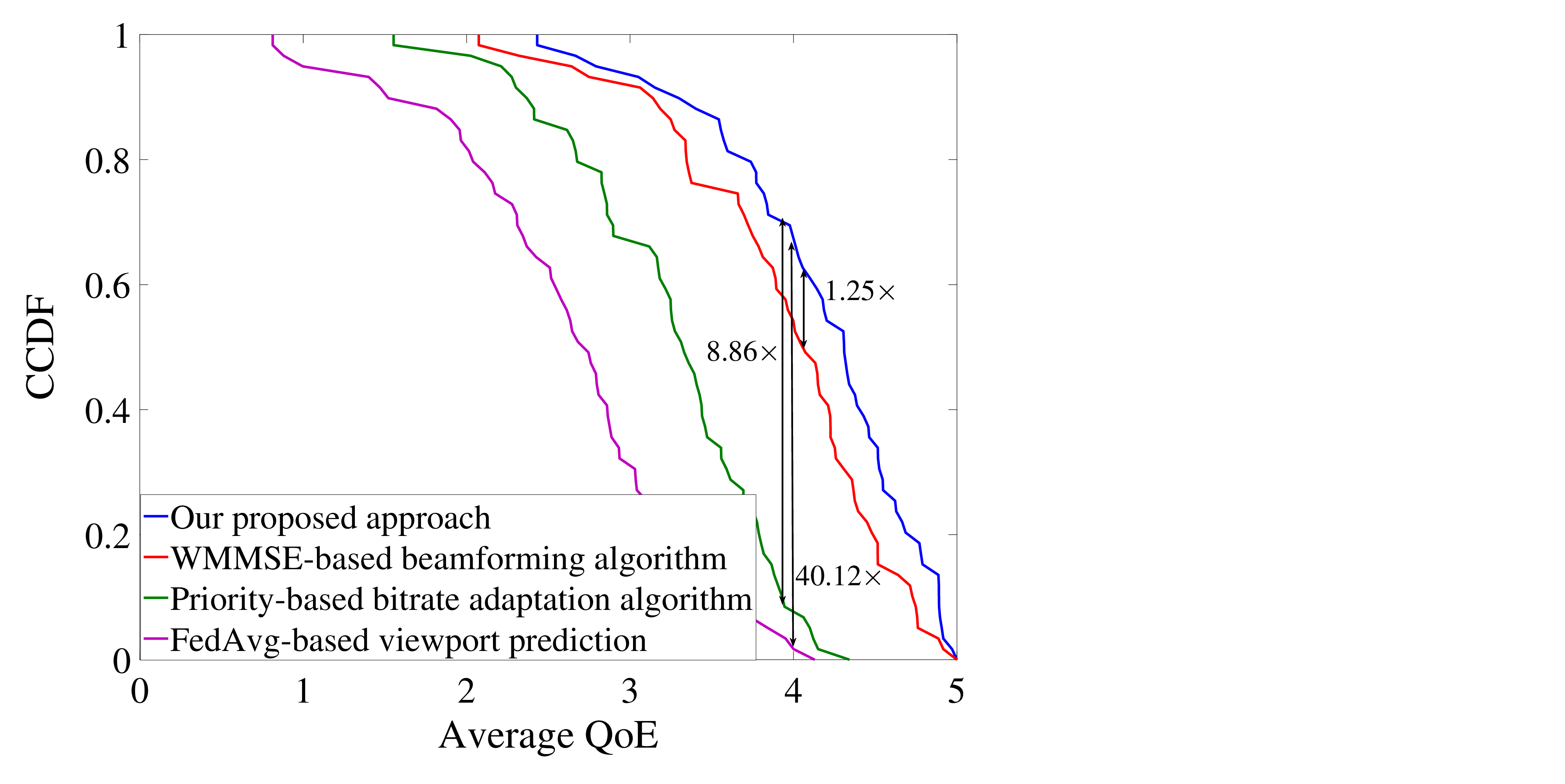}
    \caption{CCDF of the average QoE across users. We set $B=0.5$ GHz and $\max_{u\in \mathcal{U}}{B_u^\textrm{THR}} = 30$ time slots.}
    \label{CCDF_Abl}
    \vspace{-3mm}
\end{figure}

\begin{figure}[t]
    \centering
    \includegraphics[trim=0.5cm 6cm 20cm 0,clip,scale=0.25]{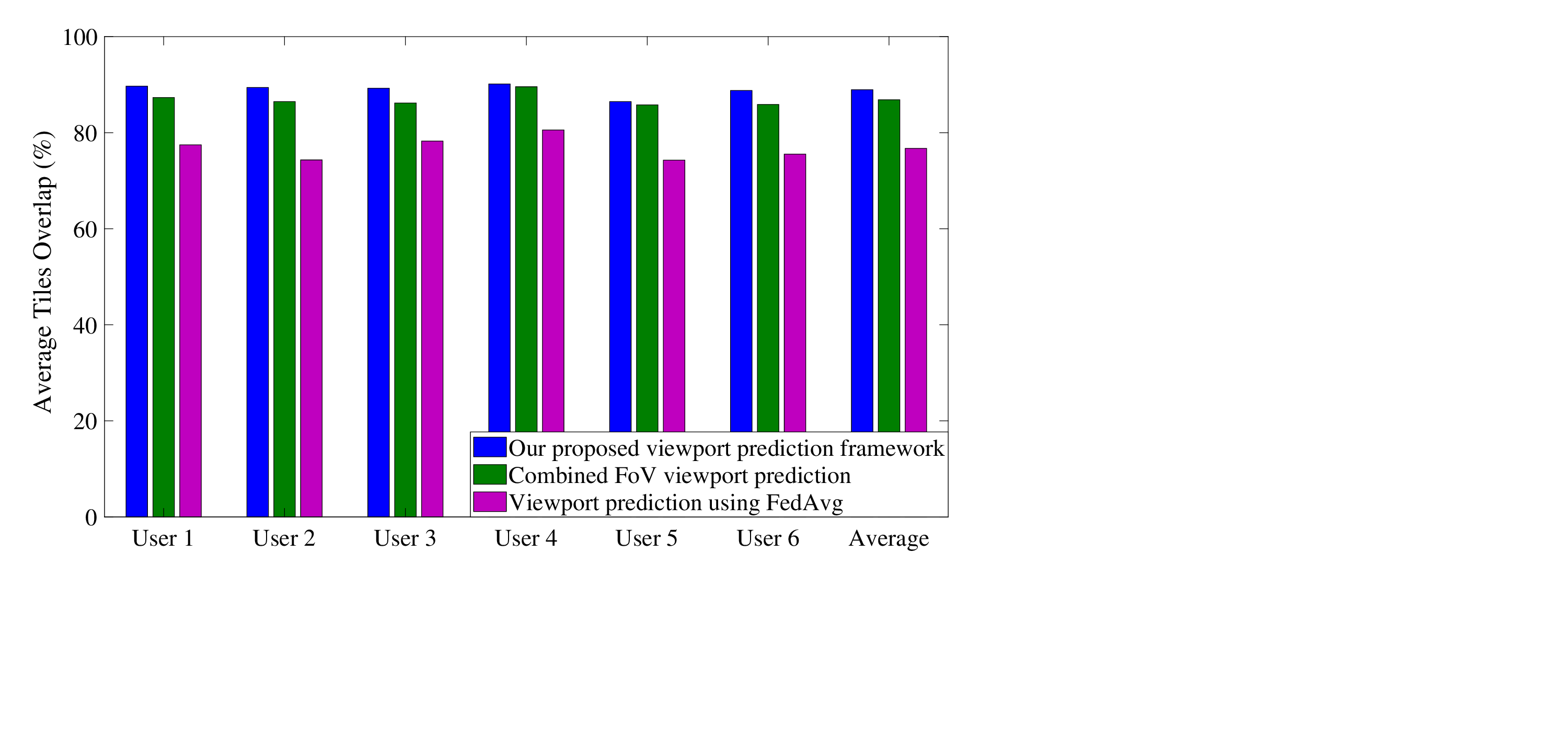}
    \caption{Average tiles overlap of our proposed viewport prediction framework for different users and the average tiles overlap obtained by averaging across users compared to the considered benchmarks.}
    \label{viewportResult}
\end{figure}

As shown in Fig.~\ref{our_approach}, our proposed 360$^{\circ}$ video streaming approach consists of three algorithms: a viewport prediction framework, a multi-agent DDPG algorithm for bitrate selection, and a multi-agent DDPG algorithm for beamforming design. To demonstrate the performance gains contributed by each algorithm in our approach, we assess their impact by fixing two of our proposed algorithms and replacing the third one with an existing algorithm from the literature. Fig.~\ref{CCDF_Abl} illustrates the CCDF of the average QoE for our proposed approach compared with three benchmarks. In the first benchmark, we replace the DDPG-based beamforming design algorithm with a WMMSE-based beamforming algorithm. In the second benchmark, we replace the DDPG-based bitrate selection algorithm with a priority-based bitrate adaptation algorithm. In the third benchmark, our proposed viewport prediction framework is replaced by a FedAvg-based viewport prediction algorithm. As shown in Fig.~\ref{CCDF_Abl}, the probability of achieving an average QoE above $4$ with our proposed approach is $1.25$ times higher than with the WMMSE-based beamforming algorithm, $8.86$ times higher than with the priority-based bitrate adaptation algorithm, and $40.12$ times higher than with the FedAvg-based viewport prediction. Thus, each of the algorithms comprising our proposed approach definitely has its own merits on providing the users with an immersive viewing experience.\par

In Fig. \ref{viewportResult}, we compare the performance of our proposed viewport prediction framework with the algorithms proposed in \cite{yaqoob2021combined} and \cite{liu2021learning1}. The results in Fig.~\ref{viewportResult} show that our proposed content-based viewport prediction framework can achieve an average tiles overlap which is on average $2.41 \%$ and $15.93\%$ higher than that of the viewport prediction algorithms proposed in \cite{yaqoob2021combined} and \cite{liu2021learning1}, respectively. \par

\begin{figure}[t]
    \centering
    \includegraphics[trim=0.5cm 6cm 19cm 0,clip,scale=0.2]{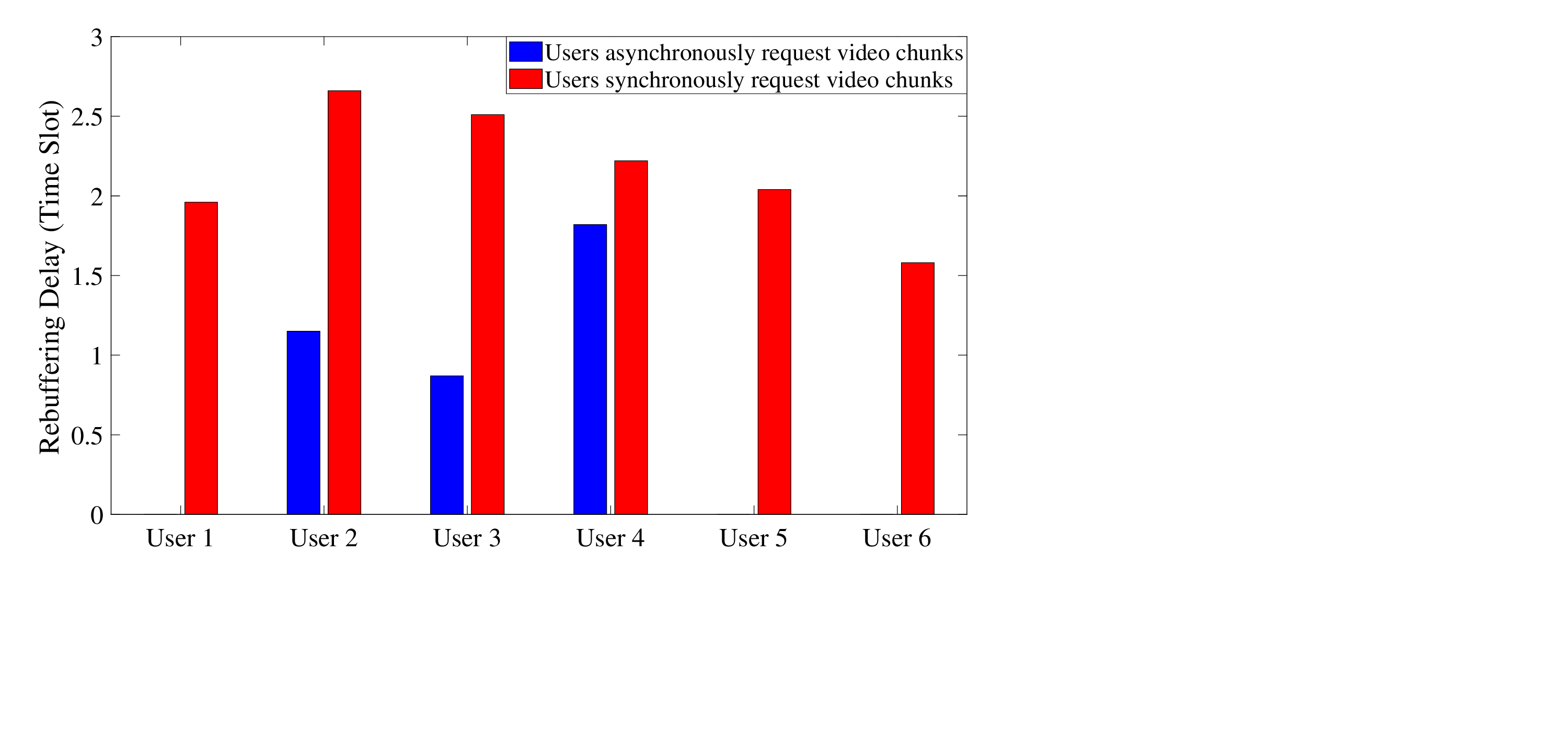}
    \caption{Rebuffering delay of each user. We set $B=0.5$ GHz.}
    \label{AsynchBuff}
\end{figure}

In Fig.~\ref{AsynchBuff}, we show the rebuffering delay of each user for two scenarios. The first scenario is when users request their next video chunks asynchronously. The second scenario is when they request them synchronously. The results in Fig. 15 illustrate that asynchronous video chunk requests can provide a lower rebuffering delay for each user (even zero rebuffering delay for users 1, 5, and 6) compared with synchronous video chunk requests. This is due to the fact that in synchronous video chunk request: (a)~each user may need to wait an extra amount of time during which the user watches the prefetched video tiles in its buffer, leading to a lower value of $B_{u}\left(\tau_{u,c,v}^\textrm{REQ}\right)$; and (b)~all users request their next video chunk simultaneously and compete for the limited wireless resources, leading to a higher transmission delay $\tau_{u,c,v}^\textrm{TD}$.

\section{Conclusion}\label{conclusin_Sec}
In this paper, we proposed a content-based viewport prediction framework for 360$^{\circ}$ videos. To address users’ privacy concerns and the data heterogeneity issue, our proposed framework employs a PFL algorithm for training the users' head movement prediction models. We applied the proposed viewport prediction framework in a THz-enabled 360$^{\circ}$ video streaming system to determine which video tiles should be transmitted to the users. We modeled the bitrate selection for video tiles and the design of beamforming vectors for the APs, as a MacDec-POMDP. To solve this problem, we proposed a hierarchical DRL framework consisting of two multi-agent DDPG algorithms. We determined a policy for tile bitrate selection using the Mac-IAICC approach and a policy for beamforming design using the Prim-CAC approach. The results on a public 360$^{\circ}$ video dataset showed that our proposed video streaming approach provides a higher QoE for the users compared with three benchmark algorithms. One direction for future work is to employ our proposed video streaming approach in Metaverse systems, where users are mobile and uplink transmission is used to send their motion tracking data.

\bibliographystyle{IEEEtran}
\footnotesize
\bibliography{ref}

\begin{thebibliography}{10}
\providecommand{\url}[1]{#1}
\csname url@samestyle\endcsname
\providecommand{\newblock}{\relax}
\providecommand{\bibinfo}[2]{#2}
\providecommand{\BIBentrySTDinterwordspacing}{\spaceskip=0pt\relax}
\providecommand{\BIBentryALTinterwordstretchfactor}{4}
\providecommand{\BIBentryALTinterwordspacing}{\spaceskip=\fontdimen2\font plus
\BIBentryALTinterwordstretchfactor\fontdimen3\font minus
  \fontdimen4\font\relax}
\providecommand{\BIBforeignlanguage}[2]{{%
\expandafter\ifx\csname l@#1\endcsname\relax
\typeout{** WARNING: IEEEtran.bst: No hyphenation pattern has been}%
\typeout{** loaded for the language `#1'. Using the pattern for}%
\typeout{** the default language instead.}%
\else
\language=\csname l@#1\endcsname
\fi
#2}}
\providecommand{\BIBdecl}{\relax}
\BIBdecl

\bibitem{setayesh2023PredFramework}
M.~Setayesh and V.~W.S.~Wong, ``A content-based viewport prediction framework
  for 360$^{\circ}$ video using personalized federated learning and fusion
  techniques,'' in \emph{Proc. of Int'l Conf. on Multimedia and Expo (ICME)},
  Brisbane, Australia, Jul. 2023.

\bibitem{mehdiICC}
------, ``Asynchronous {DRL}-based bitrate selection for 360-degree video
  streaming over {THz} wireless systems,'' in \emph{Proc. of IEEE Int'l Conf.
  Commun. (ICC)}, Denver, CO, Jun. 2024.

\bibitem{zhang2021buffer}
R.~Zhang, J.~Liu, F.~Liu, T.~Huang, Q.~Tang, S.~Wang, and F.~R. Yu,
  ``{Buffer-aware} virtual reality video streaming with personalized and
  private viewport prediction,'' \emph{IEEE J. Sel. Areas in Commun.}, vol.~40,
  no.~2, pp. 694--709, Feb. 2022.

\bibitem{chaccour2022can}
C.~Chaccour, M.~N. Soorki, W.~Saad, M.~Bennis, and P.~Popovski, ``Can
  {terahertz} provide high-rate reliable low-latency communications for
  wireless {VR}?'' \emph{IEEE Internet Things J.}, vol.~9, no.~12, pp.
  9712--9729, Jun. 2022.

\bibitem{shafie2021coverage}
A.~Shafie, N.~Yang, S.~Durrani, X.~Zhou, C.~Han, and M.~Juntti, ``Coverage
  analysis for {3D} {terahertz} communication systems,'' \emph{IEEE J. Sel.
  Areas Commun.}, vol.~39, no.~6, pp. 1817--1832, Jun. 2021.

\bibitem{shafie2021spectrum}
A.~Shafie, N.~Yang, S.~A. Alvi, C.~Han, S.~Durrani, and J.~M. Jornet,
  ``Spectrum allocation with adaptive sub-band bandwidth for {terahertz}
  communication systems,'' \emph{IEEE Trans. Commun.}, vol.~70, no.~2, pp.
  1407--1422, Feb. 2022.

\bibitem{kan2021rapt360}
N.~Kan, J.~Zou, C.~Li, W.~Dai, and H.~Xiong, ``{RAPT360}: Reinforcement
  learning-based rate adaptation for 360-degree video streaming with adaptive
  prediction and tiling,'' \emph{IEEE Trans. Circuits Syst. Video Technol.},
  vol.~32, no.~3, pp. 1607--1623, Mar. 2022.

\bibitem{yaqoob2021combined}
A.~Yaqoob and G.-M. Muntean, ``A combined field-of-view prediction-assisted
  viewport adaptive delivery scheme for 360$^{\circ}$ videos,'' \emph{IEEE
  Trans. Broadcast.}, vol.~67, no.~3, pp. 746--760, Sept. 2021.

\bibitem{li2022spherical}
J.~Li, L.~Han, C.~Zhang, Q.~Li, and Z.~Liu, ``Spherical convolution empowered
  viewport prediction in 360 video multicast with limited {FoV} feedback,''
  \emph{ACM Trans. on Multimedia Comput. Commun. Appl.}, vol.~19, no.~1, pp.
  1--23, Jan. 2023.

\bibitem{qian2018flare}
F.~Qian, B.~Han, Q.~Xiao, and V.~Gopalakrishnan, ``Flare: Practical
  viewport-adaptive 360-degree video streaming for mobile devices,'' in
  \emph{Proc. of ACM Int'l Conf. Mobile Comput. Netw. (MobiCom)}, New Delhi,
  India, Oct. 2018.

\bibitem{maniotis2019tile}
P.~Maniotis, E.~Bourtsoulatze, and N.~Thomos, ``Tile-based joint caching and
  delivery of 360 videos in heterogeneous networks,'' \emph{IEEE Trans.
  Multimedia}, vol.~22, no.~9, pp. 2382--2395, Sept. 2020.

\bibitem{xiao2022asynchronous}
Y.~Xiao, W.~Tan, and C.~Amato, ``Asynchronous actor-critic for multi-agent
  reinforcement learning,'' in \emph{Proc. of Conf. Neural Inf. Process. Syst.
  (NIPS)}, New Orleans, LA, Nov. 2022.

\bibitem{lyu2022multi}
X.~Lyu, A.~Banitalebi-Dehkordi, M.~Chen, and Y.~Zhang, ``Asynchronous,
  option-based multi-agent policy gradient: {A} conditional reasoning
  approach,'' in \emph{Proc. of IEEE/RSJ Int'l Conf. on Intell. Robots Syst.
  (IROS)}, Detroit, MI, Oct. 2023.

\bibitem{zhang2018saliency}
Z.~Zhang, Y.~Xu, J.~Yu, and S.~Gao, ``Saliency detection in 360$^{\circ}$
  videos,'' in \emph{Proc. of Eur. Conf. on Comput. Vis.}, Munich, Germany,
  Sept. 2018.

\bibitem{liu2021learning1}
X.~Liu, Y.~Deng, C.~Han, and M.~Di~Renzo, ``Learning-based prediction,
  rendering and transmission for interactive virtual reality in {RIS}-assisted
  terahertz networks,'' \emph{IEEE J. Sel. Areas in Commun.}, vol.~40, no.~2,
  pp. 710--724, Feb. 2022.

\bibitem{chao2021transformer}
F.-Y. Chao, C.~Ozcinar, and A.~Smolic, ``Transformer-based long-term viewport
  prediction in 360$^{\circ}$ video: {Scanpath} is all you need.'' in
  \emph{Proc. of Int’l Workshop Multimedia Signal Process. (MMSP)}, Tampere,
  Finland, Oct. 2021.

\bibitem{liu2021learning}
X.~Liu, X.~Li, and Y.~Deng, ``Learning-based prediction and proactive uplink
  retransmission for wireless virtual reality network,'' \emph{IEEE Trans. Veh.
  Technol.}, vol.~70, no.~10, pp. 10\,723--10\,734, Oct. 2021.

\bibitem{mcmahan2017communication}
B.~McMahan, E.~Moore, D.~Ramage, S.~Hampson, and B.~A. y~Arcas,
  ``Communication-efficient learning of deep networks from decentralized
  data,'' in \emph{Proc. of Int'l Conf. Artificial Intelligence and Statistics
  (AISTATS)}, Ft. Lauderdale, FL, Apr. 2017.

\bibitem{oh2021fedbabu}
J.~Oh, S.~Kim, and S.-Y. Yun, ``{FedBABU}: {Towards} enhanced representation
  for federated image classification,'' in \emph{Proc. of Int'l Conf. on
  Learning Representations (ICLR)}, Apr. 2022.

\bibitem{nguyen2018your}
A.~Nguyen, Z.~Yan, and K.~Nahrstedt, ``Your attention is unique: {Detecting}
  360-degree video saliency in head-mounted display for head movement
  prediction,'' in \emph{Proc. of ACM Int'l Conf. on Multimedia}, Seoul,
  Republic of Korea, Oct. 2018.

\bibitem{wu2020spherical}
C.~Wu, R.~Zhang, Z.~Wang, and L.~Sun, ``A spherical convolution approach for
  learning long term viewport prediction in {360} immersive video,'' in
  \emph{Proc. of AAAI Conf. on Artif. Intell.}, New York, NY, Feb. 2020.

\bibitem{hu2020cellular}
F.~Hu, Y.~Deng, W.~Saad, M.~Bennis, and A.~H. Aghvami, ``Cellular-connected
  wireless virtual reality: {Requirements}, challenges, and solutions,''
  \emph{IEEE Commun. Mag.}, vol.~58, no.~5, pp. 105--111, May 2020.

\bibitem{zhao2022optimization}
L.~Zhao, Y.~Cui, S.~Yang, and S.~S. Shitz, ``An optimization framework for
  general rate splitting for general multicast,'' \emph{IEEE Trans. Wireless
  Commun.}, vol.~22, no.~3, pp. 1573--1587, Mar. 2022.

\bibitem{yang2022feeling}
P.~Yang, T.~Q. Quek, J.~Chen, C.~You, and X.~Cao, ``Feeling of presence
  maximization: {mmWave}-enabled virtual reality meets deep reinforcement
  learning,'' \emph{IEEE Trans. Wireless Commun.}, vol.~21, no.~11, pp.
  10\,005--10\,019, Nov. 2022.

\bibitem{huang2022rate}
R.~Huang, V.~W.S.~Wong, and R.~Schober, ``Rate-splitting for intelligent
  reflecting surface-aided multiuser {VR} streaming,'' \emph{IEEE J. Sel. Areas
  Commun.}, vol.~41, no.~5, pp. 1516--1535, May 2023.

\bibitem{perfecto2020taming}
C.~Perfecto, M.~S. Elbamby, J.~Del~Ser, and M.~Bennis, ``Taming the latency in
  multi-user {VR} 360$^{\circ}$: {A} {QoE-aware} deep learning-aided multicast
  framework,'' \emph{IEEE Trans. Commun.}, vol.~68, no.~4, pp. 2491--2508, Apr.
  2020.

\bibitem{zhang2020adaptive}
J.~Zhang, H.~Wu, X.~Tao, and X.~Zhang, ``Adaptive bitrate video streaming in
  non-orthogonal multiple access networks,'' \emph{IEEE Trans. Veh. Technol.},
  vol.~69, no.~4, pp. 3980--3993, Apr. 2020.

\bibitem{zhang2019drl360}
Y.~Zhang, P.~Zhao, K.~Bian, Y.~Liu, L.~Song, and X.~Li, ``{DRL360}: 360-degree
  video streaming with deep reinforcement learning,'' in \emph{Proc. of IEEE
  Conf. Comput. Commun. (INFOCOM)}, Paris, France, Apr. 2019.

\bibitem{xiao2019deepvr}
G.~Xiao, M.~Wu, Q.~Shi, Z.~Zhou, and X.~Chen, ``{DeepVR}: {Deep} reinforcement
  learning for predictive panoramic video streaming,'' \emph{IEEE Trans. Cogn.
  Commun. Netw.}, vol.~5, no.~4, pp. 1167--1177, Dec. 2019.

\bibitem{jornet2011channel}
J.~M. Jornet and I.~F. Akyildiz, ``Channel modeling and capacity analysis for
  electromagnetic wireless nanonetworks in the {terahertz} band,'' \emph{IEEE
  Trans. Wireless Commun.}, vol.~10, no.~10, pp. 3211--3221, Oct. 2011.

\bibitem{mehrabian2024joint}
A.~Mehrabian and V.~W.S.~Wong, ``Joint spectrum, precoding, and phase shifts
  design for {RIS}-aided multiuser {MIMO} {THz} systems,'' \emph{IEEE Trans.
  Commun.}, vol.~72, no.~8, pp. 5087--5101, Aug. 2024.

\bibitem{hu2021virtual}
M.~Hu, X.~Luo, J.~Chen, Y.~C. Lee, Y.~Zhou, and D.~Wu, ``Virtual reality: {A}
  survey of enabling technologies and its applications in {IoT},'' \emph{J.
  Netw. Comput. Appl.}, vol. 178, p. 102970, Mar. 2021.

\bibitem{shi2011iteratively}
Q.~Shi, M.~Razaviyayn, Z.-Q. Luo, and C.~He, ``{An} iteratively weighted {MMSE}
  approach to distributed sum-utility maximization for a {MIMO} interfering
  broadcast channel,'' \emph{IEEE Trans. Signal Process.}, vol.~59, no.~9, pp.
  4331--4340, Sept. 2011.

\bibitem{zhang2021joint}
Z.~Zhang, M.~Zeng, M.~Chen, D.~Liu, W.~Saad, S.~Cui, and H.~V. Poor, ``Joint
  user grouping, version selection, and bandwidth allocation for live video
  multicasting,'' \emph{IEEE Trans. Commun.}, vol.~70, no.~1, pp. 350--365,
  Jan. 2022.

\bibitem{long2020optimal}
K.~Long, Y.~Cui, C.~Ye, and Z.~Liu, ``Optimal wireless streaming of
  multi-quality 360 {VR} video by exploiting natural, relative
  smoothness-enabled, and transcoding-enabled multicast opportunities,''
  \emph{IEEE Trans. Multimedia}, vol.~23, pp. 3670--3683, Oct. 2021.

\bibitem{han2022off}
B.~Han, Z.~Ren, Z.~Wu, Y.~Zhou, and J.~Peng, ``Off-policy reinforcement
  learning with delayed rewards,'' in \emph{Proc. of Int'l Conf. on Machine
  Learning (ICML)}, Baltimore, MD, Jul. 2022.

\bibitem{yu2023asynchronous}
C.~Yu, X.~Yang, J.~Gao, J.~Chen, Y.~Li, J.~Liu, Y.~Xiang, R.~Huang, H.~Yang,
  Y.~Wu, and Y.~Wang, ``Asynchronous multi-agent reinforcement learning for
  efficient real-time multi-robot cooperative exploration,'' in \emph{Proc. of
  Int'l Conf. on Auton. Agents and Multiagent Syst. (AAMAS)}, London, United
  Kingdom, May 2023.

\bibitem{lillicrap2015continuous}
T.~P. Lillicrap, J.~J. Hunt, A.~Pritzel, N.~Heess, T.~Erez, Y.~Tassa,
  D.~Silver, and D.~Wierstra, ``Continuous control with deep reinforcement
  learning,'' in \emph{Proc. of Int'l Conf. on Learning Representations
  (ICLR)}, San Juan, Puerto Rico, May 2016.

\bibitem{kulkarni2016hierarchical}
T.~D. Kulkarni, K.~Narasimhan, A.~Saeedi, and J.~Tenenbaum, ``Hierarchical deep
  reinforcement learning: {Integrating} temporal abstraction and intrinsic
  motivation,'' in \emph{Proc. of Conf. Neural Inf. Process. Syst. (NIPS)},
  Barcelona, Spain, Dec. 2016.

\bibitem{yun2022panoramic}
H.~Yun, S.~Lee, and G.~Kim, ``Panoramic vision transformer for saliency
  detection in 360$^{\circ}$ videos,'' in \emph{Proc. of Eur. Conf. on Comput.
  Vis. (ECCV)}, Tel Aviv, Israel, Oct. 2022.

\bibitem{kingma2014adam}
D.~P. Kingma and J.~Ba, ``Adam: {A} method for stochastic optimization,'' in
  \emph{Proc. of Int’l Conf. Learning Representations (ICLR)}, San Diego, CA,
  May 2015.

\bibitem{setayesh2023perfedmask}
M.~Setayesh, X.~Li, and V.~W.S.~Wong, ``{PerFedMask}: {Personalized} federated
  learning with optimized masking vectors,'' in \emph{Proc. of Int'l Conf. on
  Learning Representations (ICLR)}, Kigali, Rwanda, May 2023.

\bibitem{teng2021qoe}
L.~Teng, G.~Zhai, Y.~Wu, X.~Min, W.~Zhang, Z.~Ding, and C.~Xiao, ``{QoE} driven
  {VR} 360$^{\circ}$ video massive {MIMO} transmission,'' \emph{IEEE Trans.
  Wireless Commun.}, vol.~21, no.~1, pp. 18--33, Jan. 2022.

\bibitem{lowe2017multi}
R.~Lowe, Y.~Wu, A.~Tamar, J.~Harb, P.~Abbeel, and I.~Mordatch, ``Multi-agent
  actor-critic for mixed cooperative-competitive environments,'' in \emph{Proc.
  of Conf. Neural Inf. Process. Syst. (NIPS)}, Long Beach, CA, Dec. 2017.

\end{thebibliography}

\vfill

\end{document}